\documentclass[12pt,a4paper]{article}

\pdfoutput=1

\usepackage{chngcntr}
\usepackage{jheppub,psfrag,slashed,cancel,lscape,caption,array,graphicx,subcaption}
\usepackage[normalem]{ulem}
\usepackage[utf8]{inputenc}
\usepackage{amsmath}
\usepackage{nicefrac}  
\usepackage{mathtools}
\usepackage{mathrsfs}
\usepackage{multirow}
\usepackage{booktabs} 
\usepackage{nicefrac}
\usepackage{cleveref}
\usepackage{orcidlink}
\usepackage{relsize}

\def\dd{\delta\!\!{}^-\!}
\newcommand{\bse}{\begin{subequations}}
\newcommand{\ese}{\end{subequations}}

\newcommand\eps{\epsilon}

\def\cO{\mathcal{O}}

\def\cH{\mathcal{H}}

\def\eps{\epsilon}

\def\d{\mathrm{d}}

\newcommand\nn{\nonumber}

\usepackage{tikz}
\usetikzlibrary{decorations.pathmorphing}
\usetikzlibrary{decorations.markings}
\usetikzlibrary{positioning, shapes, snakes, arrows}
\usetikzlibrary{patterns}

\tikzset{
    wl/.style={line width=1pt},
    graviton/.style={line width=.8pt, -latex,decorate, decoration={snake, segment length=4pt,amplitude=1.8pt, pre length=.15cm, post length=.25cm}},
    worldlineStatic/.style={dotted, line width=1pt},
	worldline/.style={gray, line width=1pt},
	worldlineBold/.style={black, line width=.6pt},
	zUndirected/.style={line width=1pt},
     ordArrow/.style={draw=black, -{Latex[length=2.0mm,width=1.4mm]}, line width=0.6pt},
	zParticle/.style={line width=1pt,postaction={decorate},decoration={markings,mark=at position .6 with {\arrow[#1]{latex}}}},
	zParticle2/.style={line width=1pt,postaction={decorate},decoration={markings,mark=at position .7 with {\arrow[#1]{latex}}}},
	worldlineCut/.style={dotted,line width=1pt,postaction={decorate},decoration={markings,mark=at position .7 with {\arrow[#1]{latex}}}},
	worldlineCut2/.style={dotted,line width=1pt,postaction={decorate},decoration={markings,mark=at position .6 with {\arrow[#1]{latex}}}},
	zParticleF/.style={line width=1pt,postaction={decorate}},
	cscalar/.style={line width=1pt,postaction={decorate},decoration={markings,mark=at position .6 with {\arrow[#1]{latex}}}},
	cscalar2/.style={line width=1pt,postaction={decorate},decoration={markings,mark=at position .8 with {\arrow[#1]{latex}}}},
	scalar/.style={dashed,line width=1pt,postaction={decorate},decoration={markings,mark=at position .6 with {\arrow[#1]{latex}}}},
	scalar2/.style={dashed,line width=1pt},
	photon/.style={line width =.8pt, decorate, decoration={snake, segment length=4pt, amplitude=1.8pt,  pre length=.1cm, post length=.1cm}},
	photonRed/.style={red, line width =.8pt, decorate, decoration={snake, segment length=4pt, amplitude=1.8pt,  pre length=.1cm, post length=.1cm}},
	cross/.style={cross out, line width =.8pt, draw=black, minimum size=2*(#1-\pgflinewidth), inner sep=0pt, outer sep=0pt},
cross/.default={4pt}
}

\def\eqn#1{eq.~\eqref{#1}}

\def\sec#1{section~{\ref{#1}}}

\def\app#1{appendix~{\ref{#1}}}

\def\rcite#1{ref.~\cite{#1}}

\begin{document}

\begin{flushright}
\begingroup\footnotesize\ttfamily
	QMUL-PH-26-06 \\
	HU-EP-26/13
\endgroup
\end{flushright}

\vspace{15mm}

\begin{center}
{\LARGE\bfseries 
	Canonical Quantisation of \\ Bound and Unbound WQFT
\par}

\vspace{15mm}

\begingroup\scshape\large 
	Riccardo Gonzo\,\orcidlink{0000-0001-7285-6295},${}^{1}$ 
    Gustav~Mogull\,\orcidlink{0000-0003-3070-5717}${}^{1,2,3}$
\endgroup
\vspace{3mm}
					
\textit{${}^{1}$Queen Mary University of London, Mile End Road, London E1~4NS, \\ United Kingdom} \\[0.25cm]
\textit{${}^{2}$Institut f\"ur Physik, Humboldt-Universit\"at zu Berlin, 
 10099 Berlin, Germany} \\[0.25cm]
\textit{${}^{3}$Max-Planck-Institut f\"ur Gravitationsphysik
(Albert-Einstein-Institut), \\ 14476 Potsdam, Germany }

\bigskip
  
\texttt{\small\{r.gonzo@qmul.ac.uk, g.mogull@qmul.ac.uk\}}

\vspace{10mm}

\textbf{Abstract}\vspace{5mm}\par
\begin{minipage}{14.7cm}
	Using canonical quantisation, and eschewing the Schwinger-Keldysh path integral,
	we derive a version of the Worldline Quantum Field Theory (WQFT) formalism
	suitable for both scattering and bound configurations of the classical two-body problem.
	Focusing on a pair of charged particles interacting via a scalar field,
	we quantise Hamilton's equations both in flat space and around a non-zero background,
	perturbing in post-Lorentzian (PL) and self-force (SF) expansions respectively.
	Our quantisation procedure provides access to the Magnus series,
	and is perfectly suited for computing matrix elements of
	$\hat{N}(t,t_0)=- i \hbar \log\hat{U}(t,t_0)$,
	both with and without external scalar states,
	for finite time intervals (bound orbits) and infinite time intervals (scattering).
	Doing so, we provide a complete set of gauge-invariant
	matrix elements describing the 1SF scattering dynamics up to 3PL order,
	and corresponding matrix elements for bound orbits.
	We also demonstrate how $\hat{N}$-matrix elements encode physical observables, providing a unified operator-based framework for conservative and radiative dynamics of binary systems.
	The new WQFT formalism generalises naturally to both gravity and electromagnetism.
\end{minipage}\par

\end{center}
\setcounter{page}{0}
\thispagestyle{empty}
\newpage

\tableofcontents

\section{Introduction}

In the careers of most aspiring young theoretical physicists, we are taught quantum field theory (QFT) in two stages. First, we are introduced to canonical quantisation. Starting from classical field theory, we are shown how we can promote fields to operators together with Dirac brackets, and then by introducing creation operators we act on the vacuum to produce particles. We then learn how to add interactions and, working in the interaction picture (rather than Schr\"odinger or Heisenberg), we derive Dyson's formula, from which we go on to calculate $\hat{S}$-matrix elements: scattering amplitudes. In a modern QFT class, this historical approach is superseded by Feynman's path integral formalism, which works better in the context of loop amplitudes for a variety of reasons including its better handling of BRST symmetry in gauge theories. Thus, the path integral is now rightly recognised as the \emph{de facto} standard for most modern practitioners of QFT.

The question of how best to approach QFT of course depends on our goals. Recently there has been a surge of QFT applications to classical physics, and in particular the gravitational two-body problem. This is motivated by our ability to detect gravitational waves produced by the inspiral and merger of pairs of black holes and neutron stars~\cite{LIGOScientific:2016aoc,LIGOScientific:2017vwq,KAGRA:2021vkt}, and the need to prepare theoretical data for the coming third generation of gravitational wave detectors~\cite{LISA:2017pwj,Punturo:2010zz,Ballmer:2022uxx}. QFT is a convenient framework for applying different flavours of perturbation theory, with an underlying Effective Field Theory (EFT)~\cite{Goldberger:2004jt,Porto:2016pyg} idealising compact bodies as point particles (or deformations thereof). One popular approach is Non-Relativistic General Relativity (NRGR)~\cite{Goldberger:2004jt,Goldberger:2006bd,Goldberger:2009qd,Blanchet:2013haa,Foffa:2013qca}, which involves a weak-field and low-velocity post-Newtonian (PN) expansion in both Newton's constant $G$ and the relative velocity $v/c$. Here the path integral plays an important role: predictions can be made in the far zone of observation by integrating out degrees of freedom in the near zone --- in practice, by summing over collections of Feynman diagrams.

More recent work on the gravitational two-body problem has embraced alternative perturbative schemes. The post-Minkowskian (PM) expansion in $G$ is typically applied to two-body scattering encounters \cite{Bjerrum-Bohr:2022blt,Kosower:2022yvp,Buonanno:2022pgc}, and unlike the PN expansion allows for arbitrarily fast velocities. In this context, the use of QFT closely resembles the traditional $\hat{S}$-matrix theory used in collider physics. Results can be imported to the bound two-body problem directly using unbound-to-bound mappings~\cite{Kalin:2019rwq,Kalin:2019inp,Cho:2021arx,Saketh:2021sri,Gonzo:2023goe,Adamo:2024oxy,Khalaf:2025jpt} or via resummation into the strong-field regime through the effective-one-body (EOB) formalism~\cite{Damour:2017zjx,Damour:2019lcq,Antonelli:2019ytb,Khalil:2022ylj,Damour:2022ybd,Rettegno:2023ghr,Buonanno:2024vkx,Buonanno:2024byg,Damour:2025uka,Long:2025nmj,Clark:2025kvu}. However, the appearance of tails at higher PM orders --- 4PM order in the momentum impulse, or scattering angle --- implies non-locality of the underlying two-body Hamiltonian, and prevents straightforward mappings to bound motion. Nevertheless, important progress has been made on this problem~\cite{Bini:2024tft,Dlapa:2024cje,Dlapa:2025biy}.

In the context of PM scattering, significant progress has been achieved using on-shell amplitude methods, EFT and worldline-based approaches. In amplitude-based formulations, all degrees of freedom are treated within a fully second-quantised framework, with compact objects represented by massive scalar (or spinning) fields whose scattering amplitudes encode the classical two-body dynamics. Such techniques have enabled the extraction of the gravitational potential between spinless bodies~\cite{Neill:2013wsa,Cheung:2018wkq,Bern:2019crd} directly from loop-level amplitudes up to remarkably high PM orders~\cite{Bern:2021dqo,Bern:2021yeh,Bern:2022jvn,Bern:2025zno}; see also refs.~\cite{Dlapa:2021vgp,Dlapa:2022lmu}. To systematically isolate the classical contributions from quantum and superclassical terms, heavy-particle effective field theory techniques \cite{Damgaard:2019lfh,Brandhuber:2021eyq} and equivalent soft-expansion methods \cite{Bjerrum-Bohr:2021vuf,Bjerrum-Bohr:2021din,Bjerrum-Bohr:2021wwt} have been developed.
In parallel, the construction of systematic eikonal~\cite{DiVecchia:2021bdo,Cristofoli:2021jas,DiVecchia:2022nna,DiVecchia:2022piu,DiVecchia:2023frv} and observable-based formalisms, such as the Kosower–Maybee–O'Connell (KMOC) approach~\cite{Kosower:2018adc,Cristofoli:2021vyo}, have provided direct methods for extracting scattering angles, impulses and radiative observables relevant to the gravitational two-body problem from on-shell S-matrix elements. 

The localisation of massive external states onto classical trajectories in the $\hbar \to 0$ limit naturally suggests an alternative organisational principle, in which these degrees of freedom are described directly as first-quantised worldline variables, while bulk fields remain second quantised.
This hybrid description renders $\hbar$ power counting manifest in the EFT, so that classical observables arise from sums of tree-level diagrams, even when the associated integrals correspond to multi-loop topologies. This approach, which goes under the name of Worldline Quantum Field Theory (WQFT) formalism~\cite{Mogull:2020sak,Jakobsen:2023oow}, has had a run of success. For instance,  WQFT has been used to compute the state-of-the-art 5PM momentum impulse $\Delta p_i^\mu$ at first order in the mass-ratio expansion
\cite{Driesse:2024xad,Driesse:2024feo} and more recently the conservative contribution at second order~\cite{Driesse:2026qiz}.
Moreover, gravitational scattering waveforms have also been calculated~\cite{Jakobsen:2021smu,Jakobsen:2021lvp,He:2025how,Bohnenblust:2025gir}, as well as time-dependent scattering trajectories~\cite{Mogull:2025cfn} and, most recently, Compton-like gravitational wave scattering~\cite{Bjerrum-Bohr:2025bqg,Bjerrum-Bohr:2026fhs,Akpinar:2025byi,Bautista:2026qse}. Additional applications include adiabatic tides~\cite{Jakobsen:2022psy,Jakobsen:2023pvx}, scalar QED~\cite{Wang:2022ntx}, and the double copy~\cite{Shi:2021qsb,Comberiati:2022cpm}. Formal links between QFT and WQFT have also been developed~\cite{Mogull:2020sak,Damgaard:2023vnx,Ajith:2024fna,Capatti:2024bid,Du:2024rkf}, further clarifying the connection with amplitude-based techniques.

Beyond fixed-order PM scattering, worldline methods have also proven adaptable to the gravitational self-force (SF) regime, where one performs a systematic expansion in the small mass ratio $m_1/m_2$~\cite{Mino:1996nk,Poisson:2011nh,Barack:2018yvs,Gralla:2021qaf}.
In particular, a worldline-based EFT has recently been constructed that systematically integrates out the heavy degrees of freedom~\cite{Cheung:2023lnj,Cheung:2024byb}, and has subsequently been extended to incorporate spin effects~\cite{Akpinar:2025huz}. Parallel amplitude-based approaches have been developed to interface directly with the SF expansion and enable detailed comparisons~\cite{Barack:2023oqp,Khalaf:2023ozy,Adamo:2023cfp,Kosmopoulos:2023bwc,Correia:2024jgr,Bini:2024icd,Ivanov:2024sds}.  WQFT calculations in the self-force EFT have also been done on curved backgrounds, both in the probe limit~\cite{Hoogeveen:2025tew} and for wave scattering~\cite{Bjerrum-Bohr:2025bqg,Bautista:2026qse,Bjerrum-Bohr:2026fhs}.

WQFT's theoretical foundations are currently built on a modern QFT footing using the path integral~\cite{Jakobsen:2022psy}. The desire for an initial-value formulation of the dynamics necessitates use of the Schwinger-Keldysh in-in formalism~\cite{Schwinger:1960qe,Keldysh:1964ud,Galley:2009px,Kalin:2022hph,Jakobsen:2022psy}, which involves a doubling of degrees of freedom within that path integral. This setup works well for calculating one-point functions, including the momentum impulse $\Delta p_i^\mu$ or gravitational waveform, as the causality prescription for such observables involves only retarded propagators.
To compute them, one applies exactly the Feynman rules one would have in a standard ``in-out'' theory, but with retarded propagators replacing Feynman and oriented towards the single outgoing line. Although classical observables are produced by a sum of tree-level Feynman diagrams, the pattern of which follows Berends-Giele recursion~\cite{Berends:1987me}, one nevertheless encounters the Feynman integrals typically found in multi-loop QFT calculations.

Yet the Schwinger-Keldysh path integral is limited in scope. The link to the classical equations of motion is somewhat obscured --- and, crucially, Schwinger-Keldysh is not flexible enough to admit the Magnus expansion~\cite{Magnus:1954zz}, which in recent work~\cite{Kim:2024svw,Kim:2025sey,Kim:2025ebl,Brandhuber:2025igz} has been shown to play a key role in the pursuit of classical physics. While the textbook Dyson series perturbatively calculates the $\hat{S}$-matrix, Magnus instead calculates its logarithm $\hat{N}=- i \hbar \log\hat{S}$~\cite{Damgaard:2021ipf,Damgaard:2023ttc}. Recently, a precise link has been derived between the eikonal phase, radial action and the Magnusian~\cite{Kim:2025gis} --- all of which are scalars closely related to the $\hat{S}$-matrix, that encode the perturbative scattering dynamics. Not only the vacuum element of $\hat{N}$ is important: to obtain scattering observables from $\hat{N}$ including radiation-reaction, one also requires matrix elements with external gravitons~\cite{Alessio:2025flu,Damgaard:2023ttc}. In this context a crucial role is played by Poisson brackets, whose structure is directly inherited from commutators~\cite{Haddad:2025cmw,Gonzo:2024zxo,Akpinar:2025tct}. These brackets also appear in the Magnus series itself, their central role implying that a starting point based on canonical quantisation is preferable. While recent progress has been made using a classical interaction picture~\cite{Kim:2025hwi,Kim:2025hpn,Kim:2025olv}, the precise link between WQFT and Magnus remains to be fully fleshed out.

In this paper we rebuild WQFT from the ground up, using canonical quantisation and thereby avoiding the doubling of degrees of freedom inherent in the Schwinger--Keldysh path integral. This starts from a Hamiltonian description of a classical worldline theory coupled to bulk degrees of freedom. Our approach is flexible enough to handle perturbations in either PM or SF, for either bound or unbound motion with non-trivial backgrounds --- we explore in detail the formal link between these possibilities. Using a scalar toy model as a representative example, we show how the Magnus expansion emerges naturally from canonical quantisation and we can calculate generic matrix elements of $\hat{N}(t,t_0)$, which encode the classical physics we are interested in. Crucially, by allowing non-infinite time domains within $\hat{N}$, and by perturbing around a non-zero scalar vacuum, we enable direct calculations in the WQFT framework of observables involved in two-body bound orbits.

The outline for our paper is as follows. In \sec{sec:classical} we outline the scalar theory that we will use in this work, setting up the Hamiltonian formulation and Poisson brackets, and we introduce the post-Lorentzian (PL) and self-force (SF) expansions. Then, in \sec{sec:quantisation} we proceed to quantise the theory, introducing quantum fields on the bulk and worldline plus multi-particle states. We also define the Magnus series, which encodes perturbations of the $\hat{N}$ matrix. In \sec{sec:Magnus_caseA} we perform calculations of matrix elements of $\hat{N}$ in PL-scattering, then in \sec{sec:Magnus_caseB} we do the same in the SF expansion --- both for scattering and bound orbits --- providing a complete picture of the 1SF dynamics up to 3PL order. Finally, in \sec{sec:observables} we show how $\hat{N}$-matrix elements give rise to observables, again both for unbound and bound orbits.

\section{Classical theory}\label{sec:classical}

In this paper we focus on the following simple toy model:
\begin{align}\label{action}
	S=\int\!{\rm d}^Dx\left(\frac12(\partial_\mu\phi)^2-\frac{g}{3!}\phi^3\right)
	+\sum_{i=1}^2\int\!{\rm d}\tau_i\left(\frac{m_i}2(1+\dot{x}_i^2(\tau_i))-e_i\phi(x_i(\tau_i))\right)\,,
\end{align}
which consists of a pair of worldlines $i=1,2$
interacting via a real, massless scalar field in the $D$-dimensional bulk $\phi(x)$.
By varying the action one straightforwardly arrives at the 
Euler-Lagrange equations of motion: 
\begin{subequations}\label{EOMs}
\begin{align}
	\square\phi(x)&=-\frac{g}{2}\phi^2(x)
	-\sum_{i=1}^2e_i\!\int\!\d\tau_i\,\delta^D(x-x_i(\tau_i))\,,\\
	m_i\ddot{x}_i^\mu(\tau_i)&=-e_i\,\partial^\mu\phi(x_i(\tau_i))\,.
\end{align}
\end{subequations}
A similar toy model was used in ref.~\cite{Capatti:2024bid}
to analyse the classical $\hbar\to0$ limit of QFT.
The idea here is to provide the simplest possible theory consisting
of a massless field interacting with a pair of worldlines,
with a coupling (carried by $g$) that we can turn on or off as required and which controls the non-linearities in the bulk.
This theory contains all the essential features of either gravity or electrodynamics,
and our main conclusions will extend to those theories.

\subsection{Hamiltonian setup}

Our first step is moving to a Hamiltonian description of the classical theory described by the action~\eqref{action}.
This begins with the introduction of momentum variables associated with the bulk field $\phi(x)$
and the trajectory $x_i^\mu(\tau_i)$:
\begin{align}\label{momenta}
	\pi(x)&=\dot\phi(x)\,, &
	p_i^\mu(\tau_i)=m_i\dot{x}_i^\mu(\tau_i)\,,
\end{align}
where the dot denotes differentiation with respect to the global time coordinate $t=x^0$ on the bulk field $\phi$, whereas it denotes a
derivative with respect to the proper time $\tau_i$ on the worldline variables.
The equations of motion~\eqref{EOMs} are then rendered as
\begin{subequations}\label{hamiltonEqExplicit}
\begin{align}
	\dot{\phi}(x)&=\pi(x)\,, &
	\dot{\pi}(x)&=\vec{\nabla}^2\phi(x)-\frac{g}{2}\phi^2(x)
	-\sum_{i=1}^2e_i\int\!\d\tau_i\,\delta^D(x-x_i(\tau_i))\,,\\
	\dot{x}^\mu_i(\tau_i)&=\frac1{m_i}p_i^\mu\,, &
	\dot{p}_i^\mu&=-e_i\partial^\mu\phi(x_i)\,.\label{hamiltonWLEqs}
\end{align}
\end{subequations}
Each field is expanded around its own background value
\begin{align}
\begin{aligned}
\label{eq:background_field_exp}
x_i^\mu(\tau_i)
&=\bar x_i^\mu(\tau_i)+z_i^\mu(\tau_i)\,, &
p_i^\mu(\tau_i)
&=\bar p_i^\mu(\tau_i)+p_i^{\prime\mu}(\tau_i)\,, \\
\phi(x)&=\bar\phi(x)+\varphi(x)\,, &
\pi(x)&=\bar\pi(x)+\Pi(x)\,.
\end{aligned}
\end{align}
Our particular choices of the backgrounds $\bar{x}_i^\mu(\tau_i)$, $\bar{p}_i^\mu(\tau_i)$, $\bar{\phi}(x)$ and $\bar{\pi}(x)$
will depend on whether we are dealing with the free unbound motion (scattering) or with a non-trivial background solution, which can include both scattering and bound orbits.

Using these canonical momenta the action~\eqref{action} is rendered in a first-order form:
\begin{align}\label{firstOrderAction}
\begin{aligned}
	S&=\int\!{\rm d}^Dx\,\pi\dot{\phi}
	+\sum_{i=1}^2\int\!{\rm d}\tau_i\,p_i\cdot\dot{x}_i-\int\!{\rm d}t\,H(t)\,,\\
	H(t)&=H_{\rm bulk}(t)+H_1(t)+H_2(t)
	=\int\!{\rm d}^{D-1}\mathbf{x}\,\left({\mathcal H}_{\rm bulk}(x)+{\mathcal H}_1(x)+{\mathcal H}_2(x)\right)\,,
\end{aligned}
\end{align}
where $\mathcal{H}(x)$ is the Hamiltonian density.
The symplectic forms in eq.~\eqref{firstOrderAction} above imply Poisson brackets
both on the worldlines and in the bulk:
\begin{subequations}\label{poisson}
\begin{align}
    \{x_i^\mu(\tau),p_{j,\nu}(\tau)\}&=\delta_{ij}\delta^\mu_\nu\,,\label{wlPoisson}\\
    \{\phi(t,\mathbf{x}),\pi(t,\mathbf{y})\}&=\delta^{D-1}(\mathbf{x}-\mathbf{y})\,.
\end{align}
\end{subequations}
Meanwhile, the time-dependent Hamiltonian $H(t)$ is sub-divided
into parts arising from the bulk and worldlines:
\begin{subequations}\label{hamiltonians}
\begin{align}
	{\mathcal H}_{\rm bulk}(x)&=\frac12(\pi^2+(\vec{\nabla}\phi)^2)+\frac{g}{3!}\phi^3(x)\,,\\
	{\mathcal H}_{i}(x)&=\int\!{\rm d}\tau_i\,\delta^D(x-\bar{x}_i(\tau_i))\bigg[
	\frac{1}{2m_i}(p_i^2(\tau_i)-m_i^2)+e_i\phi(x+z_i(\tau_i))\bigg]\label{hamiltonianWL}\,.
\end{align}
\end{subequations}
Notice that the $D$-dimensional delta function in our expression for ${\mathcal H}_i(x)$
includes the background $\bar{x}^\mu_i(\tau_i)$ rather than the full trajectory $x^\mu_i(\tau_i)$.
This is a deliberate choice:
integration on $\tau_i$ over the delta function $\delta^D(x-\bar{x}_i(\tau_i))$ implies an exact,
nonperturbative relationship
$t=\bar{x}_i^0(\tau_i)$ between the global time coordinate $t=x^0$ and the proper time $\tau_i$
along each worldline.
So we will not need to solve for $\tau_i$ at each order in perturbation theory.
If we wanted to, we could therefore integrate $\tau_i$ out of our theory entirely ---
however, we keep it in order to maintain Lorentz invariance through subsequent steps.

The equations of motion~\eqref{hamiltonEqExplicit}
are now fully captured by Hamilton's equations:
\begin{align}\label{hamiltonEqs}
	\frac{\d\cO}{\d t}=\{\cO,H(t)\}\,,
\end{align}
where $\cO$ includes $x_i^\mu(\tau_i)$, $p_i^\mu(\tau_i)$, $\phi(x)$ and $\pi(x)$.
At first glance, one might be skeptical of eq.~\eqref{hamiltonEqs}:
does it truly apply to \emph{all} fields?
On the worldlines, the dynamical variables $x_i^\mu(\tau_i)$ and $p_i^\mu(\tau_i)$
are given in terms of the proper times $\tau_i$,
which also appear in the brackets~\eqref{wlPoisson}.
The evolution equations~\eqref{hamiltonWLEqs} are captured by
\begin{align}\label{wlEvolution}
	\dot{\cO}_i&=\frac{\d \cO_i}{\d\tau_i}=\{\cO_i,H_i(\tau_i)\}\,, &
	H_i(\tau_i)&=\frac{1}{2m_i}(p_i^2(\tau_i)-m_i^2)+e_i\phi(x_i(\tau_i))\,,
\end{align}
where $H_i(\tau_i)$ is the worldline Hamiltonian in terms of $\tau_i$,
and $\cO_i$ is either $x_i^\mu(\tau_i)$ or $p_i^\mu(\tau_i)$.
We can easily show that this viewpoint is equivalent to eq.~\eqref{hamiltonEqs}.
The Hamiltonian density $\cH_i(x)$~\eqref{hamiltonianWL}
and Hamiltonian $H_i(t)$ are related to $H_i(\tau_i)$ by
\begin{align}
\begin{aligned}
	\cH_i(x)
	&=\int\!\d\tau_i\,\delta^D(x-\bar{x}_i(\tau_i))\,H_i(\tau_i)\\
	\implies H_i(t)
	&=\int\!\d^{D-1}\mathbf{x}\,\,\cH_i(x)
	=\left.\frac1{\partial_{\tau_i}\bar{x}_i^0(\tau_i)}H_i(\tau_i)\right|_{t=\bar{x}_i^0(\tau_i)}\,,
\end{aligned}
\end{align}
where, because $\bar{x}_i^\mu$ is timelike,
we know that $\partial_{\tau_i}\bar{x}_i^0(\tau_i)>0$.
Thus, on the worldline if we assume the evolution equations~\eqref{wlEvolution}, then
\begin{align}
	\frac{\d \cO_i}{\d t}=
	\frac{\d\tau_i}{\d t}\frac{\d \cO_i}{\d\tau_i}=
	\left.\frac{1}{\partial_{\tau_i}\bar{x}_i^0(\tau_i)}\{\cO_i,H_i(\tau_i)\}\right|_{t=\bar{x}_i^0(\tau_i)}
	=\{\cO_i,H_i(t)\}\,.
\end{align}
The existence of a global time coordinate $t$ will be crucial
as we later move to quantise the system,
as it will enable us to introduce a single time-evolution operator
$\hat{U}(t,t_0)$ for the full WQFT.

In the next two subsections we examine two possible perturbations
around the background configuration in more detail.
The full Hamiltonian is broken up into background (barred)
and fluctuating (fl) parts as
\begin{align}
	\cH(x)=\bar\cH^{(\rm PL/SF)}(x)+\cH^{(\rm PL/SF)}_{\rm fl}(x)\,,
\end{align}
PL and SF referring to the post-Lorentzian and self-force expansions.
In order to avoid a cluttered notation,
we apply these labels only where there is potential ambiguity ---
in general, from our surrounding discussion it should be clear
which decomposition is being applied.
Here $\bar\cH$ depends only on the background variables $\bar\cO$,
i.e.~$\bar{x}_i^\mu(\tau_i)$, $\bar{p}_i^\mu(\tau_i)$, $\bar{\phi}(x)$ and $\bar\pi(x)$.
Hamilton's equations~\eqref{hamiltonEqs} on the background are
\begin{align}
	\frac{\d\bar{\cO}}{\d t}&=\{\bar{\cO},\bar{H}(t)\}\,,
\end{align}
with corresponding Poisson brackets
\begin{subequations}\label{poissonBG}
\begin{align}
    \{\bar x_i^\mu(\tau),\bar p_{j,\nu}(\tau)\}&=
	\delta_{ij}\delta^\mu_\nu\,, \\
    \{\bar\phi(t,\mathbf{x}),\bar\pi(t,\mathbf{y})\}&=
	\delta^{D-1}(\mathbf{x}-\mathbf{y})\,,
\end{align}
\end{subequations}
These brackets are directly inherited from those of the
complete two-body system~\eqref{poisson}.
Meanwhile, the fluctuations
$z_i^\mu(\tau_i)$, $p_i^{\prime\mu}(\tau_i)$, $\varphi(x)$ and $\Pi(x)$ obey
\begin{align}\label{hamiltonEqsFL}
	\frac{\d \cO'}{\d t}&=\{\cO',H_{\rm fl}(t)\}\,,
\end{align}
with corresponding brackets:
\begin{subequations}
\begin{align}
	\{z_i^\mu(\tau),p^\prime_{j,\nu}(\tau)\}&=
	\delta_{ij}\delta^\mu_\nu\,, \\
	\{\varphi(t,\mathbf{x}),\Pi(t,\mathbf{y})\}&=
	\delta^{D-1}(\mathbf{x}-\mathbf{y})\,.
\end{align}
\end{subequations}
Our main focus will be on these latter equations of motion involving the
fluctuating fields $\cO'$.
However, let us first examine specific solutions to the former equations
of motion for the background variables $\bar{\cO}$.

\subsection{Post-Lorentzian expansion}\label{sec:freeScattering}

We first examine the expansion around free scattering configurations, where we organize the expansion in powers of the three couplings:
\begin{align}
    e_1 &\ll 1\,, & e_2 &\ll 1\,, & g &\ll 1\,.
\end{align}
This means that the background solution is given by the trivial configuration
\begin{align}
     \bar{\phi}(x) = \bar{\pi}(x) = 0\,, \qquad \dot{\bar{x}}_i^\mu(\tau_i) = \frac{\bar{p}_i^\mu(\tau_i)}{m_i},
    \qquad
    \dot{\bar{p}}_{i,\mu}(\tau_i)
    = 0, \qquad i = 1,2  \,.
\end{align}
Throughout this subsection the ``PL'' label should be considered implicit.
We expand the first-order action in \eqn{firstOrderAction} around this solution $S= \bar{S} + S_{\rm fl}$, where 
\begin{align}
\begin{aligned}
	\bar{S} &=\sum_{i=1}^2\int\!{\rm d}\tau_i\,\bar{p}_i\cdot\dot{\bar{x}}_i-\int\!{\rm d}t\,\bar{H}(t)\,,\\
	\bar{H}(t)&=\int\!{\rm d}^{D-1}\mathbf{x}\,\left({\mathcal{\bar{H}}}_{\rm bulk}(x)+{\mathcal{\bar{H}}}_1(x)+{\mathcal{\bar{H}}}_2(x)\right)\,,
\end{aligned}
\end{align}
with the free background Hamiltonian 
\begin{align}
	{\mathcal{\bar{H}}}_{\rm bulk}(x)&=0\,,\qquad  {\mathcal{\bar{H}}}_i(x)=\int\!{\rm d}\tau_i\,\delta^D(x-\bar{x}_i(\tau_i))
	\frac{1}{2m_i}(\bar{p}_i^2(\tau_i)-m_i^2)\,.
\end{align}
On the other hand, the piece $S_{\rm fl}$ of the action involving the dynamical fields is
\begin{align}
\begin{aligned}
	S_{\rm fl} &=\int\!{\rm d}^Dx\,\Pi\dot{\varphi}+
	\sum_{i=1}^2\int\!{\rm d}\tau_i\,p'_i\cdot\dot{z}_i-\int\!{\rm d}t\,H_{\rm fl}(t)\,, \\
	H_{\rm fl}(t)&=\int\!{\rm d}^{D-1}\mathbf{x}\,\left({\mathcal{H}}_{\rm bulk, fl}(x)+{\mathcal{H}}_{1,\rm fl}(x)+{\mathcal{H}}_{2,\rm fl}(x)\right)\,,
\end{aligned}
\end{align}
written in terms of the dynamical fluctuations
\begin{align}
\begin{aligned}
	{\mathcal{H}}_{\rm bulk, fl}(x)&=\frac12(\Pi^2+(\vec{\nabla}\varphi)^2) + \frac{g}{3!}\varphi^3\,,\\
	{\mathcal{H}}_{i,\rm fl}(x)&= \int\!{\rm d}\tau_i\,\delta^D(x-\bar{x}_i(\tau_i))\Big[
	\frac{1}{2m_i}(p'_i(\tau_i))^2  + e_i \,\varphi(x+z_i(\tau_i)) \Big] \\
    &=	\int\!{\rm d}\tau_i\,\delta^D(x-\bar{x}_i(\tau_i))\Big[
	\frac{1}{2m_i}(p'_i(\tau_i))^2  + e_i \,e^{z_i\cdot\partial}\varphi(x) \Big] \,.
\end{aligned}
\end{align}

\paragraph{The straight-line scattering solution}
The background values of all fields solve the equations of motion when $g=e_i=0$:
\begin{align}
\label{eq:free_background}
	\bar\phi&=\bar\pi=0\,, &
	\bar{x}_i^\mu(\tau_i)&=b_i^\mu+\tau_iv_i^\mu\,, &
	\bar{p}_i^\mu&=m_iv_i^\mu\,.
\end{align}
Given these solutions, the background Poisson brackets~\eqref{poissonBG} are equivalent to\footnote{These
	brackets hold under the assumption that $b_i^\mu$ and $v_i^\mu$ are unconstrained.
	As discussed in \rcite{Haddad:2025cmw},
	if one imposes $(b_1-b_2)\cdot v_i=0$ then a different set of brackets is applicable,
	reflecting time-translation symmetry of the worldline Hamiltonian $H_i(\tau_i)$.
}
\begin{align}\label{bvBrackets}
	\{b_i^\mu,v_j^\nu\}=\frac{\delta_{ij}}{m_i}\eta^{\mu\nu}\,.
\end{align}
Using $\bar{x}_i^0(\tau_i)=b_i^0+\tau_iv_i^0$,
we see that the delta function constraints on the worldlines enforce 
$\tau_i=(t-b_i^0)/v_i^0$, where $t=x^0$ is the global time coordinate of the system.
So the time coordinates are related by
a simple time dilation and shift by the impact parameter.
Of course, we could therefore integrate out $\tau_i$ and write our entire Hamiltonian in terms of $t$ alone,
but by retaining both time coordinates we maintain Lorentz invariance.

\subsection{Self-force expansion}\label{sec:backgroundField}

We now consider a more generic case where we expand the
first-order action \eqref{firstOrderAction} around an integrable background solution.
To this end, we employ dimensional regularisation in the ’t Hooft–Veltman scheme: all external states, worldline degrees of freedom, and background classical fields are kept strictly four-dimensional --- where the relativistic Kepler problem is integrable --- while loop momenta and bulk fluctuations are analytically continued to $D$ spacetime dimensions. 
Throughout this subsection, the label ``SF'' should be considered implicit.

We explicitly distinguish between the light body L and the heavy body H.
Introducing the charge-to-mass ratio $q_i = e_i/m_i$, we organize the expansion in two small parameters: the self-force parameter and the bulk coupling, given respectively by
\begin{align}
    \lambda &= \frac{q_L}{q_H} \ll 1\,, &
	g &\ll 1\,.
\end{align}
At leading order in the $\lambda$ expansion, we construct the background solution
\begin{subequations}
\begin{align}
    \dot{\bar{x}}_L^\mu(\tau_L) &= \frac{\bar{p}_L^\mu(\tau_L)}{m_L}\,, &
    \dot{\bar{p}}_{L,\mu}(\tau_L) &= -e_L\,\partial_\mu \bar{\phi}\left(\bar{x}_L\right)\,,
    \label{eq:bg-eom-1}\\
    \dot{\bar{x}}_H^\mu(\tau_H) &= \frac{\bar{p}_H^\mu(\tau_H)}{m_H}\,, &
    \dot{\bar{p}}_{H,\mu}(\tau_H)  &= 0 \,.
    \label{eq:bg-eom-2} 
\end{align}
\end{subequations}
Meanwhile, the four-dimensional static background scalar field obeys
\begin{align}\label{eq:bg-phi}
    \square \bar\phi(x)
    &= -e_H \int\!\d\tau_H\,\delta^{4}\bigl(x-\bar x_H(\tau_H)\bigr)\,, &
    \bar\pi(x) &= 0\,,
\end{align}
where to preserve the integrability of the background dynamics we have constructed our solution in the strict limit $g = 0$.\footnote{In other words, we switch off bulk nonlinearities in the background solution while retaining them in the full Hamiltonian for the fluctuations. In the gravitational case only the small–mass-ratio expansion $m_1/m_2 \ll 1$ is available, and graviton nonlinearities cannot be neglected even in the probe limit.}
Thus, the heavy body sources a static Coulomb potential and
$\bar x_L^\mu(\tau_L)$, $\bar p_L^\mu(\tau_L)$ describes a relativistic
Kepler orbit, whose properties we will return to later. 

The background dynamics provided by \eqref{eq:bg-eom-1}, \eqref{eq:bg-eom-2} and \eqref{eq:bg-phi}
are a solution of the equations of motion of the following first-order action:
\begin{align}
\begin{aligned}
	\bar{S} &=\int\!{\rm d}\tau_L\,\bar{p}_L\cdot\dot{\bar{x}}_L
	+\int\!{\rm d}\tau_H\,\bar{p}_H\cdot\dot{\bar{x}}_H
	-\int\!{\rm d}t\,\bar{H}(t)\,, \\
	\bar{H}(t)&=\int\!{\rm d}^{D-1}\mathbf{x}\,
	\left({\mathcal{\bar{H}}}_{\rm bulk}(x)+{\mathcal{\bar{H}}}_L(x)+{\mathcal{\bar{H}}}_H(x)\right)\,,
    \label{eq:background_hamil_SF}
\end{aligned}
\end{align}
where we have introduced the background Hamiltonians
\begin{align}
	{\mathcal{\bar{H}}}_{\rm bulk}(x)&=\frac12 (\vec{\nabla}\bar{\phi})^2\,, \nonumber \\
	{\mathcal{\bar{H}}}_L(x)&=\int\!{\rm d}\tau_L\,\delta^D(x-\bar{x}_L(\tau_L))\left(
	\frac{1}{2m_L}(\bar{p}_L^2(\tau_L)-m_L^2)+e_L \bar{\phi}(\bar{x}_L(\tau_L))\right)\,,  \nonumber \\
    {\mathcal{\bar{H}}}_H(x)&=\int\!{\rm d}\tau_H\,\delta^D(x-\bar{x}_H(\tau_H))\left(
	\frac{1}{2m_H}(\bar{p}_H^2(\tau_H)-m_H^2)\right)\,. 
    \label{eq:background_H}
\end{align}
Note that, due to the self-force expansion, dimensional regularisation sets to zero all scaleless integrals, forcing the condition \cite{Cheung:2023lnj,Cheung:2024byb,Akpinar:2025huz}
\begin{align}
	\bar{\phi}(\bar{x}_H(\tau_H)) = 0\,.
\end{align}
We can now consider the expansion of the first order action around the background solution $S = \bar{S} + S_{\rm fl}$, where
\begin{align}
\begin{aligned}
	S_{\rm fl} &=\int\!{\rm d}^Dx\,\Pi\dot{\varphi}
	+\int\!{\rm d}\tau_L\,p'_L\cdot\dot{z}_L
	+\int\!{\rm d}\tau_H\,p'_H\cdot\dot{z}_H
	-\int\!{\rm d}t\,H_{\rm fl}(t)\,, \\
	H_{\rm fl}(t)&=\int\!{\rm d}^{D-1}\mathbf{x}\,\left({\mathcal{H}}_{\rm bulk, fl}(x)+{\mathcal{H}}_{L,\rm fl}(x)+{\mathcal{H}}_{H,\rm fl}(x)\right)\,.
\end{aligned}
\end{align}
Using the background equations of motion, we obtain the following Hamiltonian:
\begin{align}
	{\mathcal{H}}_{\rm bulk, fl}(x)&=\frac12 (\Pi^2 + |\vec{\nabla} \varphi|^2) + 
	\frac{g}{3!} \left(\varphi^3+3\bar{\phi}^2\varphi+3\bar\phi\varphi^2 + \bar\phi^3\right)\,, \\
	{\mathcal{H}}_{L,\rm fl}(x)&=\int\!{\rm d}\tau_L\,\delta^D(x-\bar{x}_L(\tau_L))\Big[
	\frac{1}{2m_L}(p'_L(\tau_L))^2  + e_L e^{z_L\cdot\partial}\varphi(x) \nonumber \\
    &\qquad \qquad \qquad \qquad \qquad + e_L (e^{z_L\cdot\partial}\bar{\phi}(x)-\bar{\phi}(x) - z_L \cdot \partial \bar{\phi}(x) ) \Big]\,,  \nonumber \\
    {\mathcal{H}}_{H,\rm fl}(x)&=\int\!{\rm d}\tau_H\,\delta^D(x-\bar{x}_H(\tau_H))\Big[
	\frac{1}{2m_H}(p'_H(\tau_H))^2  + e_H (e^{z_H\cdot\partial}\varphi(x)-\varphi(x)) \Big]\,. \nonumber
\end{align}

\paragraph{The relativistic $1/r$ background}

We discuss here the details of the relativistic Kepler dynamics encoded in the background solution.
Working in the rest frame $\bar{x}_H^\mu=v_H^\mu\tau_H=(\tau_H,\mathbf{0})$,
the heavy body sources a static Coulomb field at the origin
that governs the evolution of the light body:
\begin{align}
    \bar\phi(x)&=\frac{K}{r}\,, &
    K&=-\frac{e_H}{4\pi}\,,  &
    r &= |\mathbf{x}|\,,
\end{align}
which is the solution to \eqn{eq:bg-phi}.
Given that the potential is static and spherically symmetric,
without loss of generality we parametrise the motions of the two bodies as
\begin{align}
\begin{aligned}
  \bar{x}_L^\mu(\tau_L)
    &=\bigl(
        t(\tau_L),
        \bar{r}(\tau_L)\cos\bar{\psi}(\tau_L),
        \bar{r}(\tau_L)\sin\bar{\psi}(\tau_L),
        0  \bigr)\,, &
	\bar p_L^\mu(\tau_L) &= m_L\,\dot{\bar{x}}_L^\mu (\tau_L)\,,\\
	\bar{x}_H^\mu(\tau_H) &= (\tau_H,\mathbf{0}) \,, &
	\bar{p}_H^\mu(\tau_H) &= (m_H,\mathbf{0}) \,.
\end{aligned}
\end{align}
We drop the subscripts L on $\bar{r}$ and $\bar\psi$,
and also components of $\bar{p}_\mu$,
as these coordinates are always associated with the light body.
The two delta functions $\delta^D(x-\bar{x}_H(\tau_H))$ and 
$\delta^D(x-\bar{x}_L(\tau_L))$ in $\cH(x)$~\eqref{eq:background_H}
identify the global time coordinate as $t=x^0=\tau_H=\bar{t}_L(\tau_L)$.

Hamilton’s equations imply the existence of two conserved quantities:
the canonical energy $\bar E = \bar{p}_0$ and
the azimuthal component of the angular momentum $\bar L = \bar{p}_\psi$. 
The background Hamiltonian $\bar H(t)$~\eqref{eq:background_H} is independent of
the time coordinate $t$ and the azimuthal angle $\psi$, so
\begin{align}
\frac{\mathrm{d}\bar p_0}{\mathrm{d}t}
&=\{\bar p_0,\bar H(t)\}=0\,, &
\frac{\mathrm{d}\bar p_\psi}{\mathrm{d}t}
&=\{\bar p_\psi,\bar H(t)\}=0\,.
\end{align}
Here we used the background Poisson brackets \eqref{poissonBG}, which upon restriction to the equatorial plane reduce to the canonical three-dimensional brackets
\begin{align}\label{eq:poissonBG_3d}
\{\bar{r},\bar{p}_r\}&=1 \,, &
\{\bar{\psi},\bar{p}_\psi\}&=1 \,, &
\{\bar{t},\bar{p}_0\}&=1\,,
\end{align}
with all other Poisson brackets vanishing,
together with the scalar background being static $\partial_t \bar\phi(x)=0$
and spherically symmetric $\partial_{\psi}\bar\phi(x)=0$.
Moreover, the bulk Hamiltonian $\bar{\mathcal H}_{\rm bulk}$
depends only on the field variables $\bar\phi$ and $\bar\pi$,
so it Poisson-commutes with the particle phase-space coordinates.
We have
\begin{align}\label{eq:conserved_EL}
\bar E &= \bar p_0 = m_L\,\dot{\bar t}(\tau_L)\,, &
\bar L &= \bar p_\psi = -m_L\, \bar{r}^2\,\dot{\bar{\psi}}(\tau_L)\,,
\end{align}
when expressed in terms of the position coordinates.

We now focus on the radial component of the motion.
Using the background constraint implied by $\bar H$,
\begin{align}
0=\frac{1}{2m_L}(\bar p_L^2-m_L^2)+e_L\bar\phi(\bar x_L)
\qquad\Rightarrow\qquad
\bar p_L^2=m_L^2+\frac{2m_L e_L K}{\bar{r}},
\end{align}
together with $\bar p_L^\mu=m_L\dot{\bar x}_L^\mu$ and \eqn{eq:conserved_EL},
we obtain the radial evolution equation in the worldline parameter $\tau_L$,
\begin{align}\label{eq:Ur_scalar}
\dot{\bar{r}}(\tau_L)^2=&\frac{U_r(\bar{r}(\tau_L))}{m_L^2}, &
U_r(r):=&\,\bar E^2-m_L^2-\frac{2m_L e_L K}{r}-\frac{\bar L^2}{r^2},
\end{align}
Eliminating $\tau_L$ in favour of the azimuthal angle using
$\dot{\bar{\psi}}=\bar L/(m_L \bar{r}^2)$, one finds
\begin{align}\label{radialEquation}
\left(\frac{\mathrm{d} \bar{u}(\bar{\psi})}{\mathrm{d} \bar{\psi}}\right)^2
&=\frac{\bar E^2-m_L^2}{\bar L^2}-\bar{u}^2-\frac{2m_L e_L K}{\bar L^2}\,\bar{u}\,, &
\bar{u}(\bar{\psi}):=\frac1{\bar{r}(\bar{\psi})}\,.
\end{align}
The solution to this equation is the (non-precessing) Kepler form:\footnote{To
	derive this solution explicitly one differentiates \eqn{radialEquation} with respect to $\bar\psi$.
	Assuming $\bar{u}'(\bar{\psi}) \neq 0$, this yields
	$\bar{u}''(\bar{\psi})+\bar{u}(\bar{\psi})=-(m_L e_L K)/\bar{L}^2$.
}
\begin{align}
\bar{r}(\bar{\psi})=\frac{\bar p}{1+\bar e\cos\bar{\psi}},
\label{eq:radius_psi_scalar}
\end{align}
which does not exhibit periapsis precession.
The orbital elements are the semi-latus rectum $\bar p$ and the eccentricity $\bar e$, given by
\begin{align}\label{eq:pe_scalar}
\bar p&=-\frac{\bar L^2}{m_L e_L K}\,, &
\bar e&=\sqrt{1+\frac{\bar L^2(\bar E^2-m_L^2)}{m_L^2 e_L^2 K^2}}\,.
\end{align}
For attractive scalar interactions one has $e_LK<0$, so that $\bar p>0$.

We now distinguish between bound and scattering orbits according to the value of $\bar\Gamma := \bar E/m_L$, assuming $e_L K = -|e_L K| < 0$ to allow the existence of bound states. For bound motion ($\bar\Gamma<1$), the radial potential \eqref{eq:Ur_scalar} has two real turning points $\bar{r}_{\min}<\bar{r}_{\max}$, and the motion is periodic in the radial direction. Equivalently, the eccentricity satisfies $0\le \bar e<1$, corresponding to elliptic Kepler motion without periapsis precession. In this case the potential factorises as
\begin{align}
U_r(r)
= (m_L^2-\bar{E}^2)\,
\frac{(r-\bar{r}_{\min})(\bar{r}_{\max}-r)}{r^2}\,, \quad
\bar{r}_{\min} = \frac{\bar{p}}{1+\bar{e}}\,, \quad  \bar{r}_{\max} = \frac{\bar{p}}{1-\bar{e}}\,.
\label{eq:Ur_factorised} 
\end{align}
where $\bar{r}_{\rm min/max}$ are derived from the Kepler form~\eqref{eq:radius_psi_scalar}.
With our parametrisation, the radial period $\bar{T}_r^<$ is given by
\begin{align}\label{eq:Tr_scalar}
\bar T_r^{<}
&= \oint_{\mathcal{C}^{<}} \frac{\mathrm{d}t}{\mathrm{d} r}\,\mathrm{d} r
= \oint_{\mathcal{C}^{<}} \frac{\dot{t}}{\dot{r}}\,\mathrm{d} r
= 2\bar E \int_{\bar{r}_{\min}}^{\bar{r}_{\max}} \frac{\mathrm{d} r}{\sqrt{U_r(r)}}
= \frac{2\pi\,m_L |e_L K|\,\bar E}{(m_L^2-\bar E^2)^{3/2}} \,,
\end{align}
where $\mathcal C^{<}$ denotes the closed contour from $\bar{r}_{\min}$ to $\bar{r}_{\max}$ and back.  The corresponding radial frequency is
\begin{align}
\bar{\Omega}_r^{<}
= \frac{2\pi}{\bar T_r^{<}}
= \frac{(m_L^2-\bar E^2)^{3/2}}{m_L |e_L K|\,\bar E}.
\label{eq:Omega_r_scalar}
\end{align}
It is also useful to introduce the eccentric anomaly $\bar{u}_r$, defined as
\begin{align}
\bar{u}_r := 2 \arctan\left[\sqrt{\frac{1-\bar e}{1+\bar e}} \tan\left(\frac{\bar{\psi}}{2}\right)\right]\,.
\label{eq:psi-to-ur}
\end{align}
This allows us to parametrise the bound motion as
\begin{align}\label{eq:radius_t_bound_scalar}
\bar{r}^{<}(t)&= \bar a^{<}\bigl(1-\bar e\cos(\bar{u}_r(t))\bigr)\,, &
\bar a^{<}&=\frac{\bar p}{1-\bar e^2}\,,
\end{align}
with the Kepler equation
\begin{align}
\bar{\Omega}_r^{<}\, t = \bar{u}_r(t) - \bar e \sin(\bar{u}_r(t)) .
\label{eq:kepler_scalar}
\end{align}

Since the dynamics is separable, all gauge-invariant observables can equivalently be derived from the orbit-averaged radial action. The bound-orbit radial action is defined as the integral of the radial momentum over one full radial cycle:
\begin{align}\label{eq:Ir_scalar}
\bar I_r^{<}(\bar E,\bar L)
= \oint_{\mathcal C^{<}} \mathrm{d} r\,\bar p_r(r)
= 2 \int_{\bar{r}_{\min}}^{\bar{r}_{\max}} \mathrm{d} r\,\sqrt{U_r(r)}
= -2\pi \bar L +\frac{2\pi\,m_L |e_L K|}{\sqrt{m_L^2-\bar E^2}}\,.
\end{align}
The relations between the radial action and the orbital observables follow directly from Hamilton’s equations and the background Poisson brackets \eqref{eq:poissonBG_3d}.  In particular, using $\dot F=\{F,\bar H\}$ for any phase-space function $F$, the change in the azimuthal angle over one radial cycle is
\begin{align}
\Delta\bar\psi^{<}
= \oint_{\mathcal C^{<}} \mathrm{d} t\,\frac{\mathrm{d} \bar{\psi}}{\mathrm{d} t}
= 2 \int_{\bar{r}_{\min}}^{\bar{r}_{\max}}  \frac{\dot{\bar{\psi}}}{\dot{r}}\,\mathrm{d} r
= \oint \frac{\{\bar{\psi},\bar H\}}{\{\bar{r},\bar H\}}\,\mathrm{d} r .
\label{eq:deltapsi_back}
\end{align}
On the bound orbit the background equations of motion are subject to the on-shell condition $\bar H(\bar{r},\bar{p}_r;\bar E,\bar L)=0$, where $\bar{p}_r$ depends implicitly on $(\bar{E},\bar{L})$. Differentiating the on-shell condition at fixed $\bar{r}$ with respect to $\bar L$ gives
\begin{align}\label{eq:chain_rule}
0&= \frac{\partial \bar H}{\partial \bar{p}_r}\,
\frac{\partial \bar{p}_r}{\partial \bar L}
+ \frac{\partial \bar H}{\partial \bar L} &\implies & &\frac{\partial \bar{p}_r}{\partial \bar L}
&= -\,\frac{\partial \bar H/\partial \bar L}
        {\partial \bar H/\partial \bar{p}_r} = -\frac{\{\bar{\psi},\bar H\}}{\{\bar{r},\bar H\}}\,,  
\end{align}
having used $\{\bar{r},\bar{p}_r\}=1$ and $\{\bar{\psi},\bar{p}_\psi\}=1$ in the last step.
Therefore, combining eqs.~\eqref{eq:deltapsi_back} and \eqref{eq:chain_rule} we have
\begin{align}
\Delta\bar\psi^{<}
= -\oint_{\mathcal C^{<}} \,\frac{\partial \bar{p}_r}{\partial \bar L}  \,\mathrm{d} r
= -\frac{\partial \bar I_r^{<}}{\partial \bar L} = 2\pi\,,
\end{align}
reflecting the absence of periapsis precession for a massless scalar interaction.
Similarly, the radial period in \eqn{eq:Tr_scalar} is recovered from
\begin{align}
\bar T_r^{<}
= \oint_{\mathcal C^{<}} \frac{\mathrm{d}t}{\mathrm{d} r}\,\mathrm{d} r
= \oint_{\mathcal C^{<}} \frac{\{t,\bar H\}}{\{\bar{r},\bar H\}}\,\mathrm{d} r
= \oint_{\mathcal C^{<}} \,\frac{\partial \bar{p}_r}{\partial \bar E}\,\mathrm{d} r
= \frac{\partial \bar I_r^{<}}{\partial \bar E},
\end{align}
where in the last step we used the on-shell condition and the background brackets.

For unbound orbits ($\bar\Gamma>1$) the trajectory is hyperbolic
and the motion has a single radial turning point $\bar{r}_{\min}$,
with $r\to\infty$ in the remote past and future.
Equivalently, the eccentricity satisfies $\bar e>1$.
As in the bound case, the radial dynamics is governed by the potential $U_r(r)$~\eqref{eq:Ur_scalar},
which now admits only one real root $\bar{r}_{\rm min}$.
The unbound solutions are obtained by analytic continuation of the bound expressions,
\begin{align}\label{eq:trajectory_B2B_scalar}
&\bar{u}_r \;\to\; i\,\bar{v}_r\,, &
&\bar{\Omega}_r^{<} \;\to\; -\,i\,\bar{\Omega}_r^{>}\,,
\end{align}
leading to the time-dependent radial trajectory
\begin{align}\label{eq:radius_t_scatt_scalar}
\bar r^{>}(t)
&= \bar a^{>}\bigl(\bar e\cosh\bar{v}_r - 1\bigr)\,, &
\bar a^{>}=\frac{\bar p}{\bar e^2-1}
&= -\,\bar a^{<}\,,
\end{align}
together with the hyperbolic Kepler equation
\begin{align}\label{eq:kepler_scatt_scalar}
\bar{\Omega}_r^{>}\, t
= \bar e\,\sinh \bar{v}_r - \bar{v}_r \,.
\end{align}
As in the bound case, the dynamics is encoded in the radial action $\bar{I}_r^>$.
For unbound motion, however, the radial action is infrared divergent due to the long-range nature of the scalar interaction and must be regularized. Introducing an upper radial cutoff $R$, we define
\begin{align}\label{eq:Ir_scatt_def_scalar}
\bar I_r^{>}(\bar E,\bar L;R)
&:= \int_{\mathcal C^{>}} \mathrm{d} r\,\bar p_r(r)
= 2 \int_{\bar{r}_{\min}}^{R} \mathrm{d} r\,\sqrt{U_r(r)}\,,
\end{align}
where $\mathcal C^{>}$ denotes the open contour from $\bar{r}_{\min}$ to $R$ and back,
and $\bar{r}_{\min}$ is the same as in the bound case~\eqref{eq:Ur_factorised}.
Evaluating the integral for large $R$ yields,
\begin{align}
\hspace{-17pt}\bar I_r^{>}(\bar E,\bar L;R)
=
- \frac{2 m_L |e_L K|}{\sqrt{\bar E^2-m_L^2}}
\log\left(\frac{|e_L K| \bar{e}}{R}\right)
- 2\bar L\,\arccos\left(-\frac{1}{\bar e}\right)
+ \mathcal{O}\left(\frac{1}{R}\right) .
\label{eq:Ir_scatt_scalar}
\end{align}
The infrared regulator does not drop out of the coordinate-time delay, which is obtained from using the background brackets and the on-shell condition,
\begin{align}
\bar T_r^{>}(\bar E,\bar L;R)
= \frac{\partial \bar I_r^{>}}{\partial \bar E}
= -\frac{2 m_L |e_L K| \bar{E}}{(\bar E^2-m_L^2)^{3/2}}  \log\left(\frac{|e_L K| \bar{e}}{R}\right)\,.
\label{eq:Tr_scatt_scalar}
\end{align}
By contrast, the scattering angle is finite and regulator-independent:
\begin{align}
\Delta\bar\psi^{>}(\bar E,\bar L)
= -\,\frac{\partial \bar I_r^{>}}{\partial \bar L}
= 2\arccos\left(-\frac{1}{\bar e}\right) .
\label{eq:scatt_angle_scalar}
\end{align}
The standard scatter-to-bound maps apply to all the quantities considered above. In particular, at the background level there exists an exact correspondence between bound and unbound observables:\footnote{It is worth recalling the following identities for the analytic continuation (see also \rcite{Cho:2021arx}):
\begin{align*}
 \arccos \left(x\right)+\arccos \left(-x\right)=\pi, \qquad \qquad \log \left(x\right)-\log \left(-x\right)=i \pi\,.
\end{align*}
}
\begin{align}
\begin{aligned}
\bar I_r^{<}(\bar E,\bar L)
&= \bar I_r^{>}(\bar E,\bar L;R)
 - \bar I_r^{>}(\bar E,-\bar L;R)\,, \\
\bar T_r^{<}(\bar E,\bar L)
&= \bar T_r^{>}(\bar E,\bar L;R)
 - \bar T_r^{>}(\bar E,-\bar L;R)\,, \\
\Delta\bar\psi^{<}(\bar E,\bar L)
&= \Delta\bar\psi^{>}(\bar E,\bar L)
 + \Delta\bar\psi^{>}(\bar E,-\bar L)\,,
\end{aligned}
\end{align}
where the dependence on the infrared regulator $R$ cancels identically in the bound combinations. Importantly, this correspondence holds directly at the \emph{integrated level}, i.e. for the radial action and its associated action--angle observables.

\section{Canonical quantisation}\label{sec:quantisation}

In this section we quantise our system, and thus build the WQFT.
This means that we promote dynamical fields to operators,
which act on quantum states.
In the interaction picture,
Hamilton's equations for the fluctuating fields~\eqref{hamiltonEqsFL} become
\begin{align}
	i\hbar\frac{{\rm d}}{{\rm d}t}|\psi(t)\rangle&=\hat{H}_I(t)|\psi(t)\rangle\,,\label{schrodinger}\\
	i\hbar\frac{\rm d}{{\rm d}t}\hat{\cO}&=
	[\hat{\cO},\hat{H}_0]\,.\label{freeEvolution}
\end{align}
States $|\psi(t)\rangle$ are evolved by the interaction Hamiltonian $H_I$
and operators $\hat\cO$ evolve with the free Hamiltonian $H_0$,
the sum of which gives the full fluctuating Hamiltonian:
\begin{align}
	H_{\rm fl}(t) = H_0(t) + H_{I}(t)\,.
\end{align}
Meanwhile, the Poisson brackets~\eqref{poisson} are upgraded to equal-time commutation relations of the corresponding quantum operators (denoted by a hat):
\begin{subequations}\label{eq:commutators}
\begin{align}
    [\hat{z}_i^\mu(\tau),\hat{p}^\prime_{j,\nu}(\tau)]&=
	i\hbar\,\delta_{ij}\delta^\mu_\nu\,,\label{wlCommutator}\\
    [\hat{\varphi}(t,\mathbf{x}),\hat\Pi(t,\mathbf{y})]&=
    i\hbar\,\delta^{D-1}(\mathbf{x}-\mathbf{y})\,.
\end{align}
\end{subequations}
Using the background field expansions~\eqref{eq:background_field_exp}
we can also introduce similar brackets for the full dynamical variables:
\begin{subequations}
\begin{align}
    [\hat{x}_i^\mu(\tau),\hat{p}_{j,\nu}(\tau)]&=
	i\hbar\,\delta_{ij}\delta^\mu_\nu\,,\\
    [\hat{\phi}(t,\mathbf{x}),\hat\pi(t,\mathbf{y})]&=
    i\hbar\,\delta^{D-1}(\mathbf{x}-\mathbf{y})\,.
\end{align}
\end{subequations}
We continue to express the worldline fields as functions of $\tau_i$ ---
ultimately, it should be understood that all fields depend on $t$,
as we constrain $t=\bar{x}_i^0(\tau_i)$.

The split of $H_{\rm fl}$ into $H_0$ and $H_I$ is not unique,
so for convenience we adopt a choice that leaves all propagators
identical in either the PL or SF expansions,
while encoding background-dependent effects entirely through interaction vertices.
The free Hamiltonian $H_0$ is always defined as
\begin{align}
H_0(t)
=
H_{0,\mathrm{bulk}}(t)
+
H_{0,1}(t)
+
H_{0,2}(t),
\qquad
[\hat{H}_{0,\mathrm{bulk}}(t),\hat{H}_{0,i}(t')]=0,
\label{eq:H0_split}
\end{align}
where $H(t)=\int\!\d^{D-1}\mathbf{x}\,\cH(x)$ and
\begin{align}
\cH_{0,\mathrm{bulk}}(x)
=
\frac12\Big(\Pi^2+(\vec\nabla\varphi)^2\Big)\,,\quad
\cH_{0,i}(x)
=
\int\!\mathrm{d}\tau_i\,
\delta^D\!\bigl(x-\bar x_i(\tau_i)\bigr)\,
\frac{\bigl(p_i'(\tau_i)\bigr)^2}{2m_i}\,.
\end{align}
This describes a free massless scalar field together with free fluctuations
$z_i^\mu(\tau_i)$ around prescribed background worldlines
$\bar x_i^\mu(\tau_i)$. The explicit form of the background trajectories
$\bar x_i^\mu(\tau_i)$ differs depending on which perturbative
expansion we are using, as discussed in the
previous section and given in Eqs.~\eqref{eq:free_background} and
\eqref{eq:bg-eom-2}--\eqref{eq:bg-phi}, respectively; however, the functional
form of $H_0$ is the same in both cases.
Seeing as the bulk and worldline parts of $\hat{H}_0$ commute,
in the next two subsections we quantise these two parts of the system separately.
Then, in \sec{sec:interactions} we will discuss interactions,
and finally in \sec{sec:magnus} the Magnus series.

\subsection{Bulk fields}

The scalar field $\hat\varphi(x)$ and its conjugate momentum $\hat\Pi(x)$ are expanded in creation and annihilation operators as 
\begin{subequations}\label{modeDecomp}
\begin{align}
\hat{\varphi}(x)
&=
\int_\mathbf{p}
\left[
\hat{a}(p)e^{-ip\cdot x/\hbar}
+\hat{a}^{\dagger}(p)e^{+ip\cdot x/\hbar}
\right],
\\
\hat{\Pi}(x)
&=
\frac{1}{\hbar}\int_\mathbf{p}
\left[
(-i|\mathbf{p}|)\hat{a}(p)e^{-ip\cdot x/\hbar}
+(i|\mathbf{p}|)\hat{a}^{\dagger}(p)e^{+ip\cdot x/\hbar}
\right],
\end{align}
\end{subequations}
with Lorentz--invariant measure
\begin{align}
\int_\mathbf{p}
:=
\int\!\frac{\mathrm{d}^Dp}{(2\pi \hbar)^{D-1}}\,
\theta(p^0)\,\delta(p^2)
=
\int\!\frac{\mathrm{d}^{D-1}\mathbf{p}}{(2\pi \hbar)^{D-1}\,2|\mathbf{p}|}.
\end{align}
The operators satisfy
\begin{align}\label{raisingLowering}
[\hat a(p),\hat a^\dagger(k)]
=
2 \hbar |\mathbf{p}|\,\dd(\mathbf{p}-\mathbf{k})\,,
\end{align}
where $\dd(\mathbf{k}) := (2\pi \hbar)^{D-1} \delta^{D-1}(\mathbf{k})$. Using the mode decomposition~\eqref{modeDecomp},
one can straightforwardly derive the retarded and advanced propagators:
\begin{subequations}\label{bulkProps}
\begin{align}
	\begin{tikzpicture}[baseline={(current bounding box.center)}]
		\coordinate (x) at (-.7,0);
		\coordinate (y) at (0.5,0);
		\draw [draw=black, scalar] (x) -- (y);
		\draw [fill] (x) circle (.06) node [below] {$y$};
		\draw [fill] (y) circle (.06) node [below] {$x$};
	\end{tikzpicture}&=
	\Delta_R(x-y):=\theta(x^0-y^0)[\hat\varphi(x),\hat\varphi(y)]\nn\\
	&=\int_k e^{-ik\cdot(x-y)/\hbar}\frac{i \hbar }{k^2+{\rm sgn}(k^0)i0}\,,\label{propPhiR}\\
	\begin{tikzpicture}[baseline={(current bounding box.center)}]
		\coordinate (x) at (-.7,0);
		\coordinate (y) at (0.5,0);
		\draw [draw=black, scalar] (y) -- (x);
		\draw [fill] (x) circle (.06) node [below] {$y$};
		\draw [fill] (y) circle (.06) node [below] {$x$};
	\end{tikzpicture}&=
	\Delta_A(x-y):=-\theta(y^0-x^0)[\hat\varphi(x),\hat\varphi(y)]\nn\\
	&=  \int_k e^{-ik\cdot(x-y)/\hbar}\frac{i \hbar}{k^2-{\rm sgn}(k^0)i0}\,,\label{propPhiA}
\end{align}
\end{subequations}
where the $D$-dimensional integral measure is $\int_k:=\int\!\d^Dk/(2\pi \hbar)^D$.
These expressions are checked by performing
the contour integral on $k^0$,
picking up residues at $k^0=\pm|\mathbf{k}|$ on the same side of the real axis.

We define the bulk in--vacuum $|0\rangle_\varphi$
at time $t_0$ ($t_0\to-\infty$ for scattering)
via $\hat a(k)|0\rangle_\varphi=0$. 
Multi--particle bulk states are constructed by acting on
the vacuum with raising operators:
\begin{align}
|k_1 k_2 \cdots k_n\rangle_\varphi
=\hat a^\dagger(k_1)\hat a^\dagger(k_2)\cdots \hat a^\dagger(k_n)|0\rangle_\varphi.
\label{eq:multiparticle}
\end{align}
Thus, using \eqn{raisingLowering} we have
\begin{align}
	\langle 0|\hat\varphi(x)|k\rangle&=\hbar\, e^{-ik\cdot x/\hbar}\,, &
	\langle k|\hat\varphi(x)|0\rangle &=\hbar\, e^{ik\cdot x/\hbar}\,.
\end{align}
We prefer to orient our momenta as outgoing,
meaning that generally we contract states $\langle k|$ on the left.
So the Feynman rule for external scalars is
\begin{align}\label{extScalars}
\begin{tikzpicture}[baseline={(current bounding box.center)}]
		\coordinate (x) at (-.7,0);
		\coordinate (y) at (0.5,0);
		\draw [draw=black, scalar2] (x) -- (y) node [below] {$k$};
		\draw [fill] (x) circle (.06) node [below] {$x$};
\end{tikzpicture}=\hbar\, e^{ik\cdot x/\hbar}\,.
\end{align}
To contract with $|k\rangle$ simply use $e^{-ik\cdot x/\hbar}$,
which is equivalent to sending $k^\mu\to-k^\mu$.

\subsection{Worldline fields}

Since $H_{0,i}$ is quadratic in the worldline fluctuations
$\{\hat z_i^\mu,\hat p_i^{\prime\mu}\}$ and contains no background-dependent forces, the free equations of motion~\eqref{freeEvolution} are universal:
\begin{subequations}\label{eq:wlEvolution}
\begin{align}
\frac{\d}{\d\tau_i}\hat z_{i}^\mu(\tau_i)
&=
\frac{1}{m_i}\hat p_{i}^{\prime\mu}(\tau_i),
\\
\frac{\d}{\d\tau_i}\hat p_{i}^{\prime\mu}(\tau_i)
&=0\,.
\end{align}
\end{subequations}
By formulating these equations in terms of
$\tau_i$, rather than the global time coordinate $t$,
we eliminate all dependence on the background solutions $\bar{x}_i^\mu$ ---
in essence, constructing inertial frames around our particles.
These equations are solved by
\begin{subequations}\label{eq:wlEvolution_sol}
\begin{align}
	\hat{z}_i^\mu(\tau_i)&=\hat{b}_i^{\prime\mu}+\frac{\tau_i}{m_i}\hat{p}_i^{\prime\mu}\,,\label{zOperator}\\
	\hat{p}_i^{\prime\mu}(\tau_i)&=\hat{p}_i^{\prime\mu}\,,\label{wlBVec}
\end{align}
\end{subequations}
where both $\hat{b}_i^{\prime\mu}$ and $p_i^{\prime\mu}$ are time-independent.
Together, $\hat{b}_i^{\prime\mu}$ and $\hat{p}_i^{\prime\mu}$ span
the space of allowed worldline operators,
and the only non-zero bracket is
\begin{align}\label{zDecomp}
	[\hat{b}_i^{\prime\mu},\hat{p}_{j,\nu}^\prime]=i\hbar\,\delta_{ij}\delta^\mu_\nu\,.
\end{align}
This follows from the equal-time commutator~\eqref{wlCommutator}.

The free evolution~\eqref{zOperator} implies the universal non-equal-time commutator:
\begin{align}\label{eq:wlComm}
[\hat z_i^\mu(\tau_i),\hat z_i^\nu(\tau_i^\prime)]
=
-\frac{i \hbar}{m_i}\,\eta^{\mu\nu}\,(\tau_i-\tau_i^\prime).
\end{align}
From this commutator one immediately deduces the
retarded and advanced worldline propagators:\footnote{Notice
	that, unlike for the bulk fields,
	we cannot straightforwardly define a Feynman propagator
	on the worldline,
	a.k.a.~$\Delta_{i,F}^{\mu\nu}(\tau_i-\tau_i^\prime):=
	\mathcal{T}(\hat{z}_i^\mu(\tau_i)\hat{z}_i^\nu(\tau_i^\prime))$.
	We can, however, write down a time-symmetric propagator
	by averaging over advanced and retarded:
	$\Delta_{i,T}(\tau_i-\tau_i^\prime):=
	\Delta^{\mu\nu}_{i,R}(\tau_i-\tau_i^\prime)+\Delta^{\mu\nu}_{i,A}(\tau_i-\tau_i^\prime)=
	-(i\hbar/m_i)\eta^{\mu\nu}|\tau_i-\tau_i^\prime|$.
}
\begin{subequations}\label{wlProps}
\begin{align}
	\begin{tikzpicture}[baseline={([yshift=+.5ex](0,-.2))}]
    	\coordinate (in) at (-0.6,0);
    	\coordinate (out) at (1.4,0);
    	\coordinate (x) at (-.2,0);
    	\coordinate (y) at (1.0,0);
    	\draw [zParticle] (x) -- (y);
    	\draw [worldlineStatic] (in) -- (x);
    	\draw [worldlineStatic] (y) -- (out);
   		\draw [fill] (x) circle (.06) node [above] {$\nu$} node [below] {$\tau_i^\prime$};
    	\draw [fill] (y) circle (.06) node [above] {$\mu$} node [below] {$\tau_i$};
  	\end{tikzpicture}
	&=\Delta_{i,R}^{\mu\nu}(\tau_i-\tau_i^\prime)
	:=
	\theta(\tau_i-\tau_i^\prime)\,
	[\hat{z}_i^\mu(\tau_i),\hat{z}_i^\nu(\tau_i^\prime)]
	\label{propWLR}\\
	&=
	-\frac{i  \hbar}{m_i}\eta^{\mu\nu}\,\theta(\tau_i-\tau_i^\prime)(\tau_i-\tau_i^\prime)
	=\frac{i  \hbar}{m_i}
	\int_\omega
	e^{-i\omega(\tau_i-\tau_i^\prime)}
	\frac{\eta^{\mu\nu}}{(\omega+i0)^2},
	\nn\\
	\begin{tikzpicture}[baseline={([yshift=+.5ex](0,-.2))}]
    	\coordinate (in) at (-0.6,0);
    	\coordinate (out) at (1.4,0);
    	\coordinate (x) at (-.2,0);
    	\coordinate (y) at (1.0,0);
    	\draw [zParticle] (y) -- (x);
    	\draw [worldlineStatic] (in) -- (x);
    	\draw [worldlineStatic] (y) -- (out);
   		\draw [fill] (x) circle (.06) node [above] {$\nu$} node [below] {$\tau_i^\prime$};
    	\draw [fill] (y) circle (.06) node [above] {$\mu$} node [below] {$\tau_i$};
  	\end{tikzpicture}
	&=\Delta_{i,A}^{\mu\nu}(\tau_i-\tau_i^\prime)
	:=
	-\theta(\tau_i^\prime-\tau_i)\,
	[\hat{z}_i^\mu(\tau_i),\hat{z}_i^\nu(\tau_i^\prime)]\label{propWLA}\\
	&=-\frac{i  \hbar}{m_i}\eta^{\mu\nu}\,\theta(\tau_i^\prime-\tau_i)(\tau_i^\prime-\tau_i)
	=
	\frac{i  \hbar}{m_i}
	\int_\omega
	e^{-i\omega(\tau_i-\tau_i^\prime)}
	\frac{\eta^{\mu\nu}}{(\omega-i0)^2},
	\nn
\end{align}
\end{subequations}
where $\int_\omega := \int\!\d\omega/(2\pi)$.
Again, one can easily verify the momentum space expressions by performing
the appropriate contour integral.

The difference between retarded and advanced worldline propagators
defines the causal (directed) cut propagator:
\begin{align}
\begin{aligned}
	\begin{tikzpicture}[baseline={([yshift=+.5ex](0,-.2))}]
    	\coordinate (in) at (-0.6,0);
    	\coordinate (out) at (1.4,0);
    	\coordinate (x) at (-.2,0);
    	\coordinate (y) at (1.0,0);
    	\draw [worldlineCut2] (x) -- (y);
    	\draw [worldlineStatic] (in) -- (x);
    	\draw [worldlineStatic] (y) -- (out);
   		\draw [fill] (x) circle (.06) node [above] {$\nu$} node [below] {$\tau_i^\prime$};
    	\draw [fill] (y) circle (.06) node [above] {$\mu$} node [below] {$\tau_i$};
  	\end{tikzpicture}&=
	\Delta^{\mu\nu}_{i,R}(\tau_i-\tau_i^\prime)
	-
	\Delta^{\mu\nu}_{i,A}(\tau_i-\tau_i^\prime)=
	[\hat{z}_i^\mu(\tau_i),\hat{z}_i^\nu(\tau_i^\prime)]\\
	&=
	-\frac{i  \hbar}{m_i}\,\eta^{\mu\nu}\,(\tau_i-\tau_i^\prime)=
	-\,
	\hbar \frac{\eta^{\mu\nu}}{m_i}
	\int_\omega
	e^{-i\omega(\tau_i-\tau_i^\prime)}\,
	\dd'(\omega)\,,
\end{aligned}
\end{align}
where $\dd'(\omega) := 2\pi\,\delta'(\omega)$.
As discussed in \rcite{Haddad:2025cmw},
a simple consequence of this definition in the PL expansion
is that taking Poisson brackets between collections of WQFT diagrams produces cut worldline insertions:
\begin{align}\label{diagBracket}
\left\{\,\,\vcenter{\hbox{\begin{tikzpicture}[line cap=round,line join=round,x=1cm,y=1cm,scale=.6]
        \path [draw=black, worldlineStatic] (-4.7,0) -- (-3.83,0);
        \path [draw=black, worldlineStatic] (-3.17,0) -- (-2.2,0);
        \path [draw=black, worldlineStatic] (-4.7,-1.5) -- (-3.83,-1.5);
        \path [draw=black, worldlineStatic] (-3.17,-1.5) -- (-2.2,-1.5);
        \draw[pattern=north east lines] (-3.5,-0.75) ellipse (.55cm and .9cm);
        \filldraw[fill=white,color=white] (-3.5,-0.75) circle (9pt);
        \node (middle) at (-3.5,-0.75) {A}; 
    \end{tikzpicture}}}\,\,,\,\,
    \vcenter{\hbox{\begin{tikzpicture}[line cap=round,line join=round,x=1cm,y=1cm,scale=.6]
        \path [draw=black, worldlineStatic] (-4.7,0) -- (-3.83,0);
        \path [draw=black, worldlineStatic] (-3.17,0) -- (-2.2,0);
        \path [draw=black, worldlineStatic] (-4.7,-1.5) -- (-3.83,-1.5);
        \path [draw=black, worldlineStatic] (-3.17,-1.5) -- (-2.2,-1.5);
        \draw[pattern=north east lines] (-3.5,-0.75) ellipse (.55cm and .9cm);
        \filldraw[fill=white,color=white] (-3.5,-0.75) circle (9pt);
        \node (middle) at (-3.5,-0.75) {B}; 
\end{tikzpicture}}}\right\}=-\frac{i}{\hbar}\left(\,\vcenter{\hbox{\begin{tikzpicture}[line cap=round,line join=round,x=1cm,y=1cm,scale=.6]
        \path [draw=black, worldlineStatic] (-4.7,0) -- (-3.83,0);
        \path [draw=black, worldlineCut2] (-3.17,0) -- (-2.2,0);
        \path [draw=black, worldlineStatic] (-1.54,0) -- (-0.67,0);
        \path [draw=black, worldlineStatic] (-4.7,-1.5) -- (-3.83,-1.5);
        \path [draw=black, worldlineStatic] (-3.17,-1.5) -- (-2.2,-1.5);
        \path [draw=black, worldlineStatic] (-1.54,-1.5) -- (-0.67,-1.5);
        \draw[pattern=north east lines] (-3.5,-0.75) ellipse (.55cm and .9cm);
        \draw[pattern=north east lines] (-1.87,-0.75) ellipse (.55cm and .9cm);
        \filldraw[fill=white,color=white] (-3.5,-0.75) circle (9pt);
        \node (A) at (-3.5,-0.75) {A};
        \filldraw[fill=white,color=white] (-1.87,-0.75) circle (9pt);
        \node (B) at (-1.87,-0.75) {B}; 
    \end{tikzpicture}}}\,+\,\vcenter{\hbox{\begin{tikzpicture}[line cap=round,line join=round,x=1cm,y=1cm,scale=.6]
        \path [draw=black, worldlineStatic] (-4.7,0) -- (-3.83,0);
        \path [draw=black, worldlineStatic] (-3.17,0) -- (-2.2,0);
        \path [draw=black, worldlineStatic] (-1.54,0) -- (-0.67,0);
        \path [draw=black, worldlineStatic] (-4.7,-1.5) -- (-3.83,-1.5);
        \path [draw=black, worldlineCut2] (-3.17,-1.5) -- (-2.2,-1.5);
        \path [draw=black, worldlineStatic] (-1.54,-1.5) -- (-0.67,-1.5);
        \draw[pattern=north east lines] (-3.5,-0.75) ellipse (.55cm and .9cm);
        \draw[pattern=north east lines] (-1.87,-0.75) ellipse (.55cm and .9cm);
        \filldraw[fill=white,color=white] (-3.5,-0.75) circle (9pt);
        \node (A) at (-3.5,-0.75) {A};
        \filldraw[fill=white,color=white] (-1.87,-0.75) circle (9pt);
        \node (B) at (-1.87,-0.75) {B}; 
    \end{tikzpicture}}}\,\right)\,.
\end{align}
To obtain this relationship we have replaced the quantum commutator
with a classical Poisson bracket $[\hat{A},\hat{B}]\to i \hbar \{\bar{A},\bar{B}\}$,
which acts on the background variables
$\bar{x}_i^\mu(\tau_i)$ and $\bar{p}_i^\mu(\tau_i)$.
This follows from the background split symmetry of the PL interaction
Hamiltonian $\mathcal{H}_I^{(\rm PL)}(x)$ (to be introduced below) 
which depends on the worldline variables only through
$x_i^\mu=\bar{x}_i^\mu+z_i^\mu$ and $p_i^\mu=\bar{p}_i^\mu+p_i^{\prime\mu}$.
Thus, dependence on the operators $\hat{x}_i^\mu$ and $\hat{p}_i^\mu$ is fully captured by the corresponding background variables.
This equivalence will be revisited in more detail in the context of
observables in \sec{sec:observables},
and in \app{worldlinebrackets} ---
crucially, it does not hold in the context of SF.

Rather than introducing asymptotic worldline Fock states, we specify the worldline sector by imposing retarded boundary conditions on the fluctuation operators.
This is implemented by requiring that fluctuations vanish at
a reference time $t_{0}$:
\begin{subequations}\label{wlVacuum}
\begin{align}
\langle \psi_i(t_0) | \hat z_i^\mu(t_{0}) | \psi_i(t_0)\rangle &= 
\langle \psi_i(t_0) | \hat z_i^\mu(\tau_i) | \psi_i(t_0) \rangle = 0\,,\\
\langle \psi_i(t_0) | \hat p_i^{\prime\mu}(t_{0}) | \psi_i(t_0) \rangle &= 
\langle \psi_i(t_0) | \hat p_i^{\prime\mu}(\tau_i) | \psi_i(t_0) \rangle = 0\,.
\end{align}
\end{subequations}
This condition defines a reference state $|\psi_i(t_0)\rangle$
for the worldline fluctuations,
rather than a Fock-space vacuum,
and enforces retarded boundary conditions for their free equations of motion.
The extension to arbitrary time dependence ($\tau_i$) within the operators
$\hat z_i^\mu$ and $\hat p_i^{\prime\mu}$ above
follows from the free worldline equations and their solution in
\eqref{eq:wlEvolution}--\eqref{eq:wlEvolution_sol}.
In the case of free scattering,
one simply takes $t_0\to-\infty$,
aligning this boundary condition with that of the bulk fields.
For bound backgrounds, no asymptotic in-- or out--states exist for the worldline degrees of freedom, and so $t_{0}$ is kept finite,
corresponding to a chosen point along the background orbit.

At the free level the full Hilbert space factorises into bulk and worldline sectors, and we choose a factorised reference state:
\begin{align}
|0\rangle
&=
|0\rangle_\varphi \otimes |0\rangle_{\mathrm{wl}}\,, &
|0\rangle_{\mathrm{wl}} &:= |\psi_1(t_0)\rangle \otimes |\psi_2(t_0)\rangle .
\label{eq:vacuum_Hilbert}
\end{align}
Here $|0\rangle_\varphi$ is the standard Fock vacuum of the bulk scalar field,
while $|0\rangle_{\mathrm{wl}}$ is the reference state specifying boundary conditions~\eqref{wlVacuum} for the worldline fluctuation operators.
Notice that, as $|0\rangle_{\mathrm{wl}}$ is not a vacuum in the Fock--space sense,
it need not be time-translation invariant;
instead, it depends on the reference time $t_{0}$, 
though this dependence cancels from all orbit-averaged observables.
This situation contrasts with the typical assumption made in QFT
that eigenstates of the in-vacuum can be identified with those of the out-vacuum.

\subsection{Interactions}\label{sec:interactions}

Finally, let us introduce the different interactions present in the scalar theory.
This is where the two different possible backgrounds introduced in
\sec{sec:freeScattering} (PL) and \sec{sec:backgroundField}
(SF) play a role, leading to different interaction Hamiltonians in the PL and SF expansions.
In the PL expansion,
\begin{align}\label{eq:HintA_def}
\begin{aligned}
\cH_{I}^{(\rm PL)}(x)&=
g\,\cH_{I,\rm bulk}^{(\rm PL)}(x)
+\sum_{i=1}^2e_i\int\!\mathrm{d}\tau_i\,\delta^D(x-\bar{x}_i(\tau_i))
H^{(\rm PL)}_{I,i}(\tau_i)\,,\\
\cH^{(\rm PL)}_{I,\rm bulk}(x)&=\frac1{3!}\,\varphi^3(x)\,\qquad
H^{(\rm PL)}_{I,i}(\tau_i)=e^{z_i(\tau_i)\cdot\partial}\varphi(\bar{x}_i(\tau_i))\,,
\end{aligned}
\end{align}
corresponding to an expansion around the trivial background $\bar\phi=0$ and
straight--line trajectories $\bar{x}_i^\mu(\tau_i)=b_i^\mu+\tau_iv_i^\mu$.
From here one reads off the Feynman rules:
\begin{align}
\begin{tikzpicture}[baseline={(current bounding box.center)}]
        \draw [scalar2] (0,0) -- (1,0);
		\draw [scalar2] (0,0) -- (-.5,-.866);
		\draw [scalar2] (0,0) -- (-.5,0.866);
		\draw (0,0) node [left] {$x\,$};
		\draw [fill] (0,0) circle (.06);
    \end{tikzpicture}\,&=\,
	-\frac{ig}\hbar\int\!\d^Dx\,,\\
\begin{tikzpicture}[baseline={(current bounding box.center)},scale=1]
  		\coordinate (in) at (-1,0);
  		\coordinate (out1) at (1,0);
  		\coordinate (out2) at (1,0.5);
  		\coordinate (out3) at (1,0.9);
  		\coordinate (out4) at (1,1.4);
  		\coordinate (x) at (0,0);
		\draw (0,.3) node [left] {$x,\tau_i$};
  		\draw (out2) node [right] {$\!\!\!\vdots$};
  		\draw (out4) node [right] {$\mu_{n}$};
   		\draw (out3) node [right] {$\mu_{n-1}$};
   		\draw (out1) node [right] {$\mu_{1}$};
  		\draw [worldlineStatic] (in) -- (x);
  		\draw [zUndirected] (x) -- (out1);
  		\draw [zUndirected] (x) to[out=30,in=180] (out3);
  		\draw [zUndirected] (x) to[out=60,in=180] (out4);
		\draw [scalar2] (x) -- (0,-1);
  		\draw [fill] (x) circle (.06);
  \end{tikzpicture}\,\,&=-\frac{ie_i}\hbar\int\!\d^Dx\!\int\!\d\tau_i\,\delta^D(x-\bar{x}_i(\tau_i))
  \partial^x_{\mu_1}\partial^x_{\mu_2}\cdots\partial^x_{\mu_n}\,.
\end{align}
Notice that at worldline vertices we integrate over both a
position coordinate $x$ and a proper time coordinate $\tau_i$,
constrained by the delta function $\delta^D(x-\bar{x}_i(\tau_i))$.
This allows us to attach both bulk and worldline propagators
without ambiguity.
A similar set of configuration space WQFT Feynman rules was also introduced
in ref.~\cite{Hoogeveen:2025tew} in the context of the probe limit,
where all interactions take place on a single worldline with coordinate time $\tau$.

In the SF expansion,
the theory is expanded around a non--trivial static background scalar field
$\bar\phi(x)$, together with a Keplerian background trajectory for the light particle
$L$ and a straight background trajectory $\bar{x}_H^\mu=\tau_Hv_H^\mu$ for the heavy particle
$H$. The interaction Hamiltonian density reads, in this notation,
\begin{align}\label{eq:HintB_def}
\begin{aligned}
\cH_{I}^{(\rm SF)}(x)
&=
\frac{g}{3!}
\left(
\varphi^3
+3\bar{\phi}^2\varphi
+3\bar\phi\varphi^2
+\bar\phi^3
\right)\\
&\quad
+e_L\int\!{\rm d}\tau_L\,\delta^D(x-\bar{x}_L(\tau_L))
\Big[
e^{z_L\cdot\partial}\varphi(x)
+ \bigl(
e^{z_L\cdot\partial}\bar{\phi}
-\bar{\phi}
- z_L \cdot \partial \bar{\phi}
\bigr)
\Big]\\
&\quad
+e_H\int\!{\rm d}\tau_H\,\delta^D(x-\bar{x}_H(\tau_H))
\Big[
e^{z_H\cdot\partial}\varphi(x)-\varphi(x)
\Big]\,.
\end{aligned}
\end{align}
From this expression one reads off the corresponding configuration--space WQFT Feynman rules. As in the PL expansion, derivatives generated by the translation operators $e^{z_{L,H}\cdot\partial}$ act on the scalar line attached to the vertex. The background scalar field $\bar\phi(x)$ and the background worldlines $\bar x_{L,H}(\tau)$ are treated as fixed classical sources.

In the SF expansion $\lambda\ll1$, we parameterise the light particle coupling as $e_L=\lambda\,e_H\,m_L/m_H$ and assign the uniform scaling
\begin{align}
\varphi&=\mathcal O(\lambda)\,, &
z_L^\mu&=\mathcal O(\lambda)\,, &
z_H^\mu&=\mathcal O(\lambda)\,.
\end{align}
Keeping only terms contributing at first order in the self-force,
i.e.~terms up to $\cO(\lambda^2)$,
the interaction Hamiltonian \eqref{eq:HintB_def} reduces to
\begin{align}
\label{eq:HintB_1SF}
\cH_{I}^{(\rm SF)}(x) \Big|_{\lambda^2, g}
&=
\frac{g}{3!}
\left(
3\bar{\phi}^2\varphi
+3\bar\phi\varphi^2
+\bar\phi^3
\right) \\
&+e_L\int\!{\rm d}\tau_L\,\delta^D(x-\bar{x}_L(\tau_L))
 \varphi(x) +e_H\int\!{\rm d}\tau_H\,\delta^D(x-\bar{x}_H(\tau_H)) z_H\cdot\partial \varphi(x)
\,. \nonumber
\end{align}
Since the background is chosen to solve the equations only at $g=0$, the
terms proportional to $g$ in the fluctuating Hamiltonian are genuine
interaction insertions around this integrable background.
The corresponding Feynman rules up to first self-force order
on the worldlines are
\begin{subequations}
\begin{align}
\hspace{-10pt}
\begin{tikzpicture}[baseline={(current bounding box.center)},scale=1]
  \coordinate (in) at (-1,0);
  \coordinate (x) at (0,0);
  \coordinate (out1) at (1,0);
  \coordinate (out3) at (1,0.9);
  \coordinate (out4) at (1,1.4);
  \draw (-.2,.3) node [left] {$x,\tau_L$};
  \draw [worldlineStatic] (in) -- (x);
  \draw [worldlineStatic] (x) -- (out1);
  \draw [scalar2] (x) -- (0,-1);
  \node[draw, circle, fill=white, inner sep=1.5pt, minimum size=12pt] at (x) {$L$};
\end{tikzpicture}
\,&=
-\frac{i}{\hbar} e_L\int\!\d^Dx\!\int\!\d\tau_L\,\delta^D\!\bigl(x-\bar x_L(\tau_L)\bigr)\,, \\
\begin{tikzpicture}[baseline={(current bounding box.center)},scale=1]
  \coordinate (in) at (-1,0);
  \coordinate (x) at (0,0);
  \coordinate (out1) at (1,0);
  \draw (-.2,.3) node [left] {$x,\tau_H$};
  \draw (out1) node [right] {$\mu$};
  \draw [worldlineStatic] (in) -- (x);
  \draw [scalar2] (x) -- (0,-1);
  \draw [zUndirected] (x) -- (out1);
  \node[draw, circle, fill=white, inner sep=1.5pt, minimum size=12pt] at (x) {$H$};
\end{tikzpicture}
\,&=
-\frac{i}{\hbar} e_H\int\!\d^Dx\!\int\!\d\tau_H\,\delta^D\!\bigl(x-\bar x_H(\tau_H)\bigr)\,
\partial^x_{\mu}\,.
\label{eq:WL_vertices_caseB}
\end{align}
\end{subequations}
Notice that, seeing as the 0SF dynamics is already captured by the background motion,
there is no single-emission vertex (without a deflection mode) from the heavy body.
In the bulk,
\begin{subequations}\label{eq:bulk_vertices_caseB}
\begin{align}
\begin{tikzpicture}[baseline={(current bounding box.center)}]
  \draw [scalar2] (0,0) -- (1.25,0);
  \draw [scalar2] (0,0) -- (-1.25,0);
  \node[draw, circle, fill=white, inner sep=1.5pt, minimum size=12pt] 
       at (0,0) {\small $\mathcal{V}_{2}$};
\end{tikzpicture}
&=\;
-\frac{ig}{\hbar}\int\!\d^Dx\,\bar\phi(x)\,,
\label{eq:bulk2_caseB}
\\[1.0em]
\begin{tikzpicture}[baseline={(current bounding box.center)}]
  \draw [scalar2] (0,0) -- (1.25,0);
  \node[draw, circle, fill=white, inner sep=1.5pt, minimum size=12pt] 
       at (0,0) {\small $\mathcal{V}_{1}$};
\end{tikzpicture}
&=\;
-\frac{ig}{2\hbar}\int\!\d^Dx\,\bar\phi(x)^2\,.
\label{eq:bulk1_caseB}
\end{align}
\end{subequations}
The Feynman rules above fully specify the perturbative expansion. For notational uniformity, we denote the effective bulk insertions appearing in radiative matrix elements by $\mathcal V_i$,
with $i=1,2$ corresponding to one or two radiative scalar legs respectively.
These arise from expanding the cubic interaction around the background field $\bar\phi(x)$ and represent effective background-field insertions.
Worldline vertices on $L$ and $H$ encode multipole insertions through powers of the deflection variables $z_{L,H}$.

\subsection{The Magnus series}\label{sec:magnus}

We now turn to the construction of the time--evolution operator in the interaction picture.
The interaction--picture states obey the Schr\"odinger equation
\begin{align}
	i\,\hbar \,\frac{\mathrm d}{\mathrm dt}|\psi(t)\rangle
	=
	\hat H_I(t)\,|\psi(t)\rangle\,,
	\label{eq:SchroedingerMagnus}
\end{align}
whose solution can be written in terms of the evolution operator
\begin{align}\label{eq:Udef}
	|\psi(t)\rangle
	&=
	\hat U(t,t_0)\,|\psi(t_0)\rangle\,, &
	i\hbar\,\frac{\mathrm d}{\mathrm dt}\hat U(t,t_0)
	&=
	\hat H_I(t)\,\hat U(t,t_0)\,,
\end{align}
with initial condition $\hat U(t_0,t_0)=\mathbb I$. Rather than using the Dyson series, it will be convenient to adopt an exponential
representation of the evolution operator,
\begin{align}\label{eq:MagnusDef}
	\hat U(t,t_0)
	=
	\exp\left[\frac{i}{\hbar}\hat N(t,t_0)\right],
\end{align}
where $\hat N(t,t_0)$ is a Hermitian operator, $\hat N=\hat N^\dagger$,
as required by unitarity. The operator $\hat N$ admits a perturbative expansion,
preserving unitarity order-by-order:
\begin{align}\label{magnusSeries}
	\hat N(t,t_0)
	=
	\eta\hat N^{(1)}(t,t_0)
	+
	\eta^2\hat N^{(2)}(t,t_0)
	+
	\eta^3\hat N^{(3)}(t,t_0)
	+\cdots,
\end{align}
known as the \emph{Magnus expansion}.
Here $\eta$ is a formal parameter counting powers of the interaction Hamiltonian,
which from now on we set to one.
For a general, explicitly time--dependent interaction Hamiltonian,
the first three terms are
\begin{subequations}\label{eq:MagnusTerms}
\begin{align}
	\frac{i}{\hbar}\hat N^{(1)}(t,t_0)
	&=
	-\frac{i}{\hbar}
	\int_{t_0}^{t}\!\mathrm dt_1\,
	\hat H_I(t_1),
	\label{eq:Magnus1}
	\\[0.5em]
	\frac{i}{\hbar}\hat N^{(2)}(t,t_0)
	&=
	\frac{(-i)^2}{2 \hbar^2}
	\int_{t_0}^{t}\!\mathrm dt_1
	\int_{t_0}^{t_1}\!\mathrm dt_2\,
	[\hat H_I(t_1),\hat H_I(t_2)],
	\label{eq:Magnus2}
	\\[0.5em]
	\frac{i}{\hbar}\hat N^{(3)}(t,t_0)
	&=
	\frac{(-i)^3}{6 \hbar^3}
	\int_{t_0}^{t}\!\mathrm dt_1
	\int_{t_0}^{t_1}\!\mathrm dt_2
	\int_{t_0}^{t_2}\!\mathrm dt_3
	\nonumber\\
	&\quad\times
	\Big(
	[\hat H_I(t_1),[\hat H_I(t_2),\hat H_I(t_3)]]
	+
	[\hat H_I(t_3),[\hat H_I(t_2),\hat H_I(t_1)]]
	\Big)\,.
	\label{eq:Magnus3}
\end{align}
\end{subequations}
For scattering observables, the interval is taken to cover the entire
interaction history, $t_0\to -\infty$ and $t\to +\infty$,
with the scattering $\hat{S}$-matrix:
\begin{align}
	\hat S &= \exp\left[\frac{i}{\hbar}\hat N\right]\,, &
	\hat N &:= \hat N(+\infty,-\infty).
\end{align}
In this case the Magnus expansion is more conveniently written in terms of the Hamiltonian density:
\begin{subequations}\label{Magnus}
\begin{align}
    \frac{i}{\hbar}\hat N^{(1)} &= -\frac{i}{\hbar}\int\!\d^Dx\ \hat\cH_I(x)\,, \label{eq:N1} \\
    \frac{i}{\hbar}\hat N^{(2)}&=\frac{(-i)^2}{2 \hbar^2}\int\!\d^Dx\,\d^Dy\,
    \theta_{xy}\,[\hat\cH_I(x),\hat\cH_I(y)]\,,  \label{eq:N2} \\
   	\frac{i}{\hbar}\hat N^{(3)}&=\frac{(-i)^3}{6 \hbar^3}\int\!\d^Dx\,\d^Dy\,\d^Dz\,
    \theta_{xy}\theta_{yz}\nn\\
	&\quad\times
	\left([\hat\cH_I(x),[\hat\cH_I(y),\hat\cH_I(z)]]
	+[\hat\cH_I(z),[\hat\cH_I(y),\hat\cH_I(x)]]\right)\,.
    \label{eq:N3} 
\end{align}
\end{subequations}
Here we have introduced the step functions $\theta_{xy}:=\theta(x^0-y^0)$,
which conveniently encode the boundary conditions on time coordinates
and give a fixed causal ordering.
Thus, the Magnus expansion avoids the use of time--ordered products,
making it particularly suitable for systems with retarded boundary conditions.

In the following sections,
we will see how the Magnus expansion can be directly applied
to compute matrix elements of $\hat{N}$,
which we refer to as Magnus amplitudes:
\begin{subequations}
\begin{align}
	N(t,t_0):=&\,\langle0|\hat{N}(t,t_0)|0\rangle\,,\\
	N(t,t_0;k):=&\,\langle k|\hat{N}(t,t_0)|0\rangle/\hbar\,,\\
	N(t,t_0;k_1,k_2):=&\,\langle k_1k_2|\hat{N}(t,t_0)|0\rangle/\hbar^2\,.
\end{align}
\end{subequations}
When dealing with scattering,
where $t\to+\infty$ and $t_0\to-\infty$,
we drop the explicit time dependence on $N$.
Notice that, even when dealing with bound orbits with a finite time interval,
it is still meaningful to contract with external states:
reflecting the fact that a bound system can also emit radiation.
As the $\hat{N}$ matrix enjoys crossing symmetry at tree level,
and we orient all momenta as outgoing,
by convention we consider states contracted only on the left ---
external states on the right, which are incoming,
are obtained by sending $k_i^\mu\to-k_i^\mu$.

As in conventional $\hat{S}$-matrix theory involving the Dyson series,
assembling these matrix elements explicitly using the Magnus series
rapidly grows in complexity.
It therefore becomes preferable to use the WQFT Feynman rules.
The main additional subtlety when computing matrix elements of $\hat{N}$,
as compared with elements of $\hat{S}$,
is the appearance of non-trivial weightings of diagrams
with different internal causality flows.
These arise due to the need to recombine the step functions $\theta_{xy}$
appearing in the Magnus expansion to achieve Lorentz covariance.
Yet, as has been showcased in important work by
Brandhuber, Brown, Pichini, Travaglini and Vives Matasan~\cite{Brandhuber:2025igz},
even in the comparably simple $\phi^3$ theory such recombinations are subtle.
In traditional $\hat{S}$-matrix theory involving the Dyson series,
where all propagators are time-symmetric,
there is no analog of this problem.

Fortunately, the issue of weighting diagrams with different causality
flows has already been addressed for us.
In \rcite{Kim:2024svw}
the relative weighting factors of causality prescriptions in the 
Magnus expansion were identified as Murua coefficients~\cite{murua}.
This was explicitly confirmed for several non-trivial examples in
$\phi^3$ theory in \rcite{Brandhuber:2025igz},
as high as six external legs.
The same combinatorial analysis applies to WQFT diagrams, since the Murua
coefficients depend only on the time-ordering structure of the Magnus
series.
In this paper, we require the Murua coefficients
up to four vertices
(reproduced from~\rcite{Brandhuber:2025igz}):
\begin{subequations}\label{murua}
\begin{align}
	\omega\left(
	\begin{tikzpicture}[baseline={([yshift=-.6ex](0,0))}]
        \filldraw[fill=black] (0,0) circle (.06);
    \end{tikzpicture}
	\right)&=1\,,\\
	\omega\left(
	\begin{tikzpicture}[baseline={([yshift=-.5ex](0,0))}]
        \path [draw=black, zParticle] (0,0) -- (.8,0);
        \filldraw[fill=black] (0,0) circle (.06);
		\filldraw[fill=black] (.8,0) circle (.06);
    \end{tikzpicture}
	\right)&=\frac12\,,\label{murua2pt}\\
	\omega\left(
	\begin{tikzpicture}[baseline={([yshift=-.5ex](0,0))}]
        \path [draw=black, zParticle] (0,0) -- (.8,0);
		\path [draw=black, zParticle] (.8,0) -- (1.6,0);
        \filldraw[fill=black] (0,0) circle (.06);
		\filldraw[fill=black] (.8,0) circle (.06);
		\filldraw[fill=black] (1.6,0) circle (.06);
    \end{tikzpicture}
	\right)&=\frac13\,, &
	\omega\left(
	\begin{tikzpicture}[baseline={([yshift=-.5ex](0,0))}]
        \path [draw=black, zParticle] (0,0) -- (.8,0);
		\path [draw=black, zParticle] (1.6,0) -- (.8,0);
        \filldraw[fill=black] (0,0) circle (.06);
		\filldraw[fill=black] (0.8,0) circle (.06);
		\filldraw[fill=black] (1.6,0) circle (.06);
    \end{tikzpicture}
	\right)&=\frac16\,,\label{murua3pt}\\
	\omega\left(
	\begin{tikzpicture}[baseline={([yshift=-.5ex](0,0))}]
        \path [draw=black, zParticle] (0,0) -- (0.8,0);
		\path [draw=black, zParticle] (0.8,0) -- (1.6,0);
		\path [draw=black, zParticle] (1.6,0) -- (2.4,0);
        \filldraw[fill=black] (0,0) circle (.06);
		\filldraw[fill=black] (0.8,0) circle (.06);
		\filldraw[fill=black] (1.6,0) circle (.06);
		\filldraw[fill=black] (2.4,0) circle (.06);
    \end{tikzpicture}
	\right)&=\frac14\,, &
	\omega\left(
	\begin{tikzpicture}[baseline={([yshift=-.5ex](0,0))}]
        \path [draw=black, zParticle] (0,0) -- (0.8,0);
		\path [draw=black, zParticle] (1.6,0) -- (0.8,0);
		\path [draw=black, zParticle] (1.6,0) -- (2.4,0);
        \filldraw[fill=black] (0,0) circle (.06);
		\filldraw[fill=black] (0.8,0) circle (.06);
		\filldraw[fill=black] (1.6,0) circle (.06);
		\filldraw[fill=black] (2.4,0) circle (.06);
    \end{tikzpicture}
	\right)&=
	\omega\left(
	\begin{tikzpicture}[baseline={([yshift=-.5ex](0,0))}]
        \path [draw=black, zParticle] (0,0) -- (0.8,0);
		\path [draw=black, zParticle] (0.8,0) -- (1.6,0);
		\path [draw=black, zParticle] (2.4,0) -- (1.6,0);
        \filldraw[fill=black] (0,0) circle (.06);
		\filldraw[fill=black] (0.8,0) circle (.06);
		\filldraw[fill=black] (1.6,0) circle (.06);
		\filldraw[fill=black] (2.4,0) circle (.06);
    \end{tikzpicture}
	\right)=\frac1{12}\,,\nn \\
	\omega\left(
	\begin{tikzpicture}[baseline={([yshift=-.5ex](0,0))}]
        \path [draw=black, zParticle] (1,0) -- (0.2,0);
		\path [draw=black, zParticle] (1,0) -- (1.5656,0.5656);
		\path [draw=black, zParticle] (1,0) -- (1.5656,-.5656);
        \filldraw[fill=black] (0.2,0) circle (.06);
		\filldraw[fill=black] (1,0) circle (.06);
		\filldraw[fill=black] (1.5656,0.5656) circle (.06);
		\filldraw[fill=black] (1.5656,-.5656) circle (.06);
    \end{tikzpicture}\,
	\right)&=0\,, & 
	\omega\left(
	\begin{tikzpicture}[baseline={([yshift=-.5ex](0,0))}]
        \path [draw=black, zParticle] (0.2,0) -- (1,0);
		\path [draw=black, zParticle] (1,0) -- (1.5656,0.5656);
		\path [draw=black, zParticle] (1,0) -- (1.5656,-.5656);
        \filldraw[fill=black] (0.2,0) circle (.06);
		\filldraw[fill=black] (1,0) circle (.06);
		\filldraw[fill=black] (1.5656,0.5656) circle (.06);
		\filldraw[fill=black] (1.5656,-.5656) circle (.06);
    \end{tikzpicture}\,
	\right)&=\frac16\,.\label{murua4pt}
\end{align}
\end{subequations}
These coefficients multiply Feynman diagrams with the corresponding
causality flows,
and appear together with the usual symmetry factors ---
we will see several examples in the following sections.
The Murua coefficients do not depend in any way on the details
of the underlying theory, which internal states are propagating,
or the external states a diagram may have.
For more general expressions at higher points,
we refer the interested reader to the \texttt{Mathematica}
code attached to ref.~\cite{Kim:2024svw}.

As the topic has already been covered in other literature \cite{Kim:2024svw,Brandhuber:2025igz},
we briefly remark on the Murua coefficients' main properties:
\begin{enumerate}
	\item Coefficients with all internal arrows reversed have identical values ---
	a simple consequence of $\hat{N}=\hat{N}^\dagger$.
	\item The sum over coefficients with all possible arrow directions on a given topology (i.e.~all possible variations of retarded/advanced causality flows) is unity.
	This reflects the coefficients' role as weighting factors.
	\item The sum of two coefficients that are identical --- 
	except for the direction of a single internal line ---
	gives the same coefficient but with that internal edge removed.
	A good example is the two coefficients in \eqn{murua3pt},
	which give $1/3+1/6=1/2$~\eqref{murua2pt} when added together.
\end{enumerate}
The last property, known as the \emph{edge contraction rule},
is particularly useful when considering passive propagators in WQFT.
If the internal $i0$ prescription plays no role,
meaning there is no distinction between advanced and retarded propagators,
one may simply delete the corresponding
internal line from a diagram's Murua coefficient.
Thus, Murua coefficients may be assigned to diagrams considering only the active propagators,
whose $i0$ prescription matters for the causality flow.

\section{Magnus amplitudes: post-Lorentzian expansion}
\label{sec:Magnus_caseA}

In this section we study scattering in the perturbative PL coupling expansion,
setting $\hbar=1$ for simplicity. Throughout this section the superscript $(\rm PL)$ on the Hamiltonian and on $\hat N$ is implicit.
We begin by considering radiative examples involving external scalar states $|k_1k_2\cdots k_n\rangle$,
as these involve the Magnus expansion~\eqref{Magnus} at lower perturbative orders.
Then we consider the vacuum element $\langle0|i\hat{N}|0\rangle$.

\subsection{Radiative Magnus amplitudes}

We work through the terms in the perturbative
Magnus series~\eqref{magnusSeries} one by one.

\paragraph{Matrix elements of $\hat{N}^{(1)}$}

Inserting the interaction Hamiltonian~\eqref{eq:HintA_def} into
the leading-order Magnus operator $\hat N^{(1)}$~\eqref{eq:N1} we obtain\footnote{When performing these contractions we ignore loops, which in the WQFT are associated with quantum effects.}
\begin{subequations}
\begin{align}
	\hspace{-6pt}\left.i\hat{N}^{(1)}\right|_{e_i}&=
	-i \int\!\d^Dx\!\int\!{\rm d}\tau_i\,\delta^D(x-\bar{x}_i(\tau_i))
	\hat{H}_{I,i}(\tau_i)
	=-i e_i \int\!{\rm d}\tau_i\,
	e^{\hat{z}_i(\tau_i)\cdot\partial}\hat\varphi(\bar{x}_i(\tau_i))\,,\\
    \hspace{-6pt}\left.i\hat{N}^{(1)}\right|_g&=-i \int\!\d^Dx\,\hat\cH_{I,\rm bulk}(x)=
	-i g\int\!\d^Dx\,\frac{1}{3!}\hat\varphi^3(x)\,.
\end{align}
\end{subequations}
Notice how, on the worldlines,
integration on $x$ is trivially resolved by the Dirac delta function $\delta^D(x-\bar{x}_i(\tau_i))$,
leaving behind an integration on the proper time coordinates $\tau_i$.
Using $\langle k|\hat\varphi(x)|0\rangle=e^{ik\cdot x}$,
\begin{subequations}
\begin{align}
    \label{eq:current}
	&iN^{(1)}(k):=\langle k|i\hat{N}^{(1)}|0\rangle=
	-i\sum_{i=1}^2e_i\int\!{\rm d}\tau_i\,e^{ik\cdot\bar{x}_i(\tau_i)}=
	-i\sum_{i=1}^2e_i\,e^{ik\cdot b_i}\dd(k\cdot v_i)\,,\\
	&iN^{(1)}(k_1,k_2,k_3):=\langle k_1k_2k_3|i\hat{N}^{(1)}|0\rangle\nonumber \\
    &\qquad \qquad \qquad \quad =
	-ig\int\!\d^Dx\,e^{i(k_1+k_2+k_3)\cdot x}=-ig\,\dd(k_1+k_2+k_3)\,,
\end{align}
\end{subequations}
where $\dd(k):=(2\pi)^D\delta^D(k)$.
In terms of WQFT Feynman diagrams,
\begin{align}
	iN^{(1)}(k)&=
	\begin{tikzpicture}[baseline={(current bounding box.center)},scale=1]
  		\coordinate (in) at (-1,0);
  		\coordinate (out) at (1,0);
  		\coordinate (x) at (0,0);
  		\draw [worldlineStatic] (in) -- (out);
		\draw [scalar2] (x) -- (0,-1) node [right] {$k$};
  		\draw [fill] (x) circle (.06);
  	\end{tikzpicture}
	+
	\begin{tikzpicture}[baseline={(current bounding box.center)},scale=1]
  		\coordinate (in) at (-1,0);
  		\coordinate (out) at (1,0);
  		\coordinate (x) at (0,0);
  		\draw [worldlineStatic] (in) -- (out);
		\draw [scalar2] (x) -- (0,1) node [right] {$k$};
  		\draw [fill] (x) circle (.06);
  	\end{tikzpicture}\,,
	&
	iN^{(1)}(k_1,k_2,k_3)&=
	\begin{tikzpicture}[baseline={(current bounding box.center)}]
        \draw [scalar2] (0,0) -- (1,0);
		\draw [scalar2] (0,0) -- (-.5,-.866);
		\draw [scalar2] (0,0) -- (-.5,0.866);
		\draw (-.5,-.866) node [left] {$k_1$};
		\draw (-.5,.866) node [left] {$k_2$};
		\draw (1,0) node [right] {$k_3$};
		\draw [fill] (0,0) circle (.06);
	\end{tikzpicture}\,.
\end{align}
The first case represents the sum of the static potentials of two point charges;
the second, which does not involve either worldline,
is an ordinary three-scalar amplitude.

\paragraph{Matrix elements of $\hat{N}^{(2)}$}

Next we examine the $e_1^2$ component of $\hat{N}^{(2)}$
(which is related to $e_2^2$ by symmetry):
\begin{align}
\begin{aligned}
	\left.i\hat{N}^{(2)}\right|_{e_1^2}&=\frac{(-i)^2}{2} \int\!\d^Dx\d^Dy\ \theta_{xy}\,
	[\hat\cH_{I,1}(x),\hat\cH_{I,1}(y)] \\
	&=\frac{(-i)^2}{2} \int\!\d\tau_1\d\tau_1^\prime \,\theta(\tau_1-\tau_1^\prime )
	[\hat{H}_{I,1}(\tau_1),\hat{H}_{I,1}(\tau_1^\prime )]\,.
\end{aligned}
\end{align}
We have inserted here
$\hat{\cH}_{I,i}(x)=\int\!\d\tau_i\,\delta^D(x-\bar{x}_i(\tau_i))\hat{H}_{I,i}(\tau_i)$
and integrated on $x$ and $y$, using also the identity
\begin{align}
	\theta(\bar{x}_i^0(\tau_i)-\bar{x}_i^0(\tau_i^\prime ))&=\theta(\tau_i-\tau_i^\prime )\,,
\end{align}
which holds due to $\bar{x}_i^\mu$ being timelike vectors.\footnote{
	This identity can also be derived explicitly:
	$\theta(\bar{x}_i^0(\tau_i)-\bar{x}_i^0(\tau_i^\prime ))=
	\theta(b_i^0+\tau_iv_i^0-b_i^0-\tau_i^\prime v_i^0)=
	\theta((\tau_i-\tau_i^\prime )v_i^0)=\theta(\tau_i-\tau_i^\prime )$,
	because $v_i^0>0$ in any choice of reference frame.
}
We require terms from $\hat{H}_{I,1}$ at linear order in $\hat{z}_1$:
\begin{align}
	\left.i\hat{N}^{(2)}\right|_{e_1^2}
	&=\frac{(-i)^2}{2} e_1^2\int\!\d\tau_1\d\tau_1^\prime \,\theta(\tau_1-\tau_1^\prime )
	[\hat{z}_1^\mu(\tau_1)\partial_\mu\hat\varphi(\bar{x}_1(\tau_1)),\hat{z}_1^\nu(\tau_1^\prime )\partial_\nu\hat\varphi(\bar{x}_1(\tau_1^\prime ))]+\cdots\nn\\
	&=\frac{(-i)^2}{2} e_1^2\int\!\d\tau_1\d\tau_1^\prime \,
	\partial_\mu\hat\varphi(\bar{x}_1(\tau_1))\partial_\nu\hat\varphi(\bar{x}_1(\tau_1^\prime ))
	\Delta_{1,R}^{\mu\nu}(\tau_1-\tau_1^\prime )+\cdots\,.
\end{align}
Thus, we have recovered the retarded worldline propagator~\eqref{propWLR}.
Notice that, seeing as $\Delta_{1,R}^{\mu\nu}(\tau_1-\tau_1')=0$ when $\tau_1<\tau_1'$,
the retarded propagator automatically enforces appropriate boundary condition on the time integrals.
Thus, both the $\tau_1$ and $\tau_1^\prime$ integrals may be written over the full time domain.

Taking the matrix element $iN^{(2)}(k_1,k_2)=\langle k_1k_2|i\hat{N}^{(2)}|0\rangle$,
\begin{align}
	\left.i N^{(2)}(k_1,k_2)\right|_{e_1^2}
	&=\frac{(-i)^2}{2} e_1^2\int\!\d\tau_1\d\tau_1^\prime \,
	(ik_{1,\mu})(ik_{2,\nu})
	e^{i(k_1\cdot\bar{x}_1(\tau_1)+k_2\cdot\bar{x}_1(\tau_1^\prime ))}\\
    &\qquad\qquad\qquad \times \Big[\Delta_{1,R}^{\mu\nu}(\tau_1-\tau_1^\prime )+\Delta_{1,A}^{\mu\nu}(\tau_1-\tau_1^\prime )\Big]\nn\\
	&=\frac12\, e_1^2\,k_{1,\mu}k_{2,\nu}
	\int_{\omega}e^{i(k_1+k_2)\cdot b_1}\dd(k_1\cdot v_1-\omega)\dd(k_2\cdot v_1+\omega) \nn \\
    &\qquad\qquad\qquad \times \Big[\widetilde\Delta_{1,R}^{\mu\nu}(\omega)+\widetilde\Delta_{1,A}^{\mu\nu}(\omega)\Big]\nn\\
	&=\frac{i e_1^2}{2m_1}e^{i(k_1+k_2)\cdot b_1}\dd((k_1+k_2)\cdot v_1)
	\left(\frac{k_1\cdot k_2}{(k_1\cdot v_1+i0)^2}+\frac{k_1\cdot k_2}{(k_1\cdot v_1-i0)^2}\right)\,,\nn
\end{align}
where in the second step, using the worldline propagators~\eqref{wlProps},
we have moved to momentum space.
Diagrammatically, this expression is represented as
\begin{align}
	\left.iN^{(2)}(k_1,k_2)\right|_{e_1^2}=\,\,
	\frac12\,\,\begin{tikzpicture}[baseline={([yshift=-1.5ex](0,-.2))}]
		\path [draw=black, worldlineStatic] (-1,0) -- (-.5,0);
        \path [draw=black, zParticle] (-0.5,0) -- (0.5,0);
		\path [draw=black, worldlineStatic] (0.5,0) -- (1.0,0);
		\path [draw=black, scalar2] (0.5,0) -- (.5,-.8);
		\path [draw=black, scalar2] (-0.5,0) -- (-.5,-.8);
		\draw (-.5,-.5) node [left] {$k_1$};
		\draw (.5,-.5) node [right] {$k_2$};
		\filldraw[fill=black] (0.5,0) circle (.06);
		\filldraw[fill=black] (-.5,0) circle (.06);
    \end{tikzpicture}\,\,+\,\,
	\frac12\,\,\begin{tikzpicture}[baseline={([yshift=-1.5ex](0,-.2))}]
		\path [draw=black, worldlineStatic] (-1,0) -- (-.5,0);
        \path [draw=black, zParticle] (0.5,0) -- (-0.5,0);
		\path [draw=black, worldlineStatic] (0.5,0) -- (1.0,0);
		\path [draw=black, scalar2] (0.5,0) -- (.5,-.8);
		\path [draw=black, scalar2] (-0.5,0) -- (-.5,-.8);
		\draw (-.5,-.5) node [left] {$k_1$};
		\draw (.5,-.5) node [right] {$k_2$};
		\filldraw[fill=black] (0.5,0) circle (.06);
		\filldraw[fill=black] (-.5,0) circle (.06);
    \end{tikzpicture}\,,
    \label{eq:Compton_caseA}
\end{align}
and is a WQFT-Compton amplitude.
The factors of $1/2$ are precisely the
Murua coefficient in \eqn{murua2pt}.

We also compute the same matrix element $iN(k_1,k_2)$ at order $g e_1$.
Taking one worldline coupling and one bulk interaction in \eqn{eq:N2} gives
\begin{align}
\left.i\hat N^{(2)}\right|_{g e_1}
&=\frac{(-i)^2}{2}\int\!\d^Dx\,\d^Dy\;
\theta_{xy}\,([\hat\cH_{I,\rm bulk}(x),\hat\cH_{I,1}(y)]+[\hat\cH_{I,1}(x),\hat\cH_{I,\rm bulk}(y)])\\
&=\frac{(-i)^2}{2}\int\!\d^Dx\,\d\tau_1\,
\big(\theta(x^0-\bar{x}_1^0(\tau_1))-\theta(\bar{x}_1^0(\tau_1)-x^0)\big)\,
[\hat\cH_{I,\rm bulk}(x),\hat{H}_{I,1}(\tau_1)]\,.\nn
\end{align}
Localising the $x$--integral onto the worldline and using $\theta_{xy}[\hat\varphi(x)^3,\hat\varphi(y)]=3i\Delta_R(x-y)\hat\varphi(x)^2$,
together with $\langle k_1k_2|\hat\varphi(x)^2|0\rangle=2e^{i(k_1+k_2)\cdot x}$, we obtain
\begin{align}
\left.i N^{(2)}(k_1,k_2)\right|_{g e_1}
&=\frac{(-i)^2}{2} g e_1
\int\!\d^Dx\,\d\tau_1\, e^{i(k_1+k_2)\cdot x}
\Big[\Delta_R\big(x-\bar x_1(\tau_1)\big)+\Delta_A\big(x-\bar x_1(\tau_1)\big)\Big]\nn\\
&=\frac{(-i)^2}{2} g e_1 \int_q e^{iq\cdot b_1}\dd(k_1+k_2-q)\dd(q\cdot v_1) \nn \\
&\qquad \qquad \qquad \qquad \times \left[\frac{i}{q^2+{\rm sgn}(q^0)i0}+\frac{i}{q^2-{\rm sgn}(q^0)i0}\right]\,,\label{eq:N2_gei_matrix_RA_short}
\end{align}
where the two time orderings combine into the retarded/advanced sum.
Diagrammatically, this is equivalent to
\begin{align}
\left.iN^{(2)}(k_1,k_2)\right|_{g e_1}
&=\,\,
\frac12\,\begin{tikzpicture}[baseline={([yshift=-.7ex](.5,-.5))}]
        \path [draw=black, worldlineStatic] (-0.4,0) -- (1.4,0);
		\path [draw=black, scalar] (0.5,-.6) -- (0.5,0);
		\path [draw=black, scalar2] (0.5,-.6) -- (-0.05,-1.2) node[left] {$k_1$};
		\path [draw=black, scalar2] (0.5,-.6) -- (1.05,-1.2) node[right] {$k_2$};
        \filldraw[fill=black] (0.5,0) circle (.06);
		\filldraw[fill=black] (0.5,-.6) circle (.06);
    \end{tikzpicture}
\;\,+\,\;
\frac12\,\begin{tikzpicture}[baseline={([yshift=-.7ex](.5,-.5))}]
        \path [draw=black, worldlineStatic] (-0.4,0) -- (1.4,0);
		\path [draw=black, scalar] (0.5,0) -- (0.5,-.6);
		\path [draw=black, scalar2] (0.5,-.6) -- (-0.05,-1.2) node[left] {$k_1$};
		\path [draw=black, scalar2] (0.5,-.6) -- (1.05,-1.2) node[right] {$k_2$};
        \filldraw[fill=black] (0.5,0) circle (.06);
		\filldraw[fill=black] (0.5,-.6) circle (.06);
    \end{tikzpicture}=
\begin{tikzpicture}[baseline={([yshift=-.7ex](.5,-.5))}]
        \path [draw=black, worldlineStatic] (-0.4,0) -- (1.4,0);
		\path [draw=black, scalar2] (0.5,0) -- (0.5,-.6);
		\path [draw=black, scalar2] (0.5,-.6) -- (-0.05,-1.2) node[left] {$k_1$};
		\path [draw=black, scalar2] (0.5,-.6) -- (1.05,-1.2) node[right] {$k_2$};
        \filldraw[fill=black] (0.5,0) circle (.06);
		\filldraw[fill=black] (0.5,-.6) circle (.06);
    \end{tikzpicture}\,.
\label{eq:N2_gei_diag}
\end{align}
The $ge_2$ element is related by symmetry.
Viewing this expression in momentum space,
we see that the $q^2$ propagator cannot go on-shell ---
it is a \emph{passive propagator}, whose $i0$ prescription is immaterial.
Thus, we may combine the two diagrams.

\paragraph{Matrix elements of $\hat{N}^{(3)}$}

For these more non-trivial examples
we refrain from direct calculation of $\hat{N}$-matrix elements using the Magnus expansion.
For the reader interested in how the Magnus series
can be applied directly in such cases,
we provide the worked example of $N^{(3)}(k)|_{e_1^2e_2}$
in \app{exampleMagnus}.
We also recommend looking at the examples from
$\phi^3$ theory in \rcite{Brandhuber:2025igz}.
Instead, let us now apply the WQFT Feynman rules directly.
For clarity of the exposition,
we weight diagrams with all possible combinations of retarded/advanced propagators
using the Murua coefficients~\eqref{murua} ---
yet, as many propagators are passive many of the expressions can be collapsed.

The simplest non-trivial matrix element of $\hat{N}^{(3)}$ is $N^{(3)}(k)$,
which has five non-trivial components:
{\allowdisplaybreaks
\begin{subequations}
\begin{align}
    \left.iN^{(3)}(k)\right|_{e_1^2e_2}&=\,\,
	\frac13\,\begin{tikzpicture}[baseline={([yshift=-.5ex](.5,-.5))}]
        \path [draw=black, worldlineStatic] (-0.4,0) -- (0,0);
        \path [draw=black, zParticle] (0,0) -- (0.6,0);
		\path [draw=black, worldlineStatic] (0.6,0) -- (1.0,0);
        \path [draw=black, worldlineStatic] (-0.4,-1.) -- (1.0,-1.);
        \path [draw=black, scalar] (0,-1.) -- (0,0);
		\path [draw=black, scalar2] (0.6,0) -- (1,-.5);
        \filldraw[fill=black] (0,0) circle (.06);
        \filldraw[fill=black] (0,-1.) circle (.06);
		\filldraw[fill=black] (0.6,0) circle (.06);
    \end{tikzpicture}\,\,+\,\,
	\frac13\,\begin{tikzpicture}[baseline={([yshift=-.5ex](.5,-.5))}]
        \path [draw=black, worldlineStatic] (-0.4,0) -- (0,0);
        \path [draw=black, zParticle] (0.6,0) -- (0,0);
		\path [draw=black, worldlineStatic] (0.6,0) -- (1.0,0);
        \path [draw=black, worldlineStatic] (-0.4,-1.) -- (1.0,-1.);
        \path [draw=black, scalar] (0,0) -- (0,-1);
		\path [draw=black, scalar2] (0.6,0) -- (1,-.5);
        \filldraw[fill=black] (0,0) circle (.06);
        \filldraw[fill=black] (0,-1.) circle (.06);
		\filldraw[fill=black] (0.6,0) circle (.06);
    \end{tikzpicture}\,\,+\,\,
	\frac16\,\begin{tikzpicture}[baseline={([yshift=-.5ex](.5,-.5))}]
        \path [draw=black, worldlineStatic] (-0.4,0) -- (0,0);
        \path [draw=black, zParticle] (0.6,0) -- (0,0);
		\path [draw=black, worldlineStatic] (0.6,0) -- (1.0,0);
        \path [draw=black, worldlineStatic] (-0.4,-1.) -- (1.0,-1.);
        \path [draw=black, scalar] (0,-1.) -- (0,0);
		\path [draw=black, scalar2] (0.6,0) -- (1,-.5);
        \filldraw[fill=black] (0,0) circle (.06);
        \filldraw[fill=black] (0,-1.) circle (.06);
		\filldraw[fill=black] (0.6,0) circle (.06);
    \end{tikzpicture}\,\,+\,\,
	\frac16\,\begin{tikzpicture}[baseline={([yshift=-.5ex](.5,-.5))}]
        \path [draw=black, worldlineStatic] (-0.4,0) -- (0,0);
        \path [draw=black, zParticle] (0,0) -- (0.6,0);
		\path [draw=black, worldlineStatic] (0.6,0) -- (1.0,0);
        \path [draw=black, worldlineStatic] (-0.4,-1.) -- (1.0,-1.);
        \path [draw=black, scalar] (0,0) -- (0,-1);
		\path [draw=black, scalar2] (0.6,0) -- (1,-.5);
        \filldraw[fill=black] (0,0) circle (.06);
        \filldraw[fill=black] (0,-1.) circle (.06);
		\filldraw[fill=black] (0.6,0) circle (.06);
    \end{tikzpicture}\nn\\
	&=\,\,
	\frac12\,\begin{tikzpicture}[baseline={([yshift=-.5ex](.5,-.5))}]
        \path [draw=black, worldlineStatic] (-0.4,0) -- (0,0);
        \path [draw=black, zParticle] (0,0) -- (0.6,0);
		\path [draw=black, worldlineStatic] (0.6,0) -- (1.0,0);
        \path [draw=black, worldlineStatic] (-0.4,-1.) -- (1.0,-1.);
        \path [draw=black, scalar2] (0,-1.) -- (0,0);
		\path [draw=black, scalar2] (0.6,0) -- (1,-.5);
        \filldraw[fill=black] (0,0) circle (.06);
        \filldraw[fill=black] (0,-1.) circle (.06);
		\filldraw[fill=black] (0.6,0) circle (.06);
    \end{tikzpicture}\,\,+\,\,
	\frac12\,\begin{tikzpicture}[baseline={([yshift=-.5ex](.5,-.5))}]
        \path [draw=black, worldlineStatic] (-0.4,0) -- (0,0);
        \path [draw=black, zParticle] (0.6,0) -- (0,0);
		\path [draw=black, worldlineStatic] (0.6,0) -- (1.0,0);
        \path [draw=black, worldlineStatic] (-0.4,-1.) -- (1.0,-1.);
        \path [draw=black, scalar2] (0,0) -- (0,-1);
		\path [draw=black, scalar2] (0.6,0) -- (1,-.5);
        \filldraw[fill=black] (0,0) circle (.06);
        \filldraw[fill=black] (0,-1.) circle (.06);
		\filldraw[fill=black] (0.6,0) circle (.06);
    \end{tikzpicture}
	\,\,,\label{eq:N3e1e1e2}\\
	\left.iN^{(3)}(k)\right|_{ge_1e_2}&=\,\,
	\frac13\,\begin{tikzpicture}[baseline={([yshift=-.7ex](.5,-.5))}]
        \path [draw=black, worldlineStatic] (-0.5,0) -- (0.5,0);
		\path [draw=black, worldlineStatic] (-0.5,-1.2) -- (0.5,-1.2);
		\path [draw=black, scalar] (0,-.6) -- (0,0);
		\path [draw=black, scalar] (0,-1.2) -- (0,-.6);
		\path [draw=black, scalar2] (0,-.6) -- (0.5,-.6);
        \filldraw[fill=black] (0,0) circle (.06);
        \filldraw[fill=black] (0,-.6) circle (.06);
		\filldraw[fill=black] (0,-1.2) circle (.06);
    \end{tikzpicture}\,\,\,\,\,\,+\,\,\,\,\,\,
	\frac13\,\begin{tikzpicture}[baseline={([yshift=-.7ex](.5,-.5))}]
        \path [draw=black, worldlineStatic] (-0.5,0) -- (0.5,0);
		\path [draw=black, worldlineStatic] (-0.5,-1.2) -- (0.5,-1.2);
		\path [draw=black, scalar] (0,0) -- (0,-.6);
		\path [draw=black, scalar] (0,-.6) -- (0,-1.2);
		\path [draw=black, scalar2] (0,-.6) -- (0.5,-.6);
        \filldraw[fill=black] (0,0) circle (.06);
        \filldraw[fill=black] (0,-.6) circle (.06);
		\filldraw[fill=black] (0,-1.2) circle (.06);
    \end{tikzpicture}\,\,\,\,\,\,+\,\,\,\,\,\,
	\frac16\,\begin{tikzpicture}[baseline={([yshift=-.7ex](.5,-.5))}]
        \path [draw=black, worldlineStatic] (-0.5,0) -- (0.5,0);
		\path [draw=black, worldlineStatic] (-0.5,-1.2) -- (0.5,-1.2);
		\path [draw=black, scalar] (0,-.6) -- (0,0);
		\path [draw=black, scalar] (0,-.6) -- (0,-1.2);
		\path [draw=black, scalar2] (0,-.6) -- (0.5,-.6);
        \filldraw[fill=black] (0,0) circle (.06);
        \filldraw[fill=black] (0,-.6) circle (.06);
		\filldraw[fill=black] (0,-1.2) circle (.06);
    \end{tikzpicture}\,\,\,\,\,\,+\,\,\,\,\,\,
	\frac16\,\begin{tikzpicture}[baseline={([yshift=-.7ex](.5,-.5))}]
        \path [draw=black, worldlineStatic] (-0.5,0) -- (0.5,0);
		\path [draw=black, worldlineStatic] (-0.5,-1.2) -- (0.5,-1.2);
		\path [draw=black, scalar] (0,0) -- (0,-.6);
		\path [draw=black, scalar] (0,-1.2) -- (0,-.6);
		\path [draw=black, scalar2] (0,-.6) -- (0.5,-.6);
        \filldraw[fill=black] (0,0) circle (.06);
        \filldraw[fill=black] (0,-.6) circle (.06);
		\filldraw[fill=black] (0,-1.2) circle (.06);
    \end{tikzpicture}\nn\\
	&=\,\,\,\,\,\,\,
	\begin{tikzpicture}[baseline={([yshift=-.7ex](.5,-.5))}]
        \path [draw=black, worldlineStatic] (-0.5,0) -- (0.5,0);
		\path [draw=black, worldlineStatic] (-0.5,-1.2) -- (0.5,-1.2);
		\path [draw=black, scalar2] (0,-.6) -- (0,0);
		\path [draw=black, scalar2] (0,-1.2) -- (0,-.6);
		\path [draw=black, scalar2] (0,-.6) -- (0.5,-.6);
        \filldraw[fill=black] (0,0) circle (.06);
        \filldraw[fill=black] (0,-.6) circle (.06);
		\filldraw[fill=black] (0,-1.2) circle (.06);
    \end{tikzpicture}\,\,,\label{eq:N3_ge1e2}\\
	\left.iN^{(3)}(k)\right|_{ge_1^2}&=\,\,
	\frac13\,\begin{tikzpicture}[baseline={([yshift=-.7ex](.5,-.5))}]
        \path [draw=black, worldlineStatic] (-0.4,0) -- (1.4,0);
		\path [draw=black, scalar] (0,0) -- (0.5,-.6);
		\path [draw=black, scalar] (0.5,-.6) -- (1,0);
		\path [draw=black, scalar2] (0.5,-1.2) -- (0.5,-.6);
        \filldraw[fill=black] (0,0) circle (.06);
		\filldraw[fill=black] (1,0) circle (.06);
		\filldraw[fill=black] (0.5,-.6) circle (.06);
    \end{tikzpicture}\,\,\,+\,\,\,
	\frac12\cdot\frac16\,\begin{tikzpicture}[baseline={([yshift=-.7ex](.5,-.5))}]
        \path [draw=black, worldlineStatic] (-0.4,0) -- (1.4,0);
		\path [draw=black, scalar] (0,0) -- (0.5,-.6);
		\path [draw=black, scalar] (1,0) -- (0.5,-.6);
		\path [draw=black, scalar2] (0.5,-1.2) -- (0.5,-.6);
        \filldraw[fill=black] (0,0) circle (.06);
		\filldraw[fill=black] (1,0) circle (.06);
		\filldraw[fill=black] (0.5,-.6) circle (.06);
    \end{tikzpicture}\,\,\,+\,\,\,
	\frac12\cdot\frac16\,\begin{tikzpicture}[baseline={([yshift=-.7ex](.5,-.5))}]
        \path [draw=black, worldlineStatic] (-0.4,0) -- (1.4,0);
		\path [draw=black, scalar] (0.5,-.6) -- (0,0);
		\path [draw=black, scalar] (0.5,-.6) -- (1,0);
		\path [draw=black, scalar2] (0.5,-1.2) -- (0.5,-.6);
        \filldraw[fill=black] (0,0) circle (.06);
		\filldraw[fill=black] (1,0) circle (.06);
		\filldraw[fill=black] (0.5,-.6) circle (.06);
    \end{tikzpicture}\nn\\
	&=\,\,
	\frac12\,\begin{tikzpicture}[baseline={([yshift=-.7ex](.5,-.5))}]
        \path [draw=black, worldlineStatic] (-0.4,0) -- (1.4,0);
		\path [draw=black, scalar2] (0,0) -- (0.5,-.6);
		\path [draw=black, scalar2] (0.5,-.6) -- (1,0);
		\path [draw=black, scalar2] (0.5,-1.2) -- (0.5,-.6);
        \filldraw[fill=black] (0,0) circle (.06);
		\filldraw[fill=black] (1,0) circle (.06);
		\filldraw[fill=black] (0.5,-.6) circle (.06);
    \end{tikzpicture}\,\,.\label{eq:N3_ge1e1}
\end{align}
\end{subequations}}
The last two cases $N^{(3)}(k)|_{e_1e_2^2}$ and $N^{(3)}(k)|_{ge_2^2}$
are related by symmetry.
We have written down all allowed diagrams,
with numerical weighting factors $1/3$ and $1/6$ being precisely the
Murua coefficients given in \eqn{murua3pt}.
In the last two diagrams, we also include symmetry factors of $1/2$.
Notice that, in the first two cases, these are exactly the same diagrams
appearing in the scattering waveform --- but with different causality flows.

Using Feynman rules, the explicit expressions for these
three components are
\begin{subequations}
\begin{align}
	&\left.iN^{(3)}(k)\right|_{e_1^2e_2}
	=\frac{(-i)^3}{6} e_1^2e_2 \int\!\d\tau_1\d\tau_1^\prime\d\tau_2\,
	(ik_\mu) e^{ik\cdot\bar{x}_1(\tau_1)}\label{N3e1e1e2Explicit}\\
	&\qquad\times\big[
		2\Delta_{1,R}^{\mu\nu}(\tau_1-\tau_1^\prime)
		\partial_\nu\Delta_R(\bar{x}_1(\tau_1^\prime)-\bar{x}_2(\tau_2))+
		2\Delta_{1,A}^{\mu\nu}(\tau_1-\tau_1^\prime)
		\partial_\nu\Delta_A(\bar{x}_1(\tau_1^\prime)-\bar{x}_2(\tau_2))\nn
		\\&\qquad\,\,\,\,\,+
		\Delta_{1,R}^{\mu\nu}(\tau_1-\tau_1^\prime)
		\partial_\nu\Delta_A(\bar{x}_1(\tau_1^\prime)-\bar{x}_2(\tau_2))+
		\Delta_{1,A}^{\mu\nu}(\tau_1-\tau_1^\prime)
		\partial_\nu\Delta_R(\bar{x}_1(\tau_1^\prime)-\bar{x}_2(\tau_2))
	\big]\, ,\nn\\
	&\left.iN^{(3)}(k)\right|_{ge_1e_2}=\frac{(-i)^3}{6} ge_1e_2 \int\!\d^Dx\,\d\tau_1\d\tau_2\,
	e^{ik\cdot x}\\
	&\qquad\times\big[
		2\Delta_A(x-\bar{x}_1(\tau_1))\Delta_R(x-\bar{x}_2(\tau_2))+2\Delta_R(x-\bar{x}_1(\tau_1))\Delta_A(x-\bar{x}_2(\tau_2))\nn
		\\&\qquad\,\,\,\,\,+
		\Delta_A(x-\bar{x}_1(\tau_1))\Delta_A(x-\bar{x}_2(\tau_2))+
		\Delta_R(x-\bar{x}_1(\tau_1))\Delta_R(x-\bar{x}_2(\tau_2))\nn
	\big]\, ,\nn\\
	&\left.iN^{(3)}(k)\right|_{ge_1^2}=\frac{(-i)^3}{12} g e_1^2\int\!\d^Dx\,\d\tau_1\d\tau_1^\prime\,
	e^{ik\cdot x}\big[
		4\Delta_R(x-\bar{x}_1(\tau_1))\Delta_A(x-\bar{x}_1(\tau_1^\prime))
		\\&\qquad+
		\Delta_A(x-\bar{x}_1(\tau_1))\Delta_A(x-\bar{x}_1(\tau_1^\prime))+
		\Delta_R(x-\bar{x}_1(\tau_1))\Delta_R(x-\bar{x}_1(\tau_1^\prime))\nn
	\big]\, .\nn
\end{align}
\end{subequations}
Again, $N^{(3)}(k)|_{e_1e_2^2}$ and $N^{(3)}(k)|_{ge_2^2}$ are related by symmetry,
and all boundary conditions on time integrals are enforced by the retarded and advanced propagators.

\subsection{Vacuum Magnus amplitudes}

In the conservative case, where we project $\hat{N}$ on vacuum states $|0\rangle$, we take matrix elements only on even terms in the Magnus series.

\paragraph{Matrix element of $\hat{N}^{(2)}$}

As an example of a vacuum Magnus amplitude
we take the $e_1e_2$ component of the vacuum element of $\hat{N}^{(2)}$~\eqref{eq:N2}:
\begin{align}
\begin{aligned}
	\left.i\hat{N}^{(2)}\right|_{e_1e_2}&=\frac{(-i)^2}{2} \int\!\d^Dx\d^Dy\ \theta_{xy}\,
	[\hat\cH_{I,1}(x),\hat\cH_{I,2}(y)]
	+(1\leftrightarrow2) \\
	&=\frac{(-i)^2}{2}\int\!\d\tau_1\d\tau_2\,\theta(\bar{x}_1^0(\tau_1)-\bar{x}_2^0(\tau_2))
	[\hat{H}_{I,1}(\tau_1),\hat{H}_{I,2}(\tau_2)]+(1\leftrightarrow2)\,.
\end{aligned}
\end{align}
Here $(1\leftrightarrow2)$ indicates that we add the same term with particle labels $i=1,2$ swapped.
Having inserted
$\hat{\cH}_{I,i}(x)=\int\!\d\tau_i\,\delta^D(x-\bar{x}_i^0)\hat{H}_{I,i}(\tau_i)$,
we see again how integration on position $x$ is replaced by integrations
on the proper times $\tau_i$ along each worldline.
Inserting the worldline interaction Hamiltonian $\hat{H}_{I,i}(\tau_i)$
\eqref{eq:HintA_def},
\begin{align}
	\left.i\hat{N}^{(2)}\right|_{e_1e_2}
	&=\frac{(-i)^2}{2} e_1e_2 \int\!\d\tau_1\d\tau_2\,\theta(\bar{x}_1^0(\tau_1)-\bar{x}_2^0(\tau_2))
	[\hat\varphi(\bar{x}_1(\tau_1)),\hat\varphi(\bar{x}_2(\tau_2))]+(1\leftrightarrow2)+\cdots\nn\\
	&=\frac{(-i)^2}{2} e_1e_2 \int\!\d\tau_1\d\tau_2\,
	\big(\Delta_R(\bar{x}_1(\tau_1)-\bar{x}_2(\tau_2))+(1\leftrightarrow2)\big)+\cdots\,.
\end{align}
The ellipsis $\cdots$ denotes terms carrying additional instances of $\hat{z}_i$,
which we will not need as they annihilate the vacuum.
The vacuum element $iN^{(2)}=\langle0|i\hat{N}^{(2)}|0\rangle$ is then
\begin{align}
	\hspace{-7pt}\left.i N^{(2)}\right|_{e_1e_2}
	&=\frac{(-i)^2}{2} e_1e_2 \int\!\d\tau_1\d\tau_2\,
	\big(\Delta_R(\bar{x}_1(\tau_1)-\bar{x}_2(\tau_2))+\Delta_A(\bar{x}_1(\tau_1)-\bar{x}_2(\tau_2))\big)\\
	&=\frac{(-i)^2}{2} e_1e_2 \int_qe^{iq\cdot b}\dd(q\cdot v_1)\dd(q\cdot v_2)
	\left(\frac{i}{q^2+{\rm sgn}(q^0)i0}+\frac{i}{q^2-{\rm sgn}(q^0)i0}\right)\,,\nn
\end{align}
which, using~\eqn{bulkProps},
we have written in momentum space.
This is depicted as
\begin{align}
    \left.i N^{(2)}\right|_{e_1e_2}&=\,\,
	\frac12\,\,\begin{tikzpicture}[baseline={([yshift=-.5ex](.5,-.5))}]
        \path [draw=black, worldlineStatic] (-0.6,0) -- (0,0);
        \path [draw=black, worldlineStatic] (0,0) -- (0.6,0);
        \path [draw=black, worldlineStatic] (-0.6,-1.) -- (0.6,-1.);
        \path [draw=black, scalar] (0,-1.) -- (0,0);
        \filldraw[fill=black] (0,0) circle (.06);
        \filldraw[fill=black] (0,-1.) circle (.06);
    \end{tikzpicture}\,\,+\,\,
	\frac12\,\,\begin{tikzpicture}[baseline={([yshift=-.5ex](.5,-.5))}]
        \path [draw=black, worldlineStatic] (-0.6,0) -- (0,0);
        \path [draw=black, worldlineStatic] (0,0) -- (0.6,0);
        \path [draw=black, worldlineStatic] (-0.6,-1.) -- (0.6,-1.);
        \path [draw=black, scalar] (0,0) -- (0,-1);
        \filldraw[fill=black] (0,0) circle (.06);
        \filldraw[fill=black] (0,-1.) circle (.06);
    \end{tikzpicture}
	\,\,=\,\,
	\begin{tikzpicture}[baseline={([yshift=-.5ex](.5,-.5))}]
        \path [draw=black, worldlineStatic] (-0.6,0) -- (0,0);
        \path [draw=black, worldlineStatic] (0,0) -- (0.6,0);
        \path [draw=black, worldlineStatic] (-0.6,-1.) -- (0.6,-1.);
        \path [draw=black, scalar2] (0,0) -- (0,-1);
        \filldraw[fill=black] (0,0) circle (.06);
        \filldraw[fill=black] (0,-1.) circle (.06);
    \end{tikzpicture}\,.
\end{align}
The factors of $1/2$ are precisely the Murua coefficient in~\eqn{murua2pt},
combining via the edge contraction rule as the scalar propagator is passive
(does not go on-shell).

\paragraph{Matrix element of $\hat{N}^{(4)}$}

The two relevant components of $iN^{(4)}$ are:
\begin{subequations}
\begin{align}
    \left.i N^{(4)}\right|_{e_1^2e_2^2}&=\,\,
	\frac14\,\,\begin{tikzpicture}[baseline={([yshift=-.5ex](.5,-.5))}]
        \path [draw=black, worldlineStatic] (-0.4,0) -- (0,0);
		\path [draw=black, zParticle] (0,0) -- (1,0);
        \path [draw=black, worldlineStatic] (1,0) -- (1.4,0);
        \path [draw=black, worldlineStatic] (-0.4,-1.) -- (1.4,-1.);
        \path [draw=black, scalar] (0,-1.) -- (0,0);
		\path [draw=black, scalar] (1,0) -- (1,-1.);
        \filldraw[fill=black] (0,0) circle (.06);
        \filldraw[fill=black] (0,-1.) circle (.06);
		\filldraw[fill=black] (1,0) circle (.06);
        \filldraw[fill=black] (1,-1.) circle (.06);
    \end{tikzpicture}+
	\frac1{12}\,\,\begin{tikzpicture}[baseline={([yshift=-.5ex](.5,-.5))}]
        \path [draw=black, worldlineStatic] (-0.4,0) -- (0,0);
		\path [draw=black, zParticle] (0,0) -- (1,0);
        \path [draw=black, worldlineStatic] (1,0) -- (1.4,0);
        \path [draw=black, worldlineStatic] (-0.4,-1.) -- (1.4,-1.);
        \path [draw=black, scalar] (0,-1.) -- (0,0);
		\path [draw=black, scalar] (1,-1.) -- (1,0);
        \filldraw[fill=black] (0,0) circle (.06);
        \filldraw[fill=black] (0,-1.) circle (.06);
		\filldraw[fill=black] (1,0) circle (.06);
        \filldraw[fill=black] (1,-1.) circle (.06);
    \end{tikzpicture}+
	\frac1{12}\,\,\begin{tikzpicture}[baseline={([yshift=-.5ex](.5,-.5))}]
        \path [draw=black, worldlineStatic] (-0.4,0) -- (0,0);
		\path [draw=black, zParticle] (0,0) -- (1,0);
        \path [draw=black, worldlineStatic] (1,0) -- (1.4,0);
        \path [draw=black, worldlineStatic] (-0.4,-1.) -- (1.4,-1.);
        \path [draw=black, scalar] (0,0) -- (0,-1);
		\path [draw=black, scalar] (1,0) -- (1,-1);
        \filldraw[fill=black] (0,0) circle (.06);
        \filldraw[fill=black] (0,-1.) circle (.06);
		\filldraw[fill=black] (1,0) circle (.06);
        \filldraw[fill=black] (1,-1.) circle (.06);
    \end{tikzpicture}+
	\frac1{12}\,\,\begin{tikzpicture}[baseline={([yshift=-.5ex](.5,-.5))}]
        \path [draw=black, worldlineStatic] (-0.4,0) -- (0,0);
		\path [draw=black, zParticle] (0,0) -- (1,0);
        \path [draw=black, worldlineStatic] (1,0) -- (1.4,0);
        \path [draw=black, worldlineStatic] (-0.4,-1.) -- (1.4,-1.);
        \path [draw=black, scalar] (0,0) -- (0,-1);
		\path [draw=black, scalar] (1,-1.) -- (1,0);
        \filldraw[fill=black] (0,0) circle (.06);
        \filldraw[fill=black] (0,-1.) circle (.06);
		\filldraw[fill=black] (1,0) circle (.06);
        \filldraw[fill=black] (1,-1.) circle (.06);
    \end{tikzpicture}\nn\\
	&+\,\,\,
	\frac14\,\,\begin{tikzpicture}[baseline={([yshift=-.5ex](.5,-.5))}]
        \path [draw=black, worldlineStatic] (-0.4,-1) -- (0,-1);
		\path [draw=black, zParticle] (0,-1) -- (1,-1);
        \path [draw=black, worldlineStatic] (1,-1) -- (1.4,-1);
        \path [draw=black, worldlineStatic] (-0.4,0) -- (1.4,0);
        \path [draw=black, scalar] (0,0) -- (0,-1);
		\path [draw=black, scalar] (1,-1) -- (1,0);
        \filldraw[fill=black] (0,-1) circle (.06);
        \filldraw[fill=black] (0,0) circle (.06);
		\filldraw[fill=black] (1,-1) circle (.06);
        \filldraw[fill=black] (1,0) circle (.06);
    \end{tikzpicture}+
	\frac1{12}\,\,\begin{tikzpicture}[baseline={([yshift=-.5ex](.5,-.5))}]
        \path [draw=black, worldlineStatic] (-0.4,-1) -- (0,-1);
		\path [draw=black, zParticle] (0,-1) -- (1,-1);
        \path [draw=black, worldlineStatic] (1,-1) -- (1.4,-1);
        \path [draw=black, worldlineStatic] (-0.4,0) -- (1.4,0);
        \path [draw=black, scalar] (0,0) -- (0,-1);
		\path [draw=black, scalar] (1,0) -- (1,-1);
        \filldraw[fill=black] (0,-1) circle (.06);
        \filldraw[fill=black] (0,0) circle (.06);
		\filldraw[fill=black] (1,-1) circle (.06);
        \filldraw[fill=black] (1,0) circle (.06);
    \end{tikzpicture}+
	\frac1{12}\,\,\begin{tikzpicture}[baseline={([yshift=-.5ex](.5,-.5))}]
        \path [draw=black, worldlineStatic] (-0.4,-1) -- (0,-1);
		\path [draw=black, zParticle] (0,-1) -- (1,-1);
        \path [draw=black, worldlineStatic] (1,-1) -- (1.4,-1);
        \path [draw=black, worldlineStatic] (-0.4,0) -- (1.4,0);
        \path [draw=black, scalar] (0,-1) -- (0,0);
		\path [draw=black, scalar] (1,-1) -- (1,0);
        \filldraw[fill=black] (0,-1) circle (.06);
        \filldraw[fill=black] (0,0) circle (.06);
		\filldraw[fill=black] (1,-1) circle (.06);
        \filldraw[fill=black] (1,0) circle (.06);
    \end{tikzpicture}+
	\frac1{12}\,\,\begin{tikzpicture}[baseline={([yshift=-.5ex](.5,-.5))}]
        \path [draw=black, worldlineStatic] (-0.4,-1) -- (0,-1);
		\path [draw=black, zParticle] (0,-1) -- (1,-1);
        \path [draw=black, worldlineStatic] (1,-1) -- (1.4,-1);
        \path [draw=black, worldlineStatic] (-0.4,0) -- (1.4,0);
        \path [draw=black, scalar] (0,-1) -- (0,0);
		\path [draw=black, scalar] (1,0) -- (1,-1);
        \filldraw[fill=black] (0,-1) circle (.06);
        \filldraw[fill=black] (0,0) circle (.06);
		\filldraw[fill=black] (1,-1) circle (.06);
        \filldraw[fill=black] (1,0) circle (.06);
    \end{tikzpicture}\nn\\
	&=\,\,
	\frac12\,\,\begin{tikzpicture}[baseline={([yshift=-.5ex](.5,-.5))}]
        \path [draw=black, worldlineStatic] (-0.4,0) -- (0,0);
		\path [draw=black, zParticle] (0,0) -- (1,0);
        \path [draw=black, worldlineStatic] (1,0) -- (1.4,0);
        \path [draw=black, worldlineStatic] (-0.4,-1.) -- (1.4,-1.);
        \path [draw=black, scalar2] (0,0) -- (0,-1);
		\path [draw=black, scalar2] (1,-1.) -- (1,0);
        \filldraw[fill=black] (0,0) circle (.06);
        \filldraw[fill=black] (0,-1.) circle (.06);
		\filldraw[fill=black] (1,0) circle (.06);
        \filldraw[fill=black] (1,-1.) circle (.06);
    \end{tikzpicture}+
	\frac12\,\,\begin{tikzpicture}[baseline={([yshift=-.5ex](.5,-.5))}]
        \path [draw=black, worldlineStatic] (-0.4,-1) -- (0,-1);
		\path [draw=black, zParticle] (0,-1) -- (1,-1);
        \path [draw=black, worldlineStatic] (1,-1) -- (1.4,-1);
        \path [draw=black, worldlineStatic] (-0.4,0) -- (1.4,0);
        \path [draw=black, scalar2] (0,-1) -- (0,0);
		\path [draw=black, scalar2] (1,0) -- (1,-1);
        \filldraw[fill=black] (0,-1) circle (.06);
        \filldraw[fill=black] (0,0) circle (.06);
		\filldraw[fill=black] (1,-1) circle (.06);
        \filldraw[fill=black] (1,0) circle (.06);
    \end{tikzpicture}\,,\label{eq:N4e1e1e2e2}\\
	\left.i N^{(4)}\right|_{ge_1e_2^2}&=\,\,
	\frac16\,\,\begin{tikzpicture}[baseline={([yshift=-.5ex](.5,-.5))}]
        \path [draw=black, worldlineStatic] (-0.4,0) -- (1.4,0);
        \path [draw=black, worldlineStatic] (-0.4,-1.) -- (1.4,-1.);
        \path [draw=black, scalar] (0,-1.) -- (.5,-.55);
		\path [draw=black, scalar] (.5,-.55) -- (1,-1.);
		\path [draw=black, scalar] (.5,-.55) -- (.5,0);
        \filldraw[fill=black] (0.5,0) circle (.06);
		\filldraw[fill=black] (0.5,-.55) circle (.06);
        \filldraw[fill=black] (0,-1.) circle (.06);
        \filldraw[fill=black] (1,-1.) circle (.06);
    \end{tikzpicture}+
	\frac16\,\,\begin{tikzpicture}[baseline={([yshift=-.5ex](.5,-.5))}]
        \path [draw=black, worldlineStatic] (-0.4,0) -- (1.4,0);
        \path [draw=black, worldlineStatic] (-0.4,-1.) -- (1.4,-1.);
        \path [draw=black, scalar] (0,-1.) -- (.5,-.55);
		\path [draw=black, scalar] (.5,-.55) -- (1,-1.);
		\path [draw=black, scalar] (.5,0) -- (.5,-.55);
        \filldraw[fill=black] (0.5,0) circle (.06);
		\filldraw[fill=black] (0.5,-.55) circle (.06);
        \filldraw[fill=black] (0,-1.) circle (.06);
        \filldraw[fill=black] (1,-1.) circle (.06);
    \end{tikzpicture}+
	\frac12\cdot\frac16\,\,\begin{tikzpicture}[baseline={([yshift=-.5ex](.5,-.5))}]
        \path [draw=black, worldlineStatic] (-0.4,0) -- (1.4,0);
        \path [draw=black, worldlineStatic] (-0.4,-1.) -- (1.4,-1.);
        \path [draw=black, scalar] (0,-1.) -- (.5,-.55);
		\path [draw=black, scalar] (1,-1.) -- (.5,-.55);
		\path [draw=black, scalar] (.5,-.55) -- (.5,0);
        \filldraw[fill=black] (0.5,0) circle (.06);
		\filldraw[fill=black] (0.5,-.55) circle (.06);
        \filldraw[fill=black] (0,-1.) circle (.06);
        \filldraw[fill=black] (1,-1.) circle (.06);
    \end{tikzpicture}+
	\frac12\cdot\frac16\,\,\begin{tikzpicture}[baseline={([yshift=-.5ex](.5,-.5))}]
        \path [draw=black, worldlineStatic] (-0.4,0) -- (1.4,0);
        \path [draw=black, worldlineStatic] (-0.4,-1.) -- (1.4,-1.);
        \path [draw=black, scalar] (.5,-.55) -- (0,-1.);
		\path [draw=black, scalar] (.5,-.55) -- (1,-1.);
		\path [draw=black, scalar] (.5,0) -- (.5,-.55);
        \filldraw[fill=black] (0.5,0) circle (.06);
		\filldraw[fill=black] (0.5,-.55) circle (.06);
        \filldraw[fill=black] (0,-1.) circle (.06);
        \filldraw[fill=black] (1,-1.) circle (.06);
    \end{tikzpicture}\nn\\
	&=\,\,
	\frac12\,\,\begin{tikzpicture}[baseline={([yshift=-.5ex](.5,-.5))}]
        \path [draw=black, worldlineStatic] (-0.4,0) -- (1.4,0);
        \path [draw=black, worldlineStatic] (-0.4,-1.) -- (1.4,-1.);
        \path [draw=black, scalar2] (.5,-.55) -- (0,-1.);
		\path [draw=black, scalar2] (.5,-.55) -- (1,-1.);
		\path [draw=black, scalar2] (.5,0) -- (.5,-.55);
        \filldraw[fill=black] (0.5,0) circle (.06);
		\filldraw[fill=black] (0.5,-.55) circle (.06);
        \filldraw[fill=black] (0,-1.) circle (.06);
        \filldraw[fill=black] (1,-1.) circle (.06);
    \end{tikzpicture}\,.
\end{align}
\end{subequations}
The numerical prefactors are Murua coefficients~\eqref{murua4pt},
multiplied in front by symmetry factors where appropriate.
The $ge_1^2e_2$ component is again related by symmetry.
Using the WQFT Feynman rules,
\begin{subequations}
\begin{align}
	\left.iN^{(4)}\right|_{e_1^2e_2^2}
	&=\frac{(-i)^4}{12} e_1^2e_2^2\int\!\d\tau_1\d\tau_1^\prime\d\tau_2\d\tau_2^\prime\,
	\Delta_{1,R}^{\mu\nu}(\tau_1-\tau_1^\prime)\\
	&\qquad\qquad\times\big[
		3\partial_\mu\Delta_A(\bar{x}_1(\tau_1)-\bar{x}_2(\tau_2))
		\partial_\nu\Delta_R(\bar{x}_1(\tau_1^\prime)-\bar{x}_2(\tau_2^\prime))\nn
		\\&\qquad\qquad\,\,\,\,\,\,+
		\partial_\mu\Delta_A(\bar{x}_1(\tau_1)-\bar{x}_2(\tau_2))
		\partial_\nu\Delta_A(\bar{x}_1(\tau_1^\prime)-\bar{x}_2(\tau_2^\prime))\nn
		\\&\qquad\qquad\,\,\,\,\,\,+
		\partial_\mu\Delta_R(\bar{x}_1(\tau_1)-\bar{x}_2(\tau_2))
		\partial_\nu\Delta_R(\bar{x}_1(\tau_1^\prime)-\bar{x}_2(\tau_2^\prime))\nn
		\\&\qquad\qquad\,\,\,\,\,\,+
		\partial_\mu\Delta_R(\bar{x}_1(\tau_1)-\bar{x}_2(\tau_2))
		\partial_\nu\Delta_A(\bar{x}_1(\tau_1^\prime)-\bar{x}_2(\tau_2^\prime))
	\big]+(1\leftrightarrow2)\, ,\nn\\
	\left.iN^{(4)}\right|_{ge_1e_2^2}
	&=\frac{(-i)^4}{12} ge_1e_2^2 \int\!\d^Dx\,\d\tau_1\d\tau_2\d\tau_2^\prime\\
	&\qquad\qquad\times\big[
		2\Delta_A(x-\bar{x}_1(\tau_1))
		\Delta_R(x-\bar{x}_2(\tau_2))\Delta_A(x-\bar{x}_2(\tau_2^\prime))\nn
		\\&\qquad\qquad\,\,\,+
		2\Delta_R(x-\bar{x}_1(\tau_1))
		\Delta_R(x-\bar{x}_2(\tau_2))\Delta_A(x-\bar{x}_2(\tau_2^\prime))\nn
		\\&\qquad\qquad\,\,\,\,\,\,+
		\Delta_A(x-\bar{x}_1(\tau_1))
		\Delta_R(x-\bar{x}_2(\tau_2))\Delta_R(x-\bar{x}_2(\tau_2^\prime))\nn
		\\&\qquad\qquad\,\,\,\,\,\,+
		\Delta_R(x-\bar{x}_1(\tau_1))
		\Delta_A(x-\bar{x}_2(\tau_2))\Delta_A(x-\bar{x}_2(\tau_2^\prime))
	\big]\, .\nn
\end{align}
\end{subequations}
All integrals are taken over an infinite time domain:
the retarded and advanced propagators automatically enforce the proper boundary conditions.

\subsection{Probe limit radial action}

Finally, given its relevance for the discussion of self-force effects that follows, let us briefly recall how the probe-limit (0SF) radial action $\bar{I}_r^>$~\eqref{eq:Ir_scatt_scalar} is encoded in the PL expansion of the $\hat N$-matrix.
In the probe limit, identifying the light body with particle $1$ and the heavy source with particle $2$ in the PL construction, one has
\begin{align}
	\bar{I}_r^> +\pi\bar{L} = \langle0|\hat N^{(\rm geo)}|0\rangle
	= \langle0|\hat N|0\rangle|_{\lambda}\,.
	\label{eq:Ir_from_N_probe}
\end{align}
Here $\hat N^{(\rm geo)}$ denotes the Magnus operator obtained by restricting the PL theory to the conservative probe sector at $\mathcal{O}(\lambda)$, equivalently to the dynamics generated by the probe Hamiltonian \eqref{eq:background_hamil_SF} in the fixed scalar background of the heavy particle.  
Notice that, because in the probe limit the system does not emit radiation, this captures the complete $\hat{N}$-matrix. Diagrammatically, the radial action is \cite{Gonzo:2024zxo,Haddad:2025cmw,Kim:2025gis}
\begin{align}\label{eq:0SFradialaction}
\begin{aligned}
	i\bar{I}_r^>+i\pi\bar{L}\,\,=\,\,
	\begin{tikzpicture}[baseline={([yshift=-.5ex](.5,-.5))}]
        \path [draw=black, worldlineStatic] (-0.6,0) -- (0,0);
        \path [draw=black, worldlineStatic] (0,0) -- (0.6,0);
        \path [draw=black, worldlineStatic] (-0.6,-1.) -- (0.6,-1.);
        \path [draw=black, scalar2] (0,0) -- (0,-1);
        \filldraw[fill=black] (0,0) circle (.06);
        \filldraw[fill=black] (0,-1.) circle (.06);
    \end{tikzpicture}\,\,+\,\,
	\frac12\,\,\begin{tikzpicture}[baseline={([yshift=-.5ex](.5,-.5))}]
        \path [draw=black, worldlineStatic] (-0.4,0) -- (0,0);
		\path [draw=black, zParticle] (0,0) -- (1,0);
        \path [draw=black, worldlineStatic] (1,0) -- (1.4,0);
        \path [draw=black, worldlineStatic] (-0.4,-1.) -- (1.4,-1.);
        \path [draw=black, scalar2] (0,0) -- (0,-1);
		\path [draw=black, scalar2] (1,-1.) -- (1,0);
        \filldraw[fill=black] (0,0) circle (.06);
        \filldraw[fill=black] (0,-1.) circle (.06);
		\filldraw[fill=black] (1,0) circle (.06);
        \filldraw[fill=black] (1,-1.) circle (.06);
    \end{tikzpicture}\,\,+\,\,
	\frac13\,\,\begin{tikzpicture}[baseline={([yshift=-.5ex](.5,-.5))}]
        \path [draw=black, worldlineStatic] (-0.4,0) -- (0,0);
		\path [draw=black, zParticle] (0,0) -- (1,0);
		\path [draw=black, zParticle] (1,0) -- (2,0);
        \path [draw=black, worldlineStatic] (2,0) -- (2.4,0);
        \path [draw=black, worldlineStatic] (-0.4,-1.) -- (2.4,-1.);
        \path [draw=black, scalar2] (0,0) -- (0,-1);
		\path [draw=black, scalar2] (1,0) -- (1,-1);
		\path [draw=black, scalar2] (2,0) -- (2,-1);
        \filldraw[fill=black] (0,0) circle (.06);
        \filldraw[fill=black] (0,-1) circle (.06);
		\filldraw[fill=black] (1,0) circle (.06);
        \filldraw[fill=black] (1,-1) circle (.06);
		\filldraw[fill=black] (2,0) circle (.06);
        \filldraw[fill=black] (2,-1) circle (.06);
    \end{tikzpicture}&\\
	+
	\frac12\cdot\frac16\,\,\begin{tikzpicture}[baseline={([yshift=-.5ex](.5,-.5))}]
        \path [draw=black, worldlineStatic] (-0.4,0) -- (0,0);
		\path [draw=black, zParticle] (0,0) -- (1,0);
		\path [draw=black, zParticle] (2,0) -- (1,0);
        \path [draw=black, worldlineStatic] (2,0) -- (2.4,0);
        \path [draw=black, worldlineStatic] (-0.4,-1.) -- (2.4,-1.);
        \path [draw=black, scalar2] (0,0) -- (0,-1);
		\path [draw=black, scalar2] (1,0) -- (1,-1);
		\path [draw=black, scalar2] (2,0) -- (2,-1);
        \filldraw[fill=black] (0,0) circle (.06);
        \filldraw[fill=black] (0,-1) circle (.06);
		\filldraw[fill=black] (1,0) circle (.06);
        \filldraw[fill=black] (1,-1) circle (.06);
		\filldraw[fill=black] (2,0) circle (.06);
        \filldraw[fill=black] (2,-1) circle (.06);
    \end{tikzpicture}+
	\frac12\cdot\frac16\,\,\begin{tikzpicture}[baseline={([yshift=-.5ex](.5,-.5))}]
        \path [draw=black, worldlineStatic] (-0.4,0) -- (0,0);
		\path [draw=black, zParticle] (1,0) -- (0,0);
		\path [draw=black, zParticle] (1,0) -- (2,0);
        \path [draw=black, worldlineStatic] (2,0) -- (2.4,0);
        \path [draw=black, worldlineStatic] (-0.4,-1.) -- (2.4,-1.);
        \path [draw=black, scalar2] (0,0) -- (0,-1);
		\path [draw=black, scalar2] (1,0) -- (1,-1);
		\path [draw=black, scalar2] (2,0) -- (2,-1);
        \filldraw[fill=black] (0,0) circle (.06);
        \filldraw[fill=black] (0,-1) circle (.06);
		\filldraw[fill=black] (1,0) circle (.06);
        \filldraw[fill=black] (1,-1) circle (.06);
		\filldraw[fill=black] (2,0) circle (.06);
        \filldraw[fill=black] (2,-1) circle (.06);
    \end{tikzpicture}&\,\,+\cdots\,,
\end{aligned}
\end{align}
which is a sum over the so-called ``fan'' diagrams. The Fourier transform to position space of the tree-level contribution is infrared divergent due to the long-range nature of the scalar interaction.
In the probe sector, we regulate this divergence by introducing the large-distance cutoff $R$,\footnote{Beyond the probe limit, 1SF and higher corrections are computed diagrammatically in dimensional regularisation. Since physical observables are independent of the infrared scheme and can be computed order-by-order in the SF expansion, this mixed prescription is completely consistent.} kept explicit in the classical radial action. With this prescription, the fan diagrams reproduce the PL expansion:
\begin{align}
\bar{I}_r^{>} + \pi\bar{L} &=
    - \frac{2 m_L |e_L K|}{\sqrt{\bar E^2-m_L^2}}
    \log\left(\frac{|e_L K| \bar{e}}{R}\right)
    - 2\bar L\,\arccos\left(-\frac{1}{\bar e}\right) + \pi\bar{L}\\
     &=
    - \frac{m_L |e_L e_H|}{2 \pi \sqrt{\bar E^2-m_L^2}} \left[1 +  \log\left(\frac{\bar L \sqrt{\bar E^2-m_L^2}}{m_L R}\right)\right] - \frac{ (|e_L e_H| m_L)^3}{192 \pi ^3 \bar{L}^2 (\bar{E}^2-m_L^2)^{3/2}} + \dots \nonumber \\
    &=
-\frac{|e_L e_H|}{2\pi\sqrt{\gamma^2-1}}
\left[ 1+\log\left(\frac{b(\gamma^2-1)}{R}\right) \right]
-\frac{\,|e_L e_H|^3}{192\pi^3} \frac{1}{b^2(\gamma^2-1)^{5/2}} + \dots, \nonumber
\end{align}
where $K := -e_H/(4\pi)$, $\bar E=\bar p_{L,0}$, $\bar L=\bar p_{L,\psi}$, and $\bar e$ is defined in \eqn{eq:pe_scalar}.
It is worth noting that in four spacetime dimensions the probe radial action contains only odd PL orders.
This is a special property of the $1/r$ Coulomb potential in three spatial dimensions, which possesses an enhanced symmetry generated by the Laplace–Runge–Lenz vector. The resulting $SO(1,3)$ (scattering) symmetry constrains $\bar{I}_r^{>}$ to be an odd function of $\alpha = e_L e_H$.
In $D$ dimensions,
where the potential scales as $1/r^{D-3}$ and the Runge–Lenz symmetry is absent,
even PL orders do not vanish in this way.

\section{Magnus amplitudes: self-force expansion}
\label{sec:Magnus_caseB}

In this section we compute WQFT-Magnus amplitudes in the SF expansion,
looking at both scattering motion ($>$) and bound orbits ($<$).
We write a complete set of non-zero matrix elements of $\hat{N}$ through 1SF and 3PL
order for scattering, and a corresponding set of matrix elements for bound orbits, setting $\hbar=1$ here as in section \ref{sec:Magnus_caseA}.
The superscript (SF) on both the interaction Hamiltonian~\eqref{eq:HintB_def}
and $\hat{N}$ should be considered implicit throughout.
When considering scattering, our time interval is taken to cover
the entire interaction history: $t_0\to -\infty$ and $t\to +\infty$,
\begin{align}
\hat N^{>} := \hat N^{(\rm SF)>}(+\infty,-\infty).
\label{eq:scattering_N}
\end{align}
We will demonstrate how this reproduces the results discussed in \sec{sec:Magnus_caseA},
but from a different diagrammatic (background-field) expansion. The two $\hat{N}$ matrices enjoy a simple all-order relationship:
\begin{align}\label{eq:PLSFrelationship}
\begin{aligned}
	\exp\left(i\hat{N}^{(\rm PL)}\right)&=
	\exp\left(i\hat N^{(\rm geo)}\right)\exp\left(i\hat{N}^{(\rm SF)>}\right)\\
	\implies i\hat{N}^{(\rm PL)}&=
	i\hat N^{(\rm geo)}+i\hat{N}^{(\rm SF)>}+
	\frac12\left[i\hat N^{(\rm geo)},i\hat{N}^{(\rm SF)>}\right]+\cdots\,,
\end{aligned}
\end{align}
where the geodesic operator $\hat N^{(\rm geo)}$ encodes the 0SF radial action~\eqref{eq:0SFradialaction}.
This relation \eqref{eq:PLSFrelationship} will be derived and explored further in \sec{sec:PLSFcomparison}.

For bound motion, depending on the observable of interest, one may consider either local quantities or orbit-averaged (per-cycle) observables, such as the periastron advance or the energy and angular momentum loss per orbit. The fundamental object is the finite-interval Magnus generator
\begin{align}
\hat N^{(\rm SF)<}(t,t_0), \qquad t > t_0\,,
\end{align}
from which both local radiative amplitudes and per-orbit variations may be extracted. When the motion admits well-defined radial cycles with turning points $t_n$, we introduce the cycle-restricted Magnus generator\footnote{A related construction is the secular average $\hat N^{<}_{\rm sec} := \lim_{T\to\infty}\,\hat N^{<}(t_0+T,t_0)/T$, which is useful to study the long-time behaviour of the system.}
\begin{align}
\hat N^{<}_{T_n} := \hat N^{(\rm SF)<}(t_{n+1},t_n),
\qquad
T_n := t_{n+1}-t_n.
\end{align}
Per-orbit quantities such as $\Delta\Phi_n$, $\Delta E_n$, and $\Delta L_n$ are extracted from $\hat N^{<}_{T_n}$, as will be explained in \sec{sec:observables}. In the small-mass-ratio expansion one has
\begin{align}
T_n = \bar T_r^{<}(E_n,L_n) + \mathcal O(\lambda^2),
\end{align}
where $\bar T_r^{<}$ is the geodesic radial period \eqref{eq:Tr_scalar}.
Since the radiative per-cycle losses are already $\mathcal O(\lambda^2)$,
replacing $T_n$ by $\bar T_r^{<}$ affects $\Delta E_n$, $\Delta L_n$, and $\Delta\Phi_n$
only at $\mathcal O(\lambda^3)$ (i.e.~2SF). This argument assumes a regular adiabatic regime, away from orbital resonances or near-critical configurations (such as separatrix behaviour), where the hierarchy between the orbital and inspiral timescales may break down.
Accordingly, to first order in $\lambda$, one may evaluate cycle-integrated quantities using the background radial interval $[t_n,\,t_n+\bar T^{<}_r]$, and consider only 
\begin{align}
\hat N^{<}_{\bar T} := \hat N^{(\rm SF)<}(t_{0}+\bar{T}_r^<,t_0)\,.
\label{eq:bound_N}
\end{align}
In the following, we explain how to evaluate directly these matrix elements $\hat N^{>}$ in \eqref{eq:scattering_N} and $\hat N^{<}_{\bar T}$ in \eqref{eq:bound_N} from the Magnus expansion.

\subsection{Scattering ($>$): radiative matrix elements}

As in the perturbative coupling expansion, we begin by considering the radiative matrix elements
that involve lower-order terms in the Magnus series~\eqref{Magnus}.
As we limit ourselves to 1SF order, this means a maximum of two scalar emissions is allowed.
At leading-order in the Magnus series, we also include the $g$ coupling.

\paragraph{Matrix elements of $\hat N^{>(1)}$}

We begin with the leading--order Magnus operator $i\hat N^{>(1)}$~\eqref{eq:N1},
with the 1SF interaction Hamiltonian density given in \eqn{eq:HintB_1SF}.
Separating bulk and worldline contributions and retaining only terms that can contribute to radiative matrix elements at this order, one finds
\begin{subequations}
\begin{align}
\left.i\hat N^{>(1)}\right|_{e_L}
&=
-i e_L \int\!\d\tau_L\;
\hat\varphi\left(\bar x_L(\tau_L)\right)\,,\\
\left.i\hat N^{>(1)}\right|_{e_H}
&=
-i e_H \int\!\d\tau_H\;
\hat{z}_H^\mu(\tau_H)\partial_\mu\hat\varphi\left(\bar x_H(\tau_H)\right)\,,\\
\left.i\hat N^{>(1)}\right|_{g}
&=
-i g\int\!\d^Dx\,
\left(
\bar\phi^2\hat\varphi
+\bar\phi\,\hat\varphi^2
\right)\,.
\end{align}
\end{subequations}
Using $\langle k|\hat\varphi(x)|0\rangle=e^{ik\cdot x}$, we obtain
\begin{subequations}
\begin{align}
iN^{>(1)}(k)&:=
\langle k|i\hat N^{>(1)}|0\rangle
=
-i e_L\int\!\d\tau_L\,e^{ik\cdot\bar x_L(\tau_L)}
-i\frac{g}{2}\int\!\d^Dx\,\bar\phi(x)^2\,e^{ik\cdot x}\,,\label{eq:N1_caseB_1scalar}\\
iN^{>(1)}(k_1,k_2)&:=
\langle k_1k_2|i\hat N^{>(1)}|0\rangle
=
-i\,\frac{g}{2}\int\!\d^Dx\,\bar\phi(x)\,e^{i(k_1+k_2)\cdot x}\,.\label{eq:N1_caseB_2scalar}
\end{align}
\end{subequations}
In terms of WQFT diagrams, these two contributions are depicted as
\begin{subequations}
\begin{align}
iN^{>(1)}(k)
&=
\begin{tikzpicture}[baseline={(current bounding box.center)},scale=0.95]
  \draw [worldlineStatic] (-1,0) -- (1,0);
  \draw [scalar2] (0,0) -- (0,-1) node [right] {$k$};
  \node[draw, circle, fill=white, inner sep=1.5pt, minimum size=12pt]
         at (0,0) {\small $L$};
\end{tikzpicture}
\;+\;\,
\begin{tikzpicture}[baseline={(current bounding box.center)},scale=0.95]
  \draw [scalar2] (0,0) -- (1.2,0) node [right] {$k$};
  \node[draw, circle, fill=white, inner sep=1.5pt, minimum size=12pt]
       at (0,0) {\small $\mathcal V_1$};
\end{tikzpicture}\,,\label{eq:N1_caseB_1scalarDiag}
\\[0.8em]
iN^{>(1)}(k_1,k_2)
&=
\begin{tikzpicture}[baseline={(current bounding box.center)},scale=0.95]
  \draw [scalar2] (0,0) -- (1.2,0) node [right] {$k_2$};
  \draw [scalar2] (0,0) -- (-1.2,0) node [left] {$k_1$};
  \node[draw, circle, fill=white, inner sep=1.5pt, minimum size=12pt]
       at (0,0) {\small $\mathcal V_2$};
\end{tikzpicture}\,,
\end{align}
\end{subequations}
where worldline and bulk vertices were given in \eqn{eq:WL_vertices_caseB}
and \eqn{eq:bulk_vertices_caseB} respectively.
At leading-PL order the $\mathcal{V}_1$ contribution in \eqn{eq:N1_caseB_1scalarDiag}
coincides with $i\hat N^{(3)}|_{ge_1^2}$~\eqref{eq:N3_ge1e1},
identifying particle 1 with H;
$N^{>(1)}(k_1,k_2)$ coincides
with $N^{(2)}(k_1,k_2)|_{g e_1}$~\eqref{eq:N2_gei_matrix_RA_short}.
While there is no contribution to $N^{>(1)}(k)$~\eqref{eq:N1_caseB_1scalar} from H,
as $\langle0|\hat{z}_H^\mu(\tau_H)|0\rangle=0$,
we can nevertheless form the following two non-zero matrix elements:
\begin{subequations}
\begin{align}
	\langle k|[i\hat N^{>(1)},\hat{p}_H^{\prime\mu}]|0\rangle&=
	ie_Hk^\mu\int\!\d\tau_H\, e^{ik\cdot\bar{x}_H(\tau_H)}=
	ie_Hk^\mu\dd(k\cdot v_H)\,,\\
	\langle k|[i\hat N^{>(1)},\hat{b}_H^{\prime\mu}]|0\rangle&=
	-ie_Hk^\mu\int\!\d\tau_H\,\frac{\tau_H}{m_H}e^{ik\cdot\bar{x}_H(\tau_H)}=
	-\frac{e_H}{m_H}k^\mu\dd^\prime(k\cdot v_H)\,,
\end{align}
\end{subequations}
where we used the decomposition $\hat{z}_H^\mu(\tau_H)$~\eqref{zDecomp}, $\langle k|\partial_\mu\hat\varphi(x)|0\rangle=ik_\mu e^{ik\cdot x}$ and the background solution $\bar{x}_H^\mu(\tau_H)=v_H^\mu\tau_H$.
The role of these additional matrix elements will be clarified in \sec{sec:observables},
as part of our discussion of SF observables.

The first term in \eqn{eq:N1_caseB_1scalar} represents emission from the Keplerian worldline.
We may express it as
\begin{align}
	\left.iN^{>(1)}(k)\right|_{e_L}=-ie_L J_L^>(k)\,,
\end{align}
where we have defined the Keplerian radiative current:
\begin{align}\label{eq:JL_rad_def}
\boxed{J_L^{>}(k):=\int\!\d\tau_L\,e^{ik\cdot\bar x_L(\tau_L)}\,.}
\end{align}
The non-trivial background motion breaks the straight-line worldline
translation symmetry, so the current $J_L^>(k)$~\eqref{eq:JL_rad_def} no longer collapses to a momentum–space delta function and acquires non–trivial frequency dependence. The straight-line limit is recovered by
PL-expanding the background trajectory:
\begin{align}
\bar{x}^{\mu}_L(\tau_L) &= b_L^\mu+v_L^{\mu} \tau_L+\cdots
&
&\implies
&
J_L^{>}(k) &= e^{ik\cdot b_L}\dd(v_L \cdot k) + \cdots\,,
\label{eq:PL_current}
\end{align}
in agreement with our result for $N^{(1)}(k)$ at leading-PL order~\eqref{eq:current}.
At higher PL orders, $\bar{x}_L^\mu$ is generated by a sum of diagrams
with retarded propagators oriented towards a single outgoing (uncut)
line on the light body~\cite{Mogull:2025cfn}.

The relationship between $\hat{N}^{(\rm PL)}$ and $\hat{N}^{(\rm SF)>}$~\eqref{eq:PLSFrelationship}
implies
\begin{align}\label{eq:exampleCausalityCut}
	iN(k)=iN^{>}(k)-\frac{i}2\{i(\bar I_r^>+\pi\bar L),iN^{>}(k)\}+\cdots\,,
\end{align}
having projected on a single outgoing scalar mode $|k\rangle$.
This is easily verified at the level of Feynman diagrams.
Plugging in the PL expansion of $\bar{x}_L^\mu$ into the Keplerian current~\eqref{eq:JL_rad_def},
the light-body contribution to $N^{>(1)}(k)$ is expanded as
\begin{align}\label{eq:Lsubleading}
\begin{tikzpicture}[baseline={(current bounding box.center)},scale=0.9]
  \draw [worldlineStatic] (-1,0) -- (1,0);
  \draw [scalar2] (0,0) -- (0,-1) node [right] {$k$};
  \fill (0,0) circle (.06);
  \node[draw, circle, fill=white, inner sep=1.5pt, minimum size=12pt]
         at (0,0) {\small $L$};
\end{tikzpicture}\,\,=\,\,
\begin{tikzpicture}[baseline={(current bounding box.center)},scale=1]
  		\coordinate (in) at (-.5,0);
  		\coordinate (out) at (.5,0);
  		\coordinate (x) at (0,0);
  		\draw [worldlineStatic] (in) -- (out);
		\draw [scalar2] (x) -- (0,-.7) node [right] {$k$};
  		\draw [fill] (x) circle (.06);
\end{tikzpicture}\,\,+\,\,
\begin{tikzpicture}[baseline={([yshift=-.5ex](.5,-.5))}]
        \path [draw=black, worldlineStatic] (-0.4,0) -- (0,0);
        \path [draw=black, zParticle] (0,0) -- (0.6,0);
		\path [draw=black, worldlineStatic] (0.6,0) -- (1.0,0);
        \path [draw=black, worldlineStatic] (-0.4,-1.) -- (1.0,-1.);
        \path [draw=black, scalar2] (0,-1.) -- (0,0);
		\path [draw=black, scalar2] (0.6,0) -- (1,-.5) node [right] {$k$};
        \filldraw[fill=black] (0,0) circle (.06);
        \filldraw[fill=black] (0,-1.) circle (.06);
		\filldraw[fill=black] (0.6,0) circle (.06);
\end{tikzpicture}\,\,+\,\,
\begin{tikzpicture}[baseline={([yshift=-.5ex](.5,-.5))}]
        \path [draw=black, worldlineStatic] (-1,0) -- (-.6,0);
		\path [draw=black, zParticle] (-.6,0) -- (0,0);
        \path [draw=black, zParticle] (0,0) -- (0.6,0);
		\path [draw=black, worldlineStatic] (0.6,0) -- (1.0,0);
        \path [draw=black, worldlineStatic] (-1,-1.) -- (1.0,-1.);
		\path [draw=black, scalar2] (-.6,-1.) -- (-.6,0);
        \path [draw=black, scalar2] (0,-1.) -- (0,0);
		\path [draw=black, scalar2] (0.6,0) -- (1,-.5) node [right] {$k$};
		\filldraw[fill=black] (-.6,0) circle (.06);
        \filldraw[fill=black] (-.6,-1.) circle (.06);
        \filldraw[fill=black] (0,0) circle (.06);
        \filldraw[fill=black] (0,-1.) circle (.06);
		\filldraw[fill=black] (0.6,0) circle (.06);
\end{tikzpicture}\,\,+\,\,
\frac12\,\,\begin{tikzpicture}[baseline={([yshift=-.5ex](.5,-.5))}]
        \path [draw=black, worldlineStatic] (-1,0) -- (0,0);
		\path [draw=black, zParticle] (-.6,0) to[out=45,in=135] (0.6,0);
        \path [draw=black, zParticle] (0,0) -- (0.6,0);
		\path [draw=black, worldlineStatic] (0.6,0) -- (1.0,0);
        \path [draw=black, worldlineStatic] (-1,-1.) -- (1.0,-1.);
		\path [draw=black, scalar2] (-.6,-1.) -- (-.6,0);
        \path [draw=black, scalar2] (0,-1.) -- (0,0);
		\path [draw=black, scalar2] (0.6,0) -- (1,-.5) node [right] {$k$};
		\filldraw[fill=black] (-.6,0) circle (.06);
        \filldraw[fill=black] (-.6,-1.) circle (.06);
        \filldraw[fill=black] (0,0) circle (.06);
        \filldraw[fill=black] (0,-1.) circle (.06);
		\filldraw[fill=black] (0.6,0) circle (.06);
\end{tikzpicture}\,\,+\,\,\cdots\,.
\end{align}
Meanwhile, using the diagrammatic identity in \eqn{diagBracket},
and our expression for the radial action $\bar{I}_r^>$~\eqref{eq:0SFradialaction},
at leading-PL order the causality cut~\eqref{eq:exampleCausalityCut} produces
\begin{align}\label{eq:exampleCut}
	-\frac{i}{2}\left\{
	\begin{tikzpicture}[baseline={([yshift=-.5ex](.5,-.5))}]
        \path [draw=black, worldlineStatic] (-0.6,0) -- (0,0);
        \path [draw=black, worldlineStatic] (0,0) -- (0.6,0);
        \path [draw=black, worldlineStatic] (-0.6,-1.) -- (0.6,-1.);
        \path [draw=black, scalar2] (0,0) -- (0,-1);
        \filldraw[fill=black] (0,0) circle (.06);
        \filldraw[fill=black] (0,-1.) circle (.06);
    \end{tikzpicture}
	\,\,,\,\,
	\begin{tikzpicture}[baseline={(current bounding box.center)},scale=1]
  		\coordinate (in) at (-.5,0);
  		\coordinate (out) at (.5,0);
  		\coordinate (x) at (0,0);
  		\draw [worldlineStatic] (in) -- (out);
		\draw [scalar2] (x) -- (0,-.7) node [right] {$k$};
  		\draw [fill] (x) circle (.06);
	\end{tikzpicture}
	\right\}
	\,\,=\,\,
	-\frac12\left(
	\begin{tikzpicture}[baseline={([yshift=-.5ex](.5,-.5))}]
        \path [draw=black, worldlineStatic] (-0.4,0) -- (0,0);
        \path [draw=black, zParticle] (0,0) -- (0.6,0);
		\path [draw=black, worldlineStatic] (0.6,0) -- (1.0,0);
        \path [draw=black, worldlineStatic] (-0.4,-1.) -- (1.0,-1.);
        \path [draw=black, scalar2] (0,-1.) -- (0,0);
		\path [draw=black, scalar2] (0.6,0) -- (1,-.5) node [right] {$k$};
        \filldraw[fill=black] (0,0) circle (.06);
        \filldraw[fill=black] (0,-1.) circle (.06);
		\filldraw[fill=black] (0.6,0) circle (.06);
    \end{tikzpicture}\,\,-\,\,
	\begin{tikzpicture}[baseline={([yshift=-.5ex](.5,-.5))}]
        \path [draw=black, worldlineStatic] (-0.4,0) -- (0,0);
        \path [draw=black, zParticle] (0.6,0) -- (0,0);
		\path [draw=black, worldlineStatic] (0.6,0) -- (1.0,0);
        \path [draw=black, worldlineStatic] (-0.4,-1.) -- (1.0,-1.);
        \path [draw=black, scalar2] (0,0) -- (0,-1);
		\path [draw=black, scalar2] (0.6,0) -- (1,-.5) node [right] {$k$};
        \filldraw[fill=black] (0,0) circle (.06);
        \filldraw[fill=black] (0,-1.) circle (.06);
		\filldraw[fill=black] (0.6,0) circle (.06);
    \end{tikzpicture}
	\right)\,.
\end{align}
When added to the light vertex~\eqref{eq:Lsubleading},
this contribution converts the retarded worldline propagator to
an average over retarded and advanced propagators ---
exactly as seen in our earlier
expression for $N^{(3)}(k)|_{e_1^2e_2}$~\eqref{eq:N3e1e1e2}.

\paragraph{Matrix elements of $\hat N^{>(2)}$}

Since the heavy-worldline vertex is already linear in the deflection operator $\hat z_H$,
the first Magnus operator $\hat N^{>(1)}$ does not produce a radiative contribution from $H$.
At second order, however, two such insertions can be contracted,
yielding a non-vanishing Compton-like matrix element
analogous to that at leading-PL order~\eqref{eq:Compton_caseA}:
\begin{align}\label{eq:N2_caseB_Compton_H_short}
\left.i N^{>(2)}(k_1,k_2)\right|_{\lambda^2g^0}
&=
\frac{(-i)^2}{2} e_H^2\,
\int\!\d\tau_H\d\tau_H'\;
(ik_{1\mu})(ik_{2\nu})\,
e^{i(k_1\cdot\bar x_H(\tau_H)+k_2\cdot\bar x_H(\tau_H'))} \nn \\
&\qquad \qquad \times \Big[
\Delta^{\mu\nu}_{H,R}(\tau_H-\tau_H')
+\Delta^{\mu\nu}_{H,A}(\tau_H-\tau_H')
\Big]
\\
&=
\frac12 \,e_H^2\,k_{1\mu}k_{2\nu}
\int_{\omega}\,
\dd(k_1\cdot v_H-\omega)\,
\dd(k_2\cdot v_H+\omega)\, \nn \nonumber \\
&\qquad \qquad \times \Big[
\widetilde\Delta^{\mu\nu}_{H,R}(\omega)
+\widetilde\Delta^{\mu\nu}_{H,A}(\omega)
\Big]
\nn\\
&=
\frac{i e_H^2}{2m_H}
\dd\big((k_1+k_2)\cdot v_H\big)\,
\left(\frac{k_1\cdot k_2}{(k_1\cdot v_H+i0)^2}+\frac{k_1\cdot k_2}{(k_1\cdot v_H-i0)^2}\right)\,, \nn
\end{align}
with $\Delta^{\mu\nu}_{H,R/A}$ the retarded and advanced worldline propagators~\eqref{wlProps}, and where we have inserted $\bar{x}_H^\mu(\tau_H)=v_H^\mu\tau_H$.
The overall factor $1/2$ in \eqn{eq:N2_caseB_Compton_H_short} is the Murua
coefficient~\eqref{murua2pt} associated with the second Magnus term.
Diagrammatically, we have the following representation:
\begin{align}
\left.iN^{>(2)}(k_1,k_2)\right|_{\lambda^2g^0}
=
\frac12\,
\begin{tikzpicture}[baseline={(current bounding box.center)},scale=0.95]
  \draw [worldlineStatic] (-1.2,0) -- (-0.6,0);
  \draw [zParticle] (-0.6,0) -- (0.6,0);
  \draw [worldlineStatic] (0.6,0) -- (1.2,0);
  \draw [scalar2] (-0.6,0) -- (-0.6,-1.0) node[left] {$k_1$};
  \draw [scalar2] (0.6,0) -- (0.6,-1.0) node[right] {$k_2$};
  \node[draw, circle, fill=white, inner sep=1.5pt, minimum size=12pt] at (-0.6,0) {$H$};
  \node[draw, circle, fill=white, inner sep=1.5pt, minimum size=12pt] at (0.6,0) {$H$};
\end{tikzpicture}
\;+\;
\frac12\,
\begin{tikzpicture}[baseline={(current bounding box.center)},scale=0.95]
  \draw [worldlineStatic] (-1.2,0) -- (-0.6,0);
  \draw [zParticle2] (0.6,0) -- (-0.6,0);
  \draw [worldlineStatic] (0.6,0) -- (1.2,0);
  \draw [scalar2] (-0.6,0) -- (-0.6,-1.0) node[left] {$k_1$};
  \draw [scalar2] (0.6,0) -- (0.6,-1.0) node[right] {$k_2$};
  \node[draw, circle, fill=white, inner sep=1.5pt, minimum size=12pt] at (-0.6,0) {$H$};
  \node[draw, circle, fill=white, inner sep=1.5pt, minimum size=12pt] at (0.6,0) {$H$};
\end{tikzpicture}
\,.
\label{eq:N2_caseB_Compton_H}
\end{align}
This is precisely equal to the Compton amplitude~\eqn{eq:Compton_caseA},
identifying the heavy particle as particle 1.

\paragraph{Matrix elements of $\hat N^{>(3)}$}

The 1SF single-emission matrix element $N^{>(3)}(k)$ is
\begin{align}\label{eq:N3scatter}
\begin{aligned}
\left.iN^{>(3)}(k)\right|_{\lambda^2g^0}=&\,\,
	\frac13\,\begin{tikzpicture}[baseline={([yshift=-.5ex](.5,-.5))}]
        \path [draw=black, worldlineStatic] (-0.6,-1) -- (0,-1);
        \path [draw=black, zParticle] (0,-1) -- (1.2,-1);
		\path [draw=black, worldlineStatic] (0.8,-1) -- (1.8,-1);
        \path [draw=black, worldlineStatic] (-0.6,0.5) -- (1.8,0.5);
        \path [draw=black, scalar] (0,0.5) -- (0,-1);
        \draw [scalar2] (1.2,-1) -- (1.8,-0.2) node [right] {$k$};
        \node[draw, circle, fill=white, inner sep=1.5pt, minimum size=10pt] at (0,0.5) {\small $L$};
        \node[draw, circle, fill=white, inner sep=1.5pt, minimum size=10pt] at (0,-1) {\small $H$};
		\node[draw, circle, fill=white, inner sep=1.5pt, minimum size=10pt] at (1.2,-1) {\small $H$};
    \end{tikzpicture}\,\,+\,\,
	\frac13\,\begin{tikzpicture}[baseline={([yshift=-.5ex](.5,-.5))}]
        \path [draw=black, worldlineStatic] (-0.6,-1) -- (0,-1);
        \path [draw=black, zParticle] (1.2,-1) -- (0,-1);
		\path [draw=black, worldlineStatic] (0.8,-1) -- (1.8,-1);
        \path [draw=black, worldlineStatic] (-0.6,0.5) -- (1.8,0.5);
        \path [draw=black, scalar] (0,-1) -- (0,0.5);
        \draw [scalar2] (1.2,-1) -- (1.8,-0.2) node [right] {$k$};
        \node[draw, circle, fill=white, inner sep=1.5pt, minimum size=10pt] at (0,0.5) {\small $L$};
        \node[draw, circle, fill=white, inner sep=1.5pt, minimum size=10pt] at (0,-1) {\small $H$};
		\node[draw, circle, fill=white, inner sep=1.5pt, minimum size=10pt] at (1.2,-1) {\small $H$};
    \end{tikzpicture} \\
&\!\!\!\!\!+	\frac16\,\begin{tikzpicture}[baseline={([yshift=-.5ex](.5,-.5))}]
        \path [draw=black, worldlineStatic] (-0.6,-1) -- (0,-1);
        \path [draw=black, zParticle] (0,-1) -- (1.2,-1);
		\path [draw=black, worldlineStatic] (0.8,-1) -- (1.8,-1);
        \path [draw=black, worldlineStatic] (-0.6,0.5) -- (1.8,0.5);
        \path [draw=black, scalar] (0,-1) -- (0,0.5);
        \draw [scalar2] (1.2,-1) -- (1.8,-0.2) node [right] {$k$};
        \node[draw, circle, fill=white, inner sep=1.5pt, minimum size=10pt] at (0,0.5) {\small $L$};
        \node[draw, circle, fill=white, inner sep=1.5pt, minimum size=10pt] at (0,-1) {\small $H$};
		\node[draw, circle, fill=white, inner sep=1.5pt, minimum size=10pt] at (1.2,-1) {\small $H$};
    \end{tikzpicture}\,\,+\,\,
	\frac16\,\begin{tikzpicture}[baseline={([yshift=-.5ex](.5,-.5))}]
        \path [draw=black, worldlineStatic] (-0.6,-1) -- (0,-1);
        \path [draw=black, zParticle] (1.2,-1) -- (0,-1);
		\path [draw=black, worldlineStatic] (0.8,-1) -- (1.8,-1);
        \path [draw=black, worldlineStatic] (-0.6,0.5) -- (1.8,0.5);
        \path [draw=black, scalar] (0,0.5) -- (0,-1);
        \draw [scalar2] (1.2,-1) -- (1.8,-0.2) node [right] {$k$};
        \node[draw, circle, fill=white, inner sep=1.5pt, minimum size=10pt] at (0,0.5) {\small $L$};
        \node[draw, circle, fill=white, inner sep=1.5pt, minimum size=10pt] at (0,-1) {\small $H$};
		\node[draw, circle, fill=white, inner sep=1.5pt, minimum size=10pt] at (1.2,-1) {\small $H$};
    \end{tikzpicture}\,.
\end{aligned}
\end{align}
Summing the four time orderings above gives
\begin{align}
\begin{aligned}
\left.iN^{>(3)}(k)\right|_{\lambda^2g^0}
&=
\frac{(-i)^3}{6}\,e_L e_H^2
\int{\rm d}\tau_L\,{\rm d}\tau_H\,{\rm d}\tau_H'\;
(ik_\mu)\,
e^{ik\cdot \bar x_H(\tau_H)}\,\\
&\qquad \times \Big[
2\,\Delta_{H,R}^{\mu\nu}(\tau_H-\tau_H')\,
\partial_\nu \Delta_R\big(\bar x_H(\tau_H')-\bar x_L(\tau_L)\big)\\
&\qquad\,\,\,
+2\,\Delta_{H,A}^{\mu\nu}(\tau_H-\tau_H')\,
\partial_\nu \Delta_A\big(\bar x_H(\tau_H')-\bar x_L(\tau_L)\big)\\
&\qquad\,\,\,\,\,\,
+\Delta_{H,R}^{\mu\nu}(\tau_H-\tau_H')\,
\partial_\nu \Delta_A\big(\bar x_H(\tau_H')-\bar x_L(\tau_L)\big)\\
&\qquad\,\,\,\,\,\,
+\Delta_{H,A}^{\mu\nu}(\tau_H-\tau_H')\,
\partial_\nu \Delta_R\big(\bar x_H(\tau_H')-\bar x_L(\tau_L)\big)
\Big] .
\label{eq:N3_singleem_time}
\end{aligned}
\end{align}
Identifying the heavy/light bodies as $1=H$ and $2=L$,
at leading-PL order this maps directly onto the $e_1^2e_2$ component of $N^{(3)}(k)$~\eqref{eq:N3e1e1e2}.

\subsection{Scattering ($>$): vacuum matrix elements}

Next, we consider the vacuum elements $iN^{(n)}=\langle0|i\hat N^{(n)}|0\rangle$.
Only two matrix elements are required to describe the 1SF dynamics up to 3PL order ---
when comparing with the PL expansion,
we note that 0SF information is already captured by
the background radial action $\bar{I}_r^{>}$~\eqref{eq:0SFradialaction}.

\paragraph{Matrix elements of $\hat N^{>(2)}$}

The first non--trivial 1SF contribution
arises from the vacuum graphs with two insertions on the $L$ worldline:
\begin{align}
\begin{aligned}
	\left.i N^{>(2)}\right|_{\lambda^2g^0}
	&=\frac{(-i)^2}{2} e_L^2 \int\d\tau_L\d\tau_L^\prime\,
	\theta(\tau_L-\tau_L^\prime)
	[\hat{\varphi}(\bar{x}_L(\tau_L)),\hat{\varphi}(\bar{x}_L(\tau_L^\prime))]\\
	&=\frac{(-i)^2}{2} e_L^2 \int\d\tau_L\d\tau_L^\prime\,
	\Delta_R(\bar{x}_L(\tau_L)-\bar{x}_L(\tau_L^\prime)) \\
    &=-\frac{e_L^2}{2}\int_k\, J_L^{>}(-k)\,\widetilde\Delta_R(k)\,J_L^{>}(k),
    \label{eq:mushroom_N2}
\end{aligned}
\end{align}
involving the Keplerian radiative current~\eqref{eq:JL_rad_def}.
Diagrammatically,
\begin{align}
\left.iN^{>(2)}\right|_{\lambda^2g^0}
=
\frac12\,
\begin{tikzpicture}[baseline={(current bounding box.center)},scale=0.95]
  \draw [worldlineStatic] (-1.6,0) -- (-0.8,0);
  \draw [worldlineStatic] (-0.8,0) -- (0.8,0);
  \draw [worldlineStatic] (0.6,0) -- (1.6,0);
  \draw [scalar] (-0.8,0) to[out=-90,in=-90] (0.8,0);
  \node[draw, circle, fill=white, inner sep=1.5pt, minimum size=12pt] at (-0.8,0) {$L$};
  \node[draw, circle, fill=white, inner sep=1.5pt, minimum size=12pt] at (0.8,0) {$L$};
\end{tikzpicture}
\,\,.
\end{align}
In the PL expansion, this encodes the ``mushroom'' diagrams on the light worldline $L$, which arise when evaluating loop integrals in the radiative region. An analogous mushroom topology on the heavy worldline $H$ also appears at 3PL order, contributing to $i N^{>(6)}|_{\lambda^2 g^0}$,
and follows directly from crossing symmetry arguments
(i.e~ by exchanging $L \leftrightarrow H$ at the end of the calculation).

We may also consider the $e_Le_H$ component of $\hat{N}^{>(2)}$, which takes the form
\begin{align}
\begin{aligned}
	i\hat{N}^{>(2)}|_{e_Le_H}=\frac{e_Le_H}{2}\int\!\d\tau_L\d\tau_H\,\hat{z}_H^\mu(\tau_H)
	\Big[&\partial_\mu\Delta_R(\bar{x}_L(\tau_L)-\bar{x}_H(\tau_H))\\
	+
	&\partial_\mu\Delta_A(\bar{x}_L(\tau_L)-\bar{x}_H(\tau_H))\Big]\,.
\end{aligned}
\end{align}
While the vacuum matrix element vanishes, these two components are non-zero:
\begin{subequations}
\begin{align}
	\langle 0|[i\hat N^{>(2)},\hat{p}_H^{\prime\mu}]|0\rangle&=
	i\frac{e_Le_H}{2}\int\!\d\tau_L\d\tau_H\,
	\Big[\partial^\mu\Delta_R(\bar{x}_L(\tau_L)-\bar{x}_H(\tau_H))\nn\\
	&\qquad\qquad\qquad\qquad\!\!
	+\partial^\mu\Delta_A(\bar{x}_L(\tau_L)-\bar{x}_H(\tau_H))\Big]\,,\\
	\langle 0|[i\hat N^{>(2)},\hat{b}_H^{\prime\mu}]|0\rangle&=
	-i\frac{e_Le_H}{2}\int\!\d\tau_L\d\tau_H\,\frac{\tau_H}{m_H}
	\Big[\partial^\mu\Delta_R(\bar{x}_L(\tau_L)-\bar{x}_H(\tau_H))\nn\\
	&\qquad\qquad\qquad\qquad\qquad
	+\partial^\mu\Delta_A(\bar{x}_L(\tau_L)-\bar{x}_H(\tau_H))\Big]\,.
\end{align}
\end{subequations}
Here again we have used the decomposition of $\hat{z}_H^\mu$
into $b_H^{\prime\mu}$ and $p_H^{\prime\mu}$~\eqref{zDecomp}.
Using the Keplerian radiative current~\eqref{eq:JL_rad_def} we can also write
\begin{subequations}
\begin{align}
	\langle 0|[i\hat N^{>(2)},\hat{p}_H^{\prime\mu}]|0\rangle&=
	\frac{i}2e_Le_H\int_k(-ik^\mu)J_L^>(-k)\dd(k\cdot v_H)(\tilde\Delta_R(k)+\tilde\Delta_A(k))\,,\\
	\langle 0|[i\hat N^{>(2)},\hat{b}_H^{\prime\mu}]|0\rangle&=
	-\frac{e_Le_H}{2m_H}\int_k(-ik^\mu)J_L^>(-k)\dd^\prime(k\cdot v_H)(\tilde\Delta_R(k)+\tilde\Delta_A(k))\,.
\end{align}
\end{subequations}
These two matrix elements will, again, be relevant for our discussion of observables.

\paragraph{Matrix element of $\hat N^{>(4)}$}

The other contribution to the vacuum element of $\hat{N}^>$
arises from vacuum graphs with two insertions on the $L$ worldline
and two insertions on the $H$ worldline:
\begin{align}\label{eq:N4Bscatter}
\begin{aligned}
\left.iN^{>(4)}\right|_{\lambda^2g^0}&=\,\,
	\frac14\,\begin{tikzpicture}[baseline={([yshift=-.5ex](.5,-.5))}]
        \path [draw=black, worldlineStatic] (-0.6,-1) -- (0,-1);
        \path [draw=black, zParticle] (0,-1) -- (1.2,-1);
		\path [draw=black, worldlineStatic] (0.8,-1) -- (1.8,-1);
        \path [draw=black, worldlineStatic] (-0.6,0.5) -- (1.8,0.5);
        \path [draw=black, scalar] (0,0.5) -- (0,-1);
		\path [draw=black, scalar] (1.2,-1) -- (1.2,0.5);
        \node[draw, circle, fill=white, inner sep=1.5pt, minimum size=10pt] at (0,0.5) {\small $L$};
		\node[draw, circle, fill=white, inner sep=1.5pt, minimum size=10pt] at (1.2,0.5) {\small $L$};
        \node[draw, circle, fill=white, inner sep=1.5pt, minimum size=10pt] at (0,-1) {\small $H$};
		\node[draw, circle, fill=white, inner sep=1.5pt, minimum size=10pt] at (1.2,-1) {\small $H$};
    \end{tikzpicture}\,+\,
	\frac1{12}\,\begin{tikzpicture}[baseline={([yshift=-.5ex](.5,-.5))}]
        \path [draw=black, worldlineStatic] (-0.6,-1) -- (0,-1);
        \path [draw=black, zParticle] (0,-1) -- (1.2,-1);
		\path [draw=black, worldlineStatic] (0.8,-1) -- (1.8,-1);
        \path [draw=black, worldlineStatic] (-0.6,0.5) -- (1.8,0.5);
        \path [draw=black, scalar] (0,0.5) -- (0,-1);
		\path [draw=black, scalar] (1.2,0.5) -- (1.2,-1);
        \node[draw, circle, fill=white, inner sep=1.5pt, minimum size=10pt] at (0,0.5) {\small $L$};
		\node[draw, circle, fill=white, inner sep=1.5pt, minimum size=10pt] at (1.2,0.5) {\small $L$};
        \node[draw, circle, fill=white, inner sep=1.5pt, minimum size=10pt] at (0,-1) {\small $H$};
		\node[draw, circle, fill=white, inner sep=1.5pt, minimum size=10pt] at (1.2,-1) {\small $H$};
    \end{tikzpicture} \\
    &\,+\frac1{12}\,\begin{tikzpicture}[baseline={([yshift=-.5ex](.5,-.5))}]
        \path [draw=black, worldlineStatic] (-0.6,-1) -- (0,-1);
        \path [draw=black, zParticle] (0,-1) -- (1.2,-1);
		\path [draw=black, worldlineStatic] (0.8,-1) -- (1.8,-1);
        \path [draw=black, worldlineStatic] (-0.6,0.5) -- (1.8,0.5);
        \path [draw=black, scalar] (0,-1) -- (0,0.5);
		\path [draw=black, scalar] (1.2,-1) -- (1.2,0.5);
        \node[draw, circle, fill=white, inner sep=1.5pt, minimum size=10pt] at (0,0.5) {\small $L$};
		\node[draw, circle, fill=white, inner sep=1.5pt, minimum size=10pt] at (1.2,0.5) {\small $L$};
        \node[draw, circle, fill=white, inner sep=1.5pt, minimum size=10pt] at (0,-1) {\small $H$};
		\node[draw, circle, fill=white, inner sep=1.5pt, minimum size=10pt] at (1.2,-1) {\small $H$};
    \end{tikzpicture}\,+\,
	\frac1{12}\,\begin{tikzpicture}[baseline={([yshift=-.5ex](.5,-.5))}]
        \path [draw=black, worldlineStatic] (-0.6,-1) -- (0,-1);
        \path [draw=black, zParticle] (0,-1) -- (1.2,-1);
		\path [draw=black, worldlineStatic] (0.8,-1) -- (1.8,-1);
        \path [draw=black, worldlineStatic] (-0.6,0.5) -- (1.8,0.5);
        \path [draw=black, scalar] (0,-1) -- (0,0.5);
		\path [draw=black, scalar] (1.2,0.5) -- (1.2,-1);
        \node[draw, circle, fill=white, inner sep=1.5pt, minimum size=10pt] at (0,0.5) {\small $L$};
		\node[draw, circle, fill=white, inner sep=1.5pt, minimum size=10pt] at (1.2,0.5) {\small $L$};
        \node[draw, circle, fill=white, inner sep=1.5pt, minimum size=10pt] at (0,-1) {\small $H$};
		\node[draw, circle, fill=white, inner sep=1.5pt, minimum size=10pt] at (1.2,-1) {\small $H$};
    \end{tikzpicture}.
\end{aligned}
\end{align}
At leading-PL order,
identifying $1=L$ and $2=H$ these contributions give us the second line of
$iN^{(4)}|_{e_1^2e_2^2}$~\eqref{eq:N4e1e1e2e2}, i.e.~the 1SF terms.
The 0SF diagrams in the first line are instead
provided by the radial action $\bar{I}_r^>$\eqref{eq:0SFradialaction}.
This is consistent with our stated relationship between $\hat{N}^{(\rm PL)}$
and $\hat{N}^{(\rm SF)>}$~\eqref{eq:PLSFrelationship},
which at leading-PL order implies that
\begin{align}
	iN^{{(\rm PL)}}=i\bar{I}_r^{>}+iN^{(\rm SF)>}+\cdots\,.
\end{align}
Here we have projected the generic formula \eqn{eq:PLSFrelationship}
between the operators $\hat N^{{(\rm PL)}}$ and $\hat N^{{(\rm SF)}}$ on the vacuum $|0\rangle$.

Unlike in the PL expansion the scalar propagators coupling to the
light vertices cannot be considered passive,
so their causality prescriptions remain significant.
In terms of couplings this contribution scales as $e_L^2 e_H^2$,
which becomes 
\begin{align}\label{eq:N4_scattering_SF}
\begin{aligned}
\left.i N^{>(4)}\right|_{\lambda^2 g^0}
&=
\frac{(-i)^4}{12}\,e_L^2e_H^2
\int\! \d\tau_L\, \d\tau_L'\, \d\tau_H\, \d\tau_H'\;
\Delta^{\mu\nu}_{H,R}(\tau_H-\tau_H')\,  \\
&\quad \times\Big[3\,\partial_\mu\Delta_R\Big(\bar x_L(\tau_L)-\bar x_H(\tau_H)\Big)\,
   \partial_\nu\Delta_A\Big(\bar x_L(\tau_L')-\bar x_H(\tau_H')\Big)\\
&\qquad+
\partial_\mu\Delta_A\Big(\bar x_L(\tau_L)-\bar x_H(\tau_H)\Big)\,
\partial_\nu\Delta_A\Big(\bar x_L(\tau_L')-\bar x_H(\tau_H')\Big)\\
&\qquad+
\partial_\mu\Delta_R\Big(\bar x_L(\tau_L)-\bar x_H(\tau_H)\Big)\,
\partial_\nu\Delta_R\Big(\bar x_L(\tau_L')-\bar x_H(\tau_H')\Big)\\
&\qquad+
\partial_\mu\Delta_A\Big(\bar x_L(\tau_L)-\bar x_H(\tau_H)\Big)\,
\partial_\nu\Delta_R\Big(\bar x_L(\tau_L')-\bar x_H(\tau_H')\Big)\Big]\,.
\end{aligned}
\end{align}
Again: all boundary conditions on time integrals are enforced by
the arguments of the retarded and advanced propagators.
Fourier transforming the bulk propagators, and then performing the $\tau_L,\tau_L'$ integrals produces the light currents $J_L^>(\pm k)$~\eqref{eq:JL_rad_def}; the $\tau_H,\tau_H'$ integrals produce the energy-conserving delta function $\dd(v_H\cdot(k_1+k_2))$.
One then finds the representation
\begin{align}
\left.i N^{>(4)}\right|_{\lambda^2 g^0}
&=
\frac{(-i)^4}{12}\,e_L^2e_H^2
\int_{k_1,k_2}
(-ik_{1\mu})(-ik_{2\nu})\,
J_L^{>}(-k_1)\,J_L^{>}(-k_2)\,
\dd\big(v_H\cdot(k_1+k_2)\big)\, \nonumber\\
&\qquad\times\widetilde\Delta^{\mu\nu}_{H,R}\big(k_1\cdot v_H\big) 
\Big[
3\,\widetilde\Delta_R(k_1)\widetilde\Delta_A(k_2)
+\widetilde\Delta_A(k_1)\widetilde\Delta_A(k_2) \nonumber \\
& \qquad \qquad \qquad \qquad \qquad+\widetilde\Delta_R(k_1)\widetilde\Delta_R(k_2)
+\widetilde\Delta_A(k_1)\widetilde\Delta_R(k_2)
\Big]\,,
\label{eq:N4_scattering_SF_final}
\end{align}
with the heavy worldline propagator evaluated at
$\omega=k_1\cdot v_H$.

\subsection{Bound orbits ($<$): radiative matrix elements}

We now turn to radiative matrix elements for bound motion in the background-field expansion,
providing counterparts to the scattering $\hat{N}$-matrix elements up to 1SF order.
Working over a finite time interval we use the cycle-restricted Magnus generator
$\hat N^{<}_{T} = \hat N^{<}(t_0+T,t_0)$. In the adiabatic approximation at 1SF order, $T$ can be replaced by an average over one geodesic radial period $\bar T_r^{<}$,
giving us $\hat{N}_{\bar T}$~\eqref{eq:bound_N}.

As in the previous examples involving scattering,
boundary conditions on time integrals are conveniently
encoded by the step functions $\theta(t_i-t_i^\prime)$ implicit in the
retarded and advanced propagators.
Thus, time integrals may always be taken to cover the full domain $[t_0,t_0+\bar{T}_r^<]$. We therefore introduce a compact notation:
\begin{align}
	\oint_{\bar T}\d t&:=\int_{t_0}^{t_0+\bar{T}_r^<}\!\d t\,, &
	\oint_{\bar T}\d\tau_L&:=\int_{t_0}^{t_0+\bar{T}_r^<}\!\d t\,\frac{\d\tau_L}{\d t}\,, &
	\oint_{\bar T}\d\tau_H&:=\int_{t_0}^{t_0+\bar{T}_r^<}\!\d t\,\frac{\d\tau_H}{\d t}\,.
\end{align}
The subscript $\bar{T}$ emphasises that each of these integrals covers the same time domain,
but using different coordinates.

\paragraph{Matrix elements of $\hat N_{\bar T}^{<(1)}$}

We begin with the leading Magnus operator $i\hat N^{<(1)}_{\bar T}$ in \eqn{Magnus},
with interaction Hamiltonian density given in \eqn{eq:HintB_def}. Separating
bulk and worldline contributions and retaining only terms that contribute to
radiative matrix elements up to 1SF order, one finds
\begin{subequations}
\begin{align}
\left.i\hat N^{<(1)}_{\bar T}\right|_{e_L}
&=
-i e_L \oint_{\bar T}\d t\int\d^{D-1}\mathbf{x}\!\int{\rm d}\tau_L\,\delta^D(x-\bar{x}_L(\tau_L))
\hat\varphi\left(\bar x_L(\tau_L)\right)\nn\\
&=
-i e_L \oint_{\bar T}\d\tau_L\;
\hat\varphi\left(\bar x_L(\tau_L)\right)\,,\\
\left.i\hat N^{<(1)}_{\bar T}\right|_{e_H}
&=
-i e_H \oint_{\bar T}\d t\int\d^{D-1}\mathbf{x}\!\int{\rm d}\tau_H\,\delta^D(x-\bar{x}_H(\tau_H))
\hat{z}_H^\mu(\tau_H)\partial_\mu\hat\varphi\left(\bar x_H(\tau_H)\right)\nn\\
&=
-i e_H \oint_{\bar T}\d\tau_H\;
\hat{z}_H^\mu(\tau_H)\partial_\mu\hat\varphi\left(\bar x_H(\tau_H)\right)\,,\\
\left.i\hat N^{<(1)}_{\bar T}\right|_{g}
&=
-\frac{i}2 g \oint_{\bar T}\d t\int\d^{D-1}\mathbf{x}\,
\left(
\bar\phi^2\hat\varphi+\bar\phi\,\hat\varphi^2
\right)\,.
\end{align}
\end{subequations}
Notice how, in the first two cases,
the worldline $\tau$ integrals over an infinite time domain
get restricted to a finite time domain once the $t$ integral is performed.
Using
$\langle k|\hat\varphi(x)|0\rangle=e^{ik\cdot x}$,
\begin{align}\label{eq:N1_caseB_1scalar_bound}
iN^{<(1)}_{\bar T}(k):=
\langle k|i\hat N^{<(1)}_{\bar T}|0\rangle=
-i e_L J^{<}_{L}(k)
-i\frac{g}{2}\oint_{\bar T}\d t\int\d^{D-1}\mathbf{x}\,
\bar\phi(x)^2\,e^{ik\cdot x}\,,
\end{align}
where we have introduced the single-period current
\begin{align}\label{eq:JL_rad_def_bound}
\boxed{J^{<}_{L}(k)
:=
\oint_{\bar T}\d\tau_L\,
e^{ik\cdot \bar x_L(\tau_L)}\,.}
\end{align}
This is the bound--orbit analogue of the scattering current \eqref{eq:JL_rad_def}.
Similarly, the two-scalar matrix element is
\begin{align}
iN^{<(1)}_{\bar T}(k_1,k_2)
&:=
\langle k_1k_2|i\hat N^{<(1)}_{\bar T}|0\rangle
=
-i\,g\oint_{\bar T}\d t\int\d^{D-1}\mathbf{x}\,\,
\bar\phi(x)\,e^{i(k_1+k_2)\cdot x}\,.
\label{eq:N1_caseB_2scalar_bound}
\end{align}
In terms of WQFT Feynman diagrams, the bound--orbit analogue of the radiative
emission off the light worldline is obtained by replacing the straight static
worldline by the periodic background trajectory,
which we depict as looping back on itself:
\begin{align}
iN^{<(1)}_{\bar T}(k)
&=\,\,
\begin{tikzpicture}[baseline={(current bounding box.center)},scale=1]
    \path[draw=black, worldlineStatic] (0,0) circle (1);
    \path (1,0) coordinate (x);
    \draw [scalar2] (x) -- (2,0) node [right] {$k$};
    \node[draw, circle, fill=white, inner sep=1.5pt, minimum size=12pt]
         at (1,0) {$L$};
\end{tikzpicture}
\;+\;
\begin{tikzpicture}[baseline={(current bounding box.center)},scale=0.9]
  \draw [scalar2] (0,0) -- (1.2,0) node [right] {$k$};
  \node[draw, circle, fill=white, inner sep=1.5pt, minimum size=12pt]
       at (0,0) {\small $\mathcal V_{1}^{\bar T}$};
\end{tikzpicture}\,,
\\[0.8em]
iN^{<(1)}_{\bar T}(k_1,k_2)
&=
\begin{tikzpicture}[baseline={(current bounding box.center)},scale=0.9]
  \draw [scalar2] (0,0) -- (1.2,0) node [right] {$k_2$};
  \draw [scalar2] (0,0) -- (-1.2,0) node [left] {$k_1$};
  \node[draw, circle, fill=white, inner sep=1.5pt, minimum size=12pt]
       at (0,0) {\small $\mathcal V_{2}^{\bar{T}}$};
\end{tikzpicture}\,.
\end{align}
In analogy with the scattering case, we denote the effective bulk insertions appearing in radiative matrix elements by $\mathcal V_i^{\bar{T}}$,
where $\bar{T}$ indicates that the same local interaction vertices are evaluated on a finite time interval $\bar{T}$, which coincides with the period of the background trajectory.

Finally, and also in analogy with the scattering case, we introduce
\begin{subequations}
\begin{align}
	\langle k|[i\hat N_{\bar T}^{<(1)},\hat{p}_H^{\prime\mu}]|0\rangle&=
	ie_Hk^\mu\oint\!\d\tau_H\, e^{ik\cdot\bar{x}_H(\tau_H)}=
	ie_Hk^\mu I^{<}_{H,\bar T}(k\cdot v_H)\,,\\
	\langle k|[i\hat N_{\bar T}^{<(1)},\hat{b}_H^{\prime\mu}]|0\rangle&=
	-ie_Hk^\mu\oint\!\d\tau_H\,\frac{\tau_H}{m_H}e^{ik\cdot\bar{x}_H(\tau_H)}=
	-\frac{e_H}{m_H}k^\mu I^{<\prime}_{H,\bar T}(k\cdot v_H)\,,
\end{align}
\end{subequations}
where for the heavy body $H$ we have introduced
\begin{align}\label{heavyI}
I^{<}_{H,\bar T}(\omega):=\oint_{\bar T}\d\tau_H\,e^{i\omega\tau_H} =
\frac2\omega
e^{i\omega(\tau_{H,0}+\tau_{H,\bar T}/2)}
\sin\left(\frac{\omega\tau_{H,\bar T}}2\right) \,,
\end{align}
and its $\omega$-derivative $I^{<\prime}_{H,\bar T}(\omega)$.
In the scattering case \eqref{eq:N2_caseB_Compton_H_short}, the corresponding infinite-time integral yields the energy-conserving factor $\dd(\omega)$.
In the present bound-orbit formulation this delta function is therefore replaced by the finite-time interval
\begin{align}
\dd(\omega)
 \longrightarrow 
I^{<}_{H,\bar T}(\omega)\,,
\end{align}
the rest being identical to the scattering case.

\paragraph{Matrix elements of $\hat N_{\bar T}^{<(2)}$}

We next consider the second Magnus operator on a finite time interval:
\begin{align}
i\hat N^{<(2)}_{\bar T}
=\frac{(-i)^2}{2}\oint_{\bar T}\d t
\oint_{\bar T}\d t^\prime\;
\theta(t-t^\prime)\,[\hat H_I(t),\hat H_I(t')]\,,
\label{eq:N2_caseB_master_bound}
\end{align}
with interaction Hamiltonian density given in \eqn{eq:HintB_1SF}.
As in the scattering case, the leading worldline contribution arises from the heavy particle:
\begin{align}\label{eq:N2_caseB_Compton_H_short_bound}
\left.i N^{<(2)}_{\bar T}(k_1,k_2)\right|_{\lambda^2g^0}
&= \frac{(-i)^2}{2} e_H^2\,
\oint_{\bar T} \d\tau_H
\oint_{\bar T} \d\tau_H'\;
(ik_{1\mu})(ik_{2\nu})\,
e^{i(k_1\cdot\bar x_H(\tau_H)+k_2\cdot\bar x_H(\tau_H'))}\nn\\
&\qquad\times
\Big[ \Delta^{\mu\nu}_{H,R}(\tau_H-\tau_H') + \Delta^{\mu\nu}_{H,A}(\tau_H-\tau_H') \Big]\\
&=
\frac12\,e_H^2\,k_{1\mu}k_{2\nu}
\int_{\omega}\,
I_{H,\bar T}^{<}(k_1\cdot v_H-\omega)\,
I_{H,\bar T}^{<}(k_2\cdot v_H+\omega)\, \nn \\
&\qquad \times \Big[
\widetilde\Delta^{\mu\nu}_{H,R}(\omega)
+\widetilde\Delta^{\mu\nu}_{H,A}(\omega)
\Big]\,,\nn
\end{align}
where $I_{H,\bar{T}}^<$ was introduced in \eqn{heavyI}.
The diagrams are now evaluated over the finite time interval defining
$\hat N_{\bar T}^{<}$:
\begin{align}
\left.iN_{\bar T}^{<(2)}(k_1,k_2)\right|_{\lambda^2g^0}
=
\frac12\,
\begin{tikzpicture}[baseline={([yshift=1.7ex](.5,-.5))},scale=0.95]
  \path[draw=black, worldlineStatic] (0,0) circle (0.9);
   \path[draw=black, line width=1pt, postaction={decorate},
      decoration={markings, mark=at position 0.55 with {\arrow{latex}}}]
      (-0.9,0) arc (-180:0:0.9);
  \draw [scalar2] (-0.9,0) -- (-0.9,1.2) node[left] {$k_1$};
  \draw [scalar2] (0.9,0) -- (0.9,1.2) node[right] {$k_2$};
  \node[draw, circle, fill=white, inner sep=1.5pt, minimum size=12pt] at (-0.9,0) {\small $H$};
  \node[draw, circle, fill=white, inner sep=1.5pt, minimum size=12pt] at (0.9,0) {\small $H$};
\end{tikzpicture}
\;+\;
\frac12\,
\begin{tikzpicture}[baseline={([yshift=1.7ex](.5,-.5))},scale=0.95]
    \path[draw=black, worldlineStatic] (0,0) circle (0.9);
  \path[draw=black, line width=1pt, postaction={decorate},
      decoration={markings, mark=at position 0.55 with {\arrow{latex}}}]
      (0.9,0) arc (0:-180:0.9);
  \draw [scalar2] (-0.9,0) -- (-0.9,1.2) node[left] {$k_1$};
  \draw [scalar2] (0.9,0) -- (0.9,1.2) node[right] {$k_2$};
  \node[draw, circle, fill=white, inner sep=1.5pt, minimum size=12pt] at (-0.9,0) {\small $H$};
  \node[draw, circle, fill=white, inner sep=1.5pt, minimum size=12pt] at (0.9,0) {\small $H$};
\end{tikzpicture}
\,.
\label{eq:N2_caseB_Compton_HT}
\end{align}
Again, the background is depicted as looping back on itself as a reminder that
we are dealing with bound orbits taken over a finite time interval.

\paragraph{Matrix elements of $\hat N_{\bar T}^{<(3)}$}

At order 1SF and $g^0$, the single-emission matrix element $N^{<(3)}_{\bar T}(k)$ is given by
\begin{align}
\begin{aligned}
\left.iN^{(3)<}_{\bar T}(k)\right|_{\lambda^2g^0}
&=\,\,
\frac13\,\,\,
\begin{tikzpicture}[baseline={([yshift=1.7ex](.5,-.5))},scale=0.95]
  \path[draw=black, worldlineStatic] (0,0) circle (0.8);
  \path[draw=black, worldlineStatic] (0,0) circle (2);
  \filldraw[fill=white,color=white] (2,0) circle (9pt);
  \path (0,0.7) coordinate (Btop);
  \path (0.9,0) coordinate (Bright);
  \path (0,2.2) coordinate (A);
  \path (2.1,0.0) coordinate (R);
  \path[draw=black, zParticle, ->] (0,0.8) arc (90:0:0.8);
\path[draw=black, scalar] (A) -- (Btop);
\draw [scalar2] (Bright) -- (R) node [right] {$k$};
  \node[draw,circle,fill=white,inner sep=1.5pt,minimum size=10pt] at (0,2) {\small $L$};
  \node[draw,circle,fill=white,inner sep=1.5pt,minimum size=10pt] at (0,0.8) {\small $H$};
  \node[draw,circle,fill=white,inner sep=1.5pt,minimum size=10pt] at (0.8,0) {\small $H$};
\end{tikzpicture}
\, + \,
\frac13\,\,\,
\begin{tikzpicture}[baseline={([yshift=1.7ex](.5,-.5))},scale=0.95]
  \path[draw=black, worldlineStatic] (0,0) circle (0.8);
  \path[draw=black, worldlineStatic] (0,0) circle (2);
  \filldraw[fill=white,color=white] (2,0) circle (9pt);
  \path (0,0.7) coordinate (Btop);
  \path (0.9,0) coordinate (Bright);
  \path (0,2.2) coordinate (A);
  \path (2.1,0) coordinate (R);
  \path[draw=black, zParticle, ->] (0.8,0) arc (0:90:0.8);
\path[draw=black, scalar] (Btop) -- (A);
\draw [scalar2] (Bright) -- (R) node [right] {$k$};
  \node[draw,circle,fill=white,inner sep=1.5pt,minimum size=10pt] at (0,2) {\small $L$};
  \node[draw,circle,fill=white,inner sep=1.5pt,minimum size=10pt] at (0,0.8) {\small $H$};
  \node[draw,circle,fill=white,inner sep=1.5pt,minimum size=10pt] at (0.8,0) {\small $H$};
\end{tikzpicture}\,\\
&+ \,
\frac16\,\,\,
\begin{tikzpicture}[baseline={([yshift=1.7ex](.5,-.5))},scale=0.95]
  \path[draw=black, worldlineStatic] (0,0) circle (0.8);
  \path[draw=black, worldlineStatic] (0,0) circle (2);
  \filldraw[fill=white,color=white] (2,0) circle (9pt);
  \path (0,0.7) coordinate (Btop);
  \path (0.9,0) coordinate (Bright);
  \path (0,2.2) coordinate (A);
  \path (2.1,0.0) coordinate (R);
  \path[draw=black, zParticle, ->] (0.8,0) arc (0:90:0.8);
\path[draw=black, scalar] (A) -- (Btop);
\draw [scalar2] (Bright) -- (R) node [right] {$k$};
  \node[draw,circle,fill=white,inner sep=1.5pt,minimum size=10pt] at (0,2) {\small $L$};
  \node[draw,circle,fill=white,inner sep=1.5pt,minimum size=10pt] at (0,0.8) {\small $H$};
  \node[draw,circle,fill=white,inner sep=1.5pt,minimum size=10pt] at (0.8,0) {\small $H$};
\end{tikzpicture}
\, + \,
\frac16\,\,\,
\begin{tikzpicture}[baseline={([yshift=1.7ex](.5,-.5))},scale=0.95]
  \path[draw=black, worldlineStatic] (0,0) circle (0.8);
  \path[draw=black, worldlineStatic] (0,0) circle (2);
  \filldraw[fill=white,color=white] (2,0) circle (9pt);
  \path (0,0.7) coordinate (Btop);
  \path (0.9,0) coordinate (Bright);
  \path (0,2.2) coordinate (A);
  \path (2.1,0) coordinate (R);
  \path[draw=black, zParticle, ->] (0,0.8) arc (90:0:0.8);
\path[draw=black, scalar] (Btop) -- (A);
\draw [scalar2] (Bright) -- (R) node [right] {$k$};
  \node[draw,circle,fill=white,inner sep=1.5pt,minimum size=10pt] at (0,2) {\small $L$};
  \node[draw,circle,fill=white,inner sep=1.5pt,minimum size=10pt] at (0,0.8) {\small $H$};
  \node[draw,circle,fill=white,inner sep=1.5pt,minimum size=10pt] at (0.8,0) {\small $H$};
\end{tikzpicture}
\,\,\,,
\end{aligned}
\end{align}
which explicitly reads
\begin{align}\label{eq:N3_singleem_time_bound}
\begin{aligned}
\left.iN^{<(3)}_{\bar T}(k)\right|_{\lambda^2g^0}
&=
\frac{(-i)^3}{6}\,e_L e_H^2
\oint_{\bar T}\d\tau_L
\oint_{\bar T}\d\tau_H
\oint_{\bar T}\d\tau_H'
(ik_\mu)\,
e^{ik\cdot \bar x_H(\tau_H)}\\
&\qquad \times \Big[
2\,\Delta_{H,R}^{\mu\nu}(\tau_H-\tau_H')\,
\partial_\nu \Delta_R\big(\bar x_H(\tau_H')-\bar x_L(\tau_L)\big)\\
&\qquad\,\,\,
+2\,\Delta_{H,A}^{\mu\nu}(\tau_H-\tau_H')\,
\partial_\nu \Delta_A\big(\bar x_H(\tau_H')-\bar x_L(\tau_L)\big)\\
&\qquad\,\,\,\,\,\,
+\Delta_{H,A}^{\mu\nu}(\tau_H-\tau_H')\,
\partial_\nu \Delta_R\big(\bar x_H(\tau_H')-\bar x_L(\tau_L)\big)\\
&\qquad\,\,\,\,\,\,
+\Delta_{H,R}^{\mu\nu}(\tau_H-\tau_H')\,
\partial_\nu \Delta_A\big(\bar x_H(\tau_H')-\bar x_L(\tau_L)\big)
\Big]\,.
\end{aligned}
\end{align}
Like in the previous bound orbit examples,
this expression is essentially the same as that presented for scattering~\eqref{eq:N3_singleem_time}.
The main difference here is our restriction to integration over a finite time domain.

\subsection{Bound orbits ($<$): vacuum matrix elements}

Finally, let us consider vacuum matrix elements of $\hat{N}^{<}$ in the bound case.
Again, we provide counterparts to all scattering matrix elements at 1SF order.

\paragraph{Matrix elements of $\hat N_{\bar T}^{<(2)}$}

The simplest vacuum element is
\begin{align}
\begin{aligned}
	\left.i N_{\bar T}^{<(2)}\right|_{\lambda^2g^0}
	&=\frac{(-i)^2}{2} e_L^2 \oint_{\bar T}\d\tau_L\oint_{\bar T}\d\tau_L^\prime\,
	\theta(\tau_L-\tau_L')
	[\varphi(\bar{x}_L(\tau_L)),\varphi(\bar{x}_L(\tau_L^\prime))]\\
	&=\frac{(-i)^2}{2} e_L^2 \oint_{\bar T}\d\tau_L\oint_{\bar T}\d\tau_L^\prime\,
	\Delta_R(\bar{x}_L(\tau_L)-\bar{x}_L(\tau_L^\prime))\\
	&=-\frac{1}{2} e_L^2 \int_{k}\,
	J_L^{<}(-k)\,\widetilde\Delta_R(k)\,J_L^{<}(k)\,,
\end{aligned}
\end{align}
where in the last line we have used the definition of the bound current \eqref{eq:JL_rad_def_bound}. This expression is structurally identical to the scattering counterpart
\eqref{eq:mushroom_N2}, the only difference being that we replace
\begin{align}
J_L^{>}(k) \longrightarrow J_L^{<}(k)\,.
\end{align}
In both cases the dynamics is entirely encoded in the quadratic current structure, while the bulk dynamics remains universal. For this topology, the distinction between scattering and bound motion is therefore carried solely by the choice of worldline current, which implements the respective boundary conditions (infinite straight-line motion versus periodic radial evolution). 
Diagrammatically, this term is depicted as
\begin{align}
\left.iN_{\bar T}^{<(2)}\right|_{\lambda^2g^0}
=
\frac12\,\,
\begin{tikzpicture}[baseline={([yshift=1.7ex](.5,-.5))},scale=0.95]
  \path[draw=black, worldlineStatic] (0,0) circle (1);
  \draw [scalar] (-0.7,0) to (0.7,0);
  \node[draw, circle, fill=white, inner sep=1.5pt, minimum size=12pt] at (-1,0) {\small $L$};
  \node[draw, circle, fill=white, inner sep=1.5pt, minimum size=12pt] at (1,0) {\small $L$};
\end{tikzpicture}\,\,.
\end{align}
In other words, the light body exchanges a scalar with itself. As in the scattering scenario, there is also a related contribution to $\left.i N^{<(6)}\right|_{\lambda^2g^0}$ which can be derived from crossing symmetry arguments (i.e.~by exchanging $L \leftrightarrow H$).

We also need the mixed $e_Le_H$ component of $\hat N_{\bar T}^{<(2)}$:
\begin{align}
\begin{aligned}
	i\hat{N}_{\bar T}^{<(2)}|_{e_Le_H}=
	\frac{e_Le_H}{2}\oint_{\bar T}\d\tau_L\oint_{\bar T}\d\tau_H\,\hat{z}_H^\mu(\tau_H)
	\Big[&\partial_\mu\Delta_R(\bar{x}_L(\tau_L)-\bar{x}_H(\tau_H))\\
	+
	&\partial_\mu\Delta_A(\bar{x}_L(\tau_L)-\bar{x}_H(\tau_H))\Big]\,.
\end{aligned}
\end{align}
By analogous calculation to the scattering case, 
and using the single-period current~\eqref{eq:JL_rad_def_bound},
the two relevant matrix elements are
\begin{subequations}
\begin{align}
	\langle 0|[i\hat N_{\bar T}^{<(2)},\hat{p}_H^{\prime\mu}]|0\rangle&=
	\frac{i}2e_Le_H\int_k(-ik^\mu)J_L^<(-k)I_{H,\bar T}^<(k\cdot v_H)(\tilde\Delta_R(k)+\tilde\Delta_A(k))\,,\\
	\langle 0|[i\hat N_{\bar T}^{<(2)},\hat{b}_H^{\prime\mu}]|0\rangle&=
	-\frac{e_Le_H}{2m_H}\int_k(-ik^\mu)J_L^<(-k)I_{H,\bar T}^{<\prime}(k\cdot v_H)(\tilde\Delta_R(k)+\tilde\Delta_A(k))\,.
\end{align}
\end{subequations}
These two matrix elements will enter the SF observables discussed in  \sec{sec:observables}.

\paragraph{Matrix elements of $\hat N_{\bar T}^{<(4)}$}

Our last bound-case example is the analog of \eqn{eq:N4Bscatter}:
\begin{align}
\begin{aligned}
\left.iN^{<(4)}_{\bar T}\right|_{\lambda^2g^0}
&=\,\,
\frac14\,\,\,
\begin{tikzpicture}[baseline={([yshift=1.7ex](.5,-.5))},scale=0.95]
  \path[draw=black, worldlineStatic] (0,0) circle (0.8);
  \path[draw=black, worldlineStatic] (0,0) circle (2);
  \path (0,0.8) coordinate (Btop);
  \path (0.8,0) coordinate (Bright);
  \path (0,2.1) coordinate (A);
  \path (2.1,0) coordinate (R);
  \path[draw=black, zParticle, ->] (0,0.8) arc (90:0:0.8);
\path[draw=black, scalar] (A) -- (Btop);
\path[draw=black, scalar] (Bright) -- (R);
  \node[draw,circle,fill=white,inner sep=1.5pt,minimum size=10pt] at (0,2) {\small $L$};
    \node[draw,circle,fill=white,inner sep=1.5pt,minimum size=10pt] at (2,0) {\small $L$};
  \node[draw,circle,fill=white,inner sep=1.5pt,minimum size=10pt] at (0,0.8) {\small $H$};
  \node[draw,circle,fill=white,inner sep=1.5pt,minimum size=10pt] at (0.8,0) {\small $H$};
\end{tikzpicture}
\, + \,
\frac{1}{12}\,\,\,
\begin{tikzpicture}[baseline={([yshift=1.7ex](.5,-.5))},scale=0.95]
  \path[draw=black, worldlineStatic] (0,0) circle (0.8);
  \path[draw=black, worldlineStatic] (0,0) circle (2);
  \path (0,0.8) coordinate (Btop);
  \path (0.8,0) coordinate (Bright);
  \path (0,2.1) coordinate (A);
  \path (2.1,0) coordinate (R);
  \path[draw=black, zParticle, ->] (0,0.8) arc (90:0:0.8);
\path[draw=black, scalar] (A) -- (Btop);
\path[draw=black, scalar] (R) -- (Bright);
  \node[draw,circle,fill=white,inner sep=1.5pt,minimum size=10pt] at (0,2) {\small $L$};
    \node[draw,circle,fill=white,inner sep=1.5pt,minimum size=10pt] at (2,0) {\small $L$};
  \node[draw,circle,fill=white,inner sep=1.5pt,minimum size=10pt] at (0,0.8) {\small $H$};
  \node[draw,circle,fill=white,inner sep=1.5pt,minimum size=10pt] at (0.8,0) {\small $H$};
\end{tikzpicture}
\, \\
&+ \,
\frac{1}{12}\,\,\,
\begin{tikzpicture}[baseline={([yshift=1.7ex](.5,-.5))},scale=0.95]
  \path[draw=black, worldlineStatic] (0,0) circle (0.8);
  \path[draw=black, worldlineStatic] (0,0) circle (2);
  \path (0,0.8) coordinate (Btop);
  \path (0.8,0) coordinate (Bright);
  \path (0,2.1) coordinate (A);
  \path (2.1,0.0) coordinate (R);
  \path[draw=black, zParticle, ->] (0,0.8) arc (90:0:0.8);
\path[draw=black, scalar] (Btop) -- (A);
\path[draw=black, scalar] (Bright) -- (R);
  \node[draw,circle,fill=white,inner sep=1.5pt,minimum size=10pt] at (0,2) {\small $L$};
    \node[draw,circle,fill=white,inner sep=1.5pt,minimum size=10pt] at (2,0) {\small $L$};
  \node[draw,circle,fill=white,inner sep=1.5pt,minimum size=10pt] at (0,0.8) {\small $H$};
  \node[draw,circle,fill=white,inner sep=1.5pt,minimum size=10pt] at (0.8,0) {\small $H$};
\end{tikzpicture}
\, + \,
\frac{1}{12}\,\,\,
\begin{tikzpicture}[baseline={([yshift=1.7ex](.5,-.5))},scale=0.95]
  \path[draw=black, worldlineStatic] (0,0) circle (0.8);
  \path[draw=black, worldlineStatic] (0,0) circle (2);
  \path (0,0.8) coordinate (Btop);
  \path (0.8,0) coordinate (Bright);
  \path (0,2.1) coordinate (A);
  \path (2.1,0) coordinate (R);
  \path[draw=black, zParticle, ->] (0,0.8) arc (90:0:0.8);
\path[draw=black, scalar] (Btop) -- (A);
\path[draw=black, scalar] (R) -- (Bright);
  \node[draw,circle,fill=white,inner sep=1.5pt,minimum size=10pt] at (0,2) {\small $L$};
  \node[draw,circle,fill=white,inner sep=1.5pt,minimum size=10pt] at (2,0) {\small $L$};
  \node[draw,circle,fill=white,inner sep=1.5pt,minimum size=10pt] at (0,0.8) {\small $H$};
  \node[draw,circle,fill=white,inner sep=1.5pt,minimum size=10pt] at (0.8,0) {\small $H$};
\end{tikzpicture}
\,\,\,,
\end{aligned}
\end{align}
which becomes
\begin{align}\label{eq:N4_bound_SF_final}
\left.iN_{\bar T}^{<(4)}\right|_{\lambda^2 g^0}
&=
\frac{(-i)^4}{12}\,e_L^2e_H^2
\oint_{\bar{T}}\d\tau_L\oint_{\bar{T}}\d\tau_L'\oint_{\bar{T}}\d\tau_H\oint_{\bar{T}}\d\tau_H'\;
\Delta^{\mu\nu}_{H,R}(\tau_H-\tau_H')\, \\
&\quad \times\Big[3\,
\partial_\mu\Delta_R\Big(\bar x_L(\tau_L)-\bar x_H(\tau_H)\Big)\,
\partial_\nu\Delta_A\Big(\bar x_L(\tau_L')-\bar x_H(\tau_H')\Big)
\nonumber\\
&\qquad+
\partial_\mu\Delta_A\Big(\bar x_L(\tau_L)-\bar x_H(\tau_H)\Big)\,
\partial_\nu\Delta_A\Big(\bar x_L(\tau_L')-\bar x_H(\tau_H')\Big)
\nonumber\\
&\qquad+
\partial_\mu\Delta_R\Big(\bar x_L(\tau_L)-\bar x_H(\tau_H)\Big)\,
\partial_\nu\Delta_R\Big(\bar x_L(\tau_L')-\bar x_H(\tau_H')\Big)
\nonumber\\
&\qquad+
\partial_\mu\Delta_A\Big(\bar x_L(\tau_L)-\bar x_H(\tau_H)\Big)\,
\partial_\nu\Delta_R\Big(\bar x_L(\tau_L')-\bar x_H(\tau_H')\Big)\Big] \nonumber \\
&=
\frac{(-i)^4}{12}\,e_L^2e_H^2
\int_{k_1,k_2}
(-ik_{1\mu})(-ik_{2\nu})\,
J_L^{<}(-k_1)\,J_L^{<}(-k_2)
\nonumber\\
&\times
\int_{\omega}\,
\widetilde\Delta^{\mu\nu}_{H,R}(\omega)\,
I^{<}_{H,\bar T}\big(k_1\!\cdot v_H-\omega\big)\,
I^{<}_{H,\bar T}\big(k_2\!\cdot v_H+\omega\big)
\nonumber\\
&\times
\Big[
3\,\widetilde\Delta_R(k_1)\widetilde\Delta_A(k_2)
+\widetilde\Delta_A(k_1)\widetilde\Delta_A(k_2)
+\widetilde\Delta_R(k_1)\widetilde\Delta_R(k_2)
+\widetilde\Delta_A(k_1)\widetilde\Delta_R(k_2)
\Big]. \nn
\end{align}
Again, the diagrams mirror those of the scattering scenario, with the replacements $(J^{>}_L(k),\dd(\omega)) \rightarrow (J^{<}_L(k),I^{<}_{H,\bar T}(\omega))$:
the main practical difference is that we evaluate integrals over finite time intervals,
and with a different choice of background.

\newpage

\section{Scattering and bound observables}\label{sec:observables}

Having seen how to calculate WQFT-Magnus amplitudes $N(k_1,k_2,\ldots)$,
both in the PL and SF expansions,
let us now explore how these on-shell,
gauge-invariant building blocks encode physical observables ---
both conservative and fully dissipative.
Our discussion generalises that of Alessio, Shi and one of the authors~\cite{Alessio:2025flu},
moving from an amplitudes~\cite{Kosower:2018adc,Cristofoli:2021vyo} to a WQFT perspective and allowing
for generic time dependence in the case of bound orbits.
In the operator-based formalism, still working in the interaction picture,
the expectation value of an observable $\cO(t,\mathbf{x})$ is given by:
\begin{align}
	\langle\hat\cO(t,\mathbf{x})\rangle
	:=\langle\psi(t)|\hat\cO(t,\mathbf{x})|\psi(t)\rangle
	=\langle0|\hat{U}(t_0,t)\hat\cO(t,\mathbf{x})\hat{U}(t,t_0)|0\rangle\,.
\end{align}
Our manipulation is fully generic,
applying in both the PL and SF expansions for both bound and unbound trajectories ---
in the latter case, we simply send $t_0\to-\infty$.
Inserting the $\hat{N}$-matrix form of the time-evolution operator~\eqref{eq:MagnusDef}, this becomes
\begin{align}\label{eq:observables}
\begin{aligned}
	\langle\hat\cO(t,\mathbf{x})\rangle
	&=\langle0|e^{-\frac{i}{\hbar}\hat N(t,t_0)}\hat\cO(t,\mathbf{x})e^{\frac{i}{\hbar}\hat N(t,t_0)}|0\rangle
	=\langle0|e^{-\frac{i}{\hbar}[\hat{N}(t,t_0),\bullet]}\hat\cO(t,\mathbf{x})|0\rangle\\
	&=\sum_{n=0}^\infty\frac{1}{n!}\Big(\frac{-i}{\hbar}\Big)^n\langle0|
		[\hat{N}(t,t_0)^{\odot n},\hat\cO(t,\mathbf{x})]	
	|0\rangle\,.
\end{aligned}
\end{align}
Our task is to evaluate the nested commutators in the above formula explicitly,
before projecting on the vacuum state $|0\rangle$.

Following \rcite{Alessio:2025flu}, we
rely on a coherent-state decomposition of the $\hat{N}$-operator:
\begin{align}\label{coherent}
	\hat{N}(t,t_0)&=
	\hat{N}_\varphi(t,t_0)+
	\int_\mathbf{k}\left(\hat{N}_\varphi(t,t_0;k)\hat{a}^\dagger(k)+
	\hat{N}_\varphi(t,t_0;-k)\hat{a}(k)\right)\\
	&\quad+\frac{1}{2}\int_{\mathbf{k}_1,\mathbf{k}_2}\Big(
		\hat{N}_\varphi(t,t_0;k_1,k_2)\hat{a}^\dagger(k_1)\hat{a}^\dagger(k_2)+
		\hat{N}_\varphi(t,t_0;-k_1,-k_2)\hat{a}(k_1)\hat{a}(k_2)\nn\\
		&\quad\qquad\qquad+
		\hat{N}_\varphi(t,t_0;k_1,-k_2)\hat{a}^\dagger(k_1)\hat{a}(k_2)
	\Big)+\cdots\,.\nn
\end{align}
This decomposition provides a natural splitting between the conservative and radiative parts of $\hat{N}$, the conservative part being captured by $\hat{N}_\varphi(t,t_0)$ where radiation is neither emitted nor absorbed.
In the presence of radiation reaction this definition does not coincide with the one traditionally adopted in the literature, where ``conservative'' refers to the time-symmetric component of the dynamics. The operators are defined as
\begin{align}
	\hat{N}_\varphi(t,t_0;k_1,\ldots,k_n):=
	{}_\varphi\langle k_1\cdots k_n|\hat{N}(t,t_0)|0\rangle_\varphi/\hbar^n\,.
\end{align}
which includes the vacuum element where $n=0$.
We also recall that contracting states on the right corresponds
merely to a sign flip $k_i^\mu\to-k_i^\mu$. 
By projecting only on the bulk states, we retain the operator dependence on
the worldline fluctuations. Thus the bulk-projected quantities
$\hat N_\varphi(t,t_0;k_1,\ldots,k_n)$ are still operators on the worldline
Hilbert space.

We now expand these operators in the time-independent basis
$\hat b'_{i,\alpha},\hat p'_{i,\alpha}$:
\begin{align}
\label{eq:worldline_zero_mode_expansion}
\hat N_\varphi(t,t_0;k_1,\ldots,k_n)
&=
N(t,t_0;k_1,\ldots,k_n) \\
&\qquad+
\hat b^{\prime\alpha}_i
N^{b_i}_{\alpha}(t,t_0;k_1,\ldots,k_n)
+
\hat p^{\prime\alpha}_i
N^{p_i}_{\alpha}(t,t_0;k_1,\ldots,k_n)
+\cdots\,,\nonumber 
\end{align}
where repeated worldline labels are summed over. The ellipsis denotes terms
quadratic and higher in the worldline fluctuation operators $\{\hat b'_{i,\alpha},\hat p'_{i,\alpha}\}$. The linear coefficients are extracted by taking the following commutators before projecting on the worldline reference state~\eqref{eq:vacuum_Hilbert}:
\begin{subequations}\label{eq:coefficients_WL}
\begin{align}
N^{b_i}_{\alpha}(t,t_0;k_1,\ldots,k_n)
&=
\frac{1}{i\hbar}
\left\langle 0 \,\Big|
\left[
\hat N_\varphi(t,t_0;k_1,\ldots,k_n),
(\hat p'_i)_{\alpha}
\right]
\Big|\, 0\right\rangle_{\rm wl},
\\
N^{p_i}_{\alpha}(t,t_0;k_1,\ldots,k_n)
&=
-\frac{1}{i\hbar}
\left\langle 0 \,\Big|
\left[
\hat N_\varphi(t,t_0;k_1,\ldots,k_n),
(\hat b'_i)_{\alpha}
\right]
\Big| \, 0\right\rangle_{\rm wl}.
\end{align}
\end{subequations}
Higher coefficients are obtained analogously from nested commutators.

In the PL expansion 
the coefficients $N_\alpha^{b_i}$ and $N_\alpha^{p_i}$~\eqref{eq:coefficients_WL} are not independent:
as discussed in \app{worldlinebrackets},
due to the worldline interaction Hamiltonian~\eqref{eq:HintA_def}
depending only on the combination $x_i^\mu=\bar{x}_i^\mu+z_i^\mu$ ---
referred to as the background split symmetry ---
these additional coefficients are fixed by derivatives of the
ordinary projected elements $N(k_1,\ldots,k_n)$ with respect to the
background fields.
In the SF expansion this is no longer true in general,
and so the additional matrix elements are needed. Yet at 1SF the light body is special: the vertex
that survives in the interaction Hamiltonian~\eqref{eq:HintB_1SF} is
\begin{align}
    e_L\varphi(\bar x_L)
    =
    \left.e_L\varphi(\bar x_L+z_L)\right|_{z_L=0}.
\end{align}
Therefore, when extracting the worldline coefficients,
one first lifts this vertex to its full-coordinate form and only then takes the commutator. Equivalently,
\begin{align}
\label{eq:SF_light_lift}
\left.
e_L \frac{\partial}{\partial z_L^\alpha}
\varphi(\bar x_L+z_L)
\right|_{z_L=0}
=
e_L \frac{\partial}{\partial \bar x_L^\alpha}
\varphi(\bar x_L).
\end{align}
Together with $z_L^\alpha(\tau_L)
=
b_L^{\prime\alpha}
+
(\tau_L/m_L) p_L^{\prime\alpha}$,
this fixes the light coefficients $N^{b_L}_{\alpha}$ and $N^{p_L}_{\alpha}$
from the ordinary reduced light-line matrix elements ---
in exactly the manner familiar from the PL expansion.
These coefficients are needed for light-worldline observables, but they do not require separate diagrams. The heavy line is different: the reduced 1SF Hamiltonian \eqref{eq:HintB_1SF} already contains explicit heavy-deflection vertices. Hence $N^{b_H}_{\alpha}$ and $N^{p_H}_{\alpha}$ are genuine additional entries in the SF basis: the relevant ones were computed explicitly in \sec{sec:Magnus_caseB}.

In order to define spatial projections and relative angles we fix a timelike unit vector $U_{t_0}^\mu$, constructed from the total momentum at a reference time $t_0$,
\begin{align}
U_{t_0}^\mu := \frac{\bar P^\mu(t_0)}{|\bar P(t_0)|},
\qquad
\Pi^{\mu \nu} := \eta^{\mu \nu} - U_{t_0}^\mu U^{\nu}_{t_0}.
\label{eq:comframe_t0}
\end{align}
Here $\bar P^\mu(t_0)=\bar p_1^\mu(t_0)+\bar p_2^\mu(t_0)$
is the total momentum evaluated on the background trajectories.
For scattering motion it is natural to choose $t_0\to -\infty$, so that $U_{t_0}^\mu$ coincides with the incoming center--of--mass direction. For bound motion, $t_0$ may be chosen as an arbitrary reference time (e.g.\ the first periastron), and the corresponding $U_{t_0}^\mu$ is then held fixed throughout the evolution.

\subsection{Worldline observables}

In the context of a worldline operator $\hat\cO_i$,
i.e.~$\hat{x}_i^\mu$ or $\hat{p}_i^\mu$,
eq.~\eqref{eq:observables} becomes
\begin{align}\label{wlObservable}
\begin{aligned}
    \langle\hat\cO_i(\tau_i)\rangle
	&=\langle0|\hat\cO_i(\tau_i)|0\rangle
	-\frac{i}{\hbar}\langle0|[\hat{N}(t,t_0),\hat\cO_i(\tau_i)]|0\rangle\\
	&\qquad-\frac1{2! \hbar^2}\langle0|[\hat{N}(t,t_0),[\hat{N}(t,t_0),\hat\cO_i(\tau_i)]]|0\rangle+\cdots\Big|_{t=\bar{x}_i^0(\tau_i)}\,,
\end{aligned}
\end{align}
where we identify the final time coordinate as $t=\bar{x}_i^0(\tau_i)$.
Into this formula, we insert the coherent-state decomposition~\eqref{coherent}:
\begin{subequations}
\begin{align}
	[\hat{N}(t,t_0),\hat\cO_i(\tau_i)]&=
	[\hat{N}_\varphi(t,t_0),\hat\cO_i(\tau_i)]+\cdots\,,\\
	[\hat{N}(t,t_0),[\hat{N}(t,t_0),\hat\cO_i(\tau_i)]]&=
	[\hat{N}_\varphi(t,t_0),[\hat{N}_\varphi(t,t_0),\hat\cO_i(\tau_i)]]\\
	&\quad+\hbar \int_\mathbf{k}\left(\hat{N}_\varphi(t,t_0;-k)[\hat{N}_\varphi(t,t_0;k),\hat\cO_i(\tau_i)]+\text{c.c.}\right)+\cdots\,.\nn
\end{align}
\end{subequations}
We keep only terms that will not annihilate the vacuum $|0\rangle$,
with terms carrying additional instances of the raising and lowering operators
$\hat{a}(p)$ and $\hat{a}^\dagger(p)$ hidden by the ellipsis $\cdots$.
All instances of $\hat{a}(p)$ and $\hat{a}^\dagger(p)$ have thus been absorbed
using $[\hat{a}(p),\hat{a}^\dagger(q)]=2\hbar|\mathbf{p}|\dd(\mathbf{p}-\mathbf{q})$. Plugging back into eq.~\eqref{wlObservable}, the remaining worldline dependence is captured by commutators with the worldline fluctuation operators. We therefore define
\begin{align}
\mathscr D_{\cO_i(\tau_i)}
N(t,t_0;k_1,\ldots,k_n)
:=
\frac{1}{i\hbar}
\left\langle0\left|
\left[
\hat N_\varphi(t,t_0;k_1,\ldots,k_n),
\hat\cO_i(\tau_i)
\right]\right|0
\right\rangle_{\rm wl}.
\label{eq:Dworldline}
\end{align}
In terms of the worldline expansion \eqref{eq:worldline_zero_mode_expansion}, the two elementary cases are described in \eqref{eq:coefficients_WL}. The expectation value of a worldline observable can then be written as
\begin{align}\label{wlFromN}
\begin{aligned}
\langle\hat\cO_i(\tau_i)\rangle
&=
\underbrace{
\bar\cO_i(\tau_i)
+
\sum_{n=1}^{\infty} \frac{1}{n!}
\mathscr D_{\cO_i(\tau_i)}
\bigl(N(t,t_0)^{\odot n}\bigr)
}_{\rm conservative}
\\
&\quad
\underbrace{
-\frac{i}{2!}\int_{\mathbf{k}}
\left(
N(t,t_0;-k)\,
\mathscr D_{\cO_i(\tau_i)}N(t,t_0;k)
-\text{c.c.}
\right)
+\cdots
}_{\rm radiative},
\end{aligned}
\end{align}
where again $t=\bar{x}_i^0(\tau_i)$ and we have introduced the shorthand notation
\begin{align}
\mathscr D_{\cO_i(\tau_i)}
\bigl(N(t,t_0)^{\odot n}\bigr)
:=
\frac{1}{(i\hbar)^n}
\left\langle0\left|
\left[
\hat N_\varphi(t,t_0),
\ldots
\left[
\hat N_\varphi(t,t_0),
\hat \cO_i(\tau_i)
\right]
\ldots
\right]\right|0
\right\rangle_{\rm wl}\,,
\end{align}
generalising \eqn{eq:Dworldline}.

In the PL expansion, the split symmetry discussed in \app{worldlinebrackets} implies that the operation $\mathscr D_{\cO_i}$ is equivalently represented by a Poisson bracket \eqref{poissonBG} acting on the background data,
\begin{align}
    \mathscr D_{\cO_i}N^{(\rm PL)}
    =
    \{N^{(\rm PL)},\bar\cO_i\}\,.
    \label{eq:DPLbracket}
\end{align}
As explained in \rcite{Haddad:2025cmw}, these brackets correct the causality prescription on worldline propagators, giving the all-retarded prescription typical of scattering observables. In the SF expansion, however, the replacement $\mathscr D_{\cO_i}N^{(\rm SF)}\to\{N^{(\rm SF)},\bar\cO_i\}$ is not automatic.
The operation $\mathscr D_{\cO_i}$ should then instead be evaluated
through the worldline coefficients in \eqn{eq:coefficients_WL}:
\begin{align}
	\mathscr D_{\cO_i(\tau_i)}N^{(\rm SF)}=\frac1{i\hbar}\left(
		N^{({\rm SF})b_i}_\alpha[\hat{b}^{\prime\alpha}_i,\hat{\mathcal{O}}_i(\tau_i)]+
		N^{({\rm SF})p_i}_\alpha[\hat{p}^{\prime\alpha}_i,\hat{\mathcal{O}}_i(\tau_i)]
	\right)\,.
\end{align}
The explicit commutator-projected matrix elements were computed in \sec{sec:Magnus_caseB}.
Yet at 1SF order, on the light line as demonstrated in \eqn{eq:SF_light_lift}, this is unnecessary ---
we may therefore use an analogous version of \eqn{eq:DPLbracket},
i.e.~$\mathscr D_{\cO_L}N^{(\rm SF)}=\{N^{(\rm SF)},\bar\cO_L\}$.

We can now show that our definition of the conservative dynamics, as defined by the coherent state expansion in \eqn{coherent}, implies zero losses of linear and angular momentum due to radiation involving the scalar field $\phi$. Restricting for simplicity to the PL expansion,\footnote{
The same separation holds in the SF expansion: the conservative sector is defined by retaining the bulk-vacuum projected part of $\hat{N}$ and dropping the coherent-state elements. However, for such expansion, conservation of the total particle momentum and angular momentum need not be manifest at each truncated SF order, since the heavy source recoil is higher order in the mass-ratio expansion. For this reason, in the
SF setup it is more transparent to compute the radiated momentum and angular
momentum in terms of the matrix elements with external scalar states.} this is true because
\begin{subequations}
\begin{align}
	\{N^{(\rm PL)},\bar{P}^\mu\}&=0\,, &
	\bar{P}^\mu&=\bar{p}_1^\mu+\bar{p}_2^\mu\,,\label{pBracket} \\
	\{N^{(\rm PL)},\bar{J}^\mu\}&=0\,, &
	\bar{J}^\mu&={\eps^\mu}_{\nu\rho\sigma}
	\big(\bar{x}_1^{\nu}\bar{p}_1^{\rho}+
	\bar{x}_2^{\nu}\bar{p}_2^{\rho}\big)
	\frac{\bar{P}^\sigma}{|\bar{P}|}\,,\label{jBracket}
\end{align}
\end{subequations}
where the angular momentum vector $\bar{J}^\mu$ is taken in the frame
defined by the (unit-normalised) total momentum vector $\bar{P}^\mu/|\bar{P}|$.
The bracket with $\bar{P}^\mu$~\eqref{pBracket} vanishes due to translation invariance:
$N^{(\rm PL)}$ depends on the background trajectories
$\bar{x}_i^\mu$ only via $\bar{x}_1-\bar{x}_2$,
so $\{N^{(\rm PL)},\bar{p}_1^\mu\}=-\{N^{(\rm PL)},\bar{p}_2^\mu\}$.
Similarly, the bracket with $\bar{J}^\mu$ vanishes by rotational invariance,
since $N^{(\rm PL)}$ depends on the background data only through
scalars such as $(\bar{x}_1-\bar{x}_2)\cdot\bar{p}_i$,
$(\bar{x}_1-\bar{x}_2)^2$ and $\bar{p}_1\cdot\bar{p}_2$.
Thus the conservative evolution generated by $N^{(\rm PL)}$
gives no net change in the total linear or angular momentum of the particles. Any non-zero
asymptotic change of these charges is therefore controlled by the radiative
terms involving the coherent-state elements $N^{(\rm PL)}(k_1,\dots,k_n)$.

\paragraph{Scattering scenario}

The main observable of interest is the momentum impulse: 
\begin{align}
\begin{aligned}
	\Delta p_i^\mu:=&\,\langle\hat{p}_i^\mu(+\infty)\rangle-\langle\hat{p}_i^\mu(-\infty)\rangle\\
=&\, \underbrace{
\sum_{n=1}^{\infty} \frac{1}{n!}
\mathscr D_{p_i^\mu}
\bigl(N^{\odot n}\bigr)
}_{\Delta p_{i,\rm cons}^\mu}
\underbrace{
-\frac{i}{2!}\int_{\mathbf{k}}
\left(
N(-k)\,\mathscr D_{p_i^\mu}N(k)-\text{c.c.}
\right)
+\cdots
}_{\Delta p_{i,\rm rad}^\mu}\,,
\end{aligned}
\end{align}
where $\mathscr D_{p_i^\mu}$ denotes the operator defined in
\eqref{eq:Dworldline}. In the PL expansion it reduces to the background
bracket $\mathscr D_{p_i^\mu}N^{(\rm PL)} =\{N^{(\rm PL)},\bar p_i^\mu\}$, recovering eq.(26) of \rcite{Alessio:2025flu}
\begin{align}
	\Delta p_i^\mu=&\,(e^{\{N^{\rm (PL)},\bullet\}}-1)\bar{p}_i^\mu -\frac{i}{2!}\int_{\mathbf{k}}(N^{\rm (PL)}(-k)\{N^{\rm (PL)}(k),\bar{p}_i^\mu\}-\text{c.c.})+\cdots\,.
\end{align}
For the conservative part in the PL expansion,
the fact that $p_i^2=(p_i+\Delta p_{i,\rm cons})^2$ (on-shell condition)
together with $\Delta p_{1,\rm cons}^\mu=-\Delta p_{2,\rm cons}^\mu$
(momentum conservation)
implies that the impulse has a universal form:
\begin{align}\label{angleForm}
	\Delta p_{1,\rm cons}^\mu=
	|p_\infty|\sin\theta\frac{b^\mu}{|b|}+(\cos\theta-1)p_{\infty}^\mu\,,
\end{align}
where $p_\infty^\mu:=p_{\rm rel}^\mu(-\infty)$,
the relative momentum vector $p_{\rm rel}^\mu(t)$ being defined
in the incoming center-of-mass frame \eqref{eq:comframe_t0} with $U^\mu := U_{t_0=-\infty}^\mu$:
\begin{align}
p_{\rm rel}^\mu(t)
:=
\frac{E_2(t)\,p_1^\mu(t)-E_1(t)\,p_2^\mu(t)}{|P(t)|},
\qquad
E_i(t):=U_\nu p_i^\nu(t)\,.
\end{align}
By construction $U_\mu p_{\rm rel}^\mu=0$,
so $p_{\rm rel}^\mu$ lies in the $U$--orthogonal hyperplane.
The formula~\eqref{angleForm} represents a simple rotation
by angle $\theta$ in the plane defined by $b^\mu$ and $p_\infty^\mu$.
The scattering angle is given by~\cite{Haddad:2025cmw}
\begin{align}
	\theta=-\frac{\partial N^{(\rm PL)}}{\partial L}\,,
    \label{eq:cons}
\end{align}
where $L=|p_\infty||b|$. This implies that $N^{(\rm PL)}$ may be identified with the radial action.

In the presence of radiation the total momentum of the system need not be conserved between asymptotic times,
$P^\mu(+\infty) \neq P^\mu(-\infty)$, so that the outgoing state generically experiences recoil.
For this reason it is convenient to define a scattering angle that is
symmetric in the two bodies and remains meaningful when $\Delta P^\mu \neq 0$.
We use the relative impulse:
\begin{align}
\Delta p_{\rm rel}^\mu
:=
p_{\rm rel}^\mu(+\infty)
-
p_{\rm rel}^\mu(-\infty),
\end{align}
where the variation includes the implicit changes in the energies $E_i$. The relative scattering angle is then defined by \cite{Dlapa:2022lmu,Buonanno:2024vkx}
\begin{align}
\cos\theta_{\rm rel}
=
\frac{
\big(p_{\infty}+\Delta p_{\rm rel}\big)
\cdot
p_\infty
}{
\big|p_\infty+\Delta p_{\rm rel}\big|
\;
\big|p_\infty\big|
}\,.
\end{align}
In the conservative case $\Delta P^\mu=0$, so that $p_{\rm rel}^\mu$ reduces to the usual center--of--mass spatial momentum and $\theta_{\rm rel}$ coincides with the scattering angle $\theta$~\eqref{eq:cons}.

In scattering, it is natural to characterise dissipation by the net change of
the total momentum and total angular momentum between asymptotic times,
\begin{align}
\Delta P^\mu:=\langle \hat P^\mu(+\infty)\rangle-\langle \hat P^\mu(-\infty)\rangle,
\qquad
\Delta J^\mu:=\langle \hat J^\mu(+\infty)\rangle-\langle \hat J^\mu(-\infty)\rangle.
\end{align}
In the PL background-bracket representation, using eqs.~\eqref{pBracket} and \eqref{jBracket}, their conservative pieces vanish identically, so that the leading contributions are radiative. At the first non-vanishing order in the coherent-state expansion one finds
\begin{align}
\Delta P^\mu
=
-\frac{i}{2}\int_{\mathbf{k}}
\Big(
N^{\rm (PL)}(-k)\,\{N^{\rm (PL)}(k),\bar P^\mu\}
-\text{c.c.}
\Big)
+\cdots\,,
\label{eq:DeltaP_rad}
\\
\Delta J^\mu
=
-\frac{i}{2}\int_{\mathbf{k}}
\Big(
N^{\rm (PL)}(-k)\,\{N^{\rm (PL)}(k),\bar J^\mu\}
-\text{c.c.}
\Big)
+\cdots\,,
\label{eq:DeltaJ_rad}
\end{align}
where $\bar P^\mu=\bar p_1^\mu+\bar p_2^\mu$ and $\bar J^\mu$ is given in \eqn{jBracket}.
These expressions make manifest that all net losses are controlled by the radiative Magnus elements $N^{\rm (PL)}(k_1,\dots,k_n)$ and their Poisson brackets with the corresponding background charges.

\paragraph{Bound scenario}

For bound motion we again work in the initial center--of--mass frame \eqref{eq:comframe_t0} defined by $U_{t_0}^\mu$, chosen at a convenient initial reference time $t_0$ of the first periastron passage.
Periastron times $\{t_n\}$ are defined as successive local minima of the projected radial separation
\begin{align}
R^\mu(t)
:=
(\Pi_{t_0})^\mu{}_\nu
\big(
\langle \hat x_1^\nu(t)\rangle
-
\langle \hat x_2^\nu(t)\rangle
\big),
\qquad
r(t) = \sqrt{-R^\mu R_\mu},
\end{align}
through the turning-point conditions
\begin{align}
\dot r(t_n)=0,
\qquad
\ddot r(t_n)>0 .
\end{align}
This definition does not assume integrability or strict periodicity, only the existence of radial turning points.
We then introduce at each periastron the relative momentum
\begin{align}
p_{\rm rel}^\mu(t_n)
:=
\frac{E_2(t_n)\,p_1^\mu(t_n)-E_1(t_n)\,p_2^\mu(t_n)}{|P(t_n)|}\,,
\qquad
E_i(t_n):=(U_{t_0})_\nu p_i^\nu(t_n)\,.
\end{align}
Defining the per-cycle relative impulse
\begin{align}
\Delta p_{{\rm rel},n}^\mu
:=
p_{\rm rel}^\mu(t_{n+1})
-
p_{\rm rel}^\mu(t_n),
\end{align}
the relative periastron advance per radial cycle is given by
\begin{align}
\cos\Delta\Phi_n^{\rm rel}
=
\frac{
\big(p_{\rm rel}(t_n)+\Delta p_{{\rm rel},n}\big)
\cdot
p_{\rm rel}(t_n)
}{
\big|p_{\rm rel}(t_n)+\Delta p_{{\rm rel},n}\big|
\;
\big|p_{\rm rel}(t_n)\big|
}.
\end{align}
In the strictly conservative limit the turning points repeat periodically, so that $\Delta\Phi_n=\Delta\Phi$ is independent of $n$.  In that case one recovers the Hamilton--Jacobi identity
\begin{align}
\Delta\Phi
=
-\frac{\partial N^{<}}{\partial L},
\end{align}
where $N^{<}=I_r^{<}(E,L)$ is the conservative bound radial action, with
$E=E(t_0)$ and $L=L(t_0)$ the invariants labelling the reference bound orbit.

For bound motion it is convenient to work with scalar energy and angular momentum defined in a fixed reference frame. We define
\begin{align}
&E(t) := U_{t_0\,\mu} P^\mu(t),\quad
J^\mu(t) := \frac12\,\varepsilon^{\mu}{}_{\nu\rho\sigma}\,U_{t_0}^\nu\,J^{\rho\sigma}(t),
\quad
L(t) := \sqrt{-J^\mu(t)J_\mu(t)} , 
\end{align}
and the per-cycle losses at successive periastra as
\begin{align}
\Delta E_n := E(t_{n+1})-E(t_n),\qquad
\Delta L_n := L(t_{n+1})-L(t_n).
\end{align}
At the reference initial time $t=t_0$, one has
\begin{align}
E(t_0)=\bar E + m_H,
\qquad
L(t_0)=\bar L,
\end{align}
where $\bar E$ and $\bar L$ denote the corresponding background quantities. Since $m_H$ is constant, the operation $\mathscr D_E$ reduces to $\mathscr D_{\bar E}$ at this order.
At leading non-vanishing order in the coherent-state expansion these losses are controlled by the radiative Magnus element $N_n(k):= N(k;t_{n+1},t_n)$, yielding
\begin{align}
\Delta E_n
&=
-\frac{i}{2}\int_{\mathbf{k}}
\Big[
N_n(-k)\,\mathscr D_{\bar E} N_n(k)
-\text{c.c.}
\Big] 
+\cdots,
\\
\Delta L_n
&=
-\frac{i}{2}\int_{\mathbf{k}}
\Big[
N_n(-k)\,\mathscr D_{\bar L} N_n(k) 
-\text{c.c.}
\Big] 
+\cdots.
\end{align}

At $0$SF the motion is strictly periodic with radial period $\bar T_r(E,L)$. When radiative effects are included, the turning points $t_n$ remain well defined by $\dot r(t_n)=0$, but the cycle duration and orbital parameters drift,
\begin{align}
T_n:=t_{n+1}-t_n=\bar T_r(E_n,L_n)+\mathcal O(\lambda^2),
\end{align}
with $(E_n,L_n)$ slowly varying from cycle to cycle. In an adiabatic regime, where the per-cycle variations are $\mathcal O(\lambda^2)$, the evolution may be regarded as a sequence of conservative bound orbits with instantaneous parameters $(E_n,L_n)$, provided that we are away from resonances and critical orbits. Accordingly,
\begin{align}
\Delta E_n=\Delta E_{\rm orb}(E_n,L_n)&+\mathcal O(\lambda^3),\qquad
\Delta L_n=\Delta L_{\rm orb}(E_n,L_n)+\mathcal O(\lambda^3),\\
&\Delta\Phi_n=\Delta\Phi_{\rm orb}(E_n,L_n)+\mathcal O(\lambda^2).
\end{align}
Beyond the adiabatic regime the exact per-cycle definitions above remain valid without assuming periodicity.

\subsection{Bulk observables}

The expectation of the on-shell scalar field $\hat{\phi}(x)$ at time $t=x^0$ is given by
\begin{align}\label{bulkObservable}
    \langle\hat \phi(x)\rangle
	=\sum_{n=0}^\infty\frac{(-i)^n}{n! \hbar^n}\langle0|
		[\hat{N}(t,t_0)^{\odot n},\hat\phi(x)]	
	|0\rangle\,,
\end{align}
which we now evaluate by decomposing $\hat{\phi}(x)$ into raising and lowering operators.
Focusing on the first relevant contribution, we have
\begin{align}
	[\hat{N}(t,t_0),\hat{\phi}(x)]&=
	-\hbar \int_\mathbf{k}\Big(e^{-ik\cdot x}\hat{N}_\varphi(t,t_0;k)-
	e^{ik\cdot x}\hat{N}_\varphi(t,t_0;-k)\Big)+\cdots\,,
\end{align}
where again the ellipsis $\cdots$ denotes additional terms carrying
$\hat{a}$ and $\hat{a}^\dagger$ that we will not need as they annihilate the vacuum.
Substituting this into \eqref{bulkObservable}, the remaining contractions are
commutators between bulk-projected Magnus elements:
\begin{align}\label{expScalar}
    \langle\hat \phi(x)\rangle
	&=\bar\phi(x)
	\\
    &+i\sum_{n=1}^\infty\frac{1}{n!}\int_\mathbf{k}
    \Bigg[
        \frac{e^{-ik\cdot x}}{(i\hbar)^{n-1}}
        \left\langle 0 \, \Big|
        \left[
            \hat N_\varphi(t,t_0)^{\odot(n-1)},
            \hat N_\varphi(t,t_0;k)
        \right]
        \Big| \,0 \, \right\rangle_{\rm wl}
        -\text{c.c.}
	\Bigg]
    +\cdots . \nonumber 
\end{align}
To extract the radiative field at large distance we evaluate the remaining
on-shell integral with $k^\mu=\omega(1,\hat{\mathbf n})$ by stationary phase.
For $x^\mu=(t,\mathbf x)$ and $r=|\mathbf x|$, the angular integral is
dominated by the saddle $\hat{\mathbf n}=\mathbf x/r$. Thus
\begin{align}
	\int_\mathbf{k}e^{ik\cdot x}
	=
	\frac{i}{4\pi r}\int_\omega e^{i\omega u}
	+\cO\left(\frac1{r^2}\right),
\end{align}
where $u=t-r$ is retarded time. The dependence of the Magnus
elements on the direction of $\mathbf k$ is therefore evaluated at
$\hat{\mathbf k}=\mathbf x/r$, while the frequency dependence is kept explicit.
Thus \eqref{expScalar} becomes, at leading order in the $1/r$ expansion,
\begin{align}\label{bulkFromN}
	\langle\hat \phi(u,\mathbf{x})\rangle
	&=\bar\phi(x)
    \\
    &
    -\frac{1}{4\pi r}
    \sum_{n=1}^\infty\frac1{n!}\int_{\omega}
	\Bigg(
        \frac{e^{-i\omega u}}{(i\hbar)^{n-1}}
        \left\langle 0 \, \Big|
        \left[
            \hat N_\varphi(t,t_0)^{\odot(n-1)},
            \hat N_\varphi(t,t_0;k)
        \right]
        \Big| \,0 \, \right\rangle_{\rm wl}
        +\text{c.c.}
    \Bigg) \nonumber
\end{align}
where the spatial direction of $k^\mu$ is fixed by the observation direction,
$\hat{\mathbf k}=\mathbf x/r$. Higher coherent-state terms are discussed in \rcite{Alessio:2025flu}. Restricting to the strictly conservative scenario, in which $N(t,t_0;k_1,\ldots,k_n)=0$ for $n\geq1$, it is clear that $\langle\hat\phi(x)\rangle=\bar\phi(x)$, i.e.~the background field remains unchanged.

In the PL case, the split symmetry allows these worldline-projected commutators to be evaluated using background Poisson brackets (see \app{worldlinebrackets}). For example,
\begin{align}
&\frac{1}{(i\hbar)^{n-1}}
\left\langle 0 \, \Big|
\left[
    (\hat N_\varphi^{(\rm PL)})^{\odot(n-1)},
    \hat N_\varphi^{(\rm PL)}(k)
\right]
\Big| \,0 \, \right\rangle_{\rm wl} =
\{(N^{(\rm PL)})^{\odot(n-1)},N^{(\rm PL)}(k)\}.
\end{align}
In the SF expansion, instead, the commutators in \eqref{expScalar} and \eqref{bulkFromN} should be evaluated directly using the full worldline operator expansion in \eqref{eq:worldline_zero_mode_expansion} and \eqref{eq:coefficients_WL}. 

\subsection{Comparing $\hat{N}$ in the PL and SF expansions}
\label{sec:PLSFcomparison}

We now derive the relation between the PL Magnus operator
$\hat N^{(\rm PL)}$ and the scattering SF Magnus operator
$\hat N^{(\rm SF)>}$, specialising to scattering over an infinite time
interval where both descriptions are applicable. The two expansions describe
the same scattering map, but choose different reference trajectories: the PL
expansion is organised around straight worldlines, while the SF expansion is
organised around the probe trajectory of the light body in the field of the
heavy source.

To compare the two descriptions, it is useful to isolate from the PL Magnus operator the conservative probe contribution $\hat N^{(\rm geo)}$, defined in \sec{sec:Magnus_caseA} and related to the geodesic radial action~\eqref{eq:Ir_from_N_probe}. In the following, however, $\hat N^{(\rm geo)}$ should be regarded as an operator before projection onto worldline states, since its commutators generate the change of variables from straight-line data to probe-background data.
We now specialise to the light-particle impulse, identifying the light body
with particle $1$. Universality of the scattering map implies that the
asymptotic light momentum computed in the PL and SF descriptions must agree:
\begin{align}\label{eq:bothExpansions}
\begin{aligned}
	\langle \hat{p}_1^{\mu}(+\infty)\rangle
	&=
	\left\langle0\left|\,
	\exp\left(-\frac{i}{\hbar}\hat N^{(\rm PL)}\right)
	\hat{p}_1^{\mu(\rm PL)}(+\infty)
	\exp\left(\frac{i}{\hbar}\hat N^{(\rm PL)}\right)
	\right|0\right\rangle
	\\
	&=
	\left\langle0\left|\,
	\exp\left(-\frac{i}{\hbar}\hat N^{(\rm SF)>}\right)
	\hat{p}_1^{\mu(\rm SF)}(+\infty)
	\exp\left(\frac{i}{\hbar}\hat N^{(\rm SF)>}\right)
	\right|0\right\rangle .
\end{aligned}
\end{align}
Here $\hat{p}_1^{(\rm PL/SF)}$ is the momentum operator,
which differs between the two cases due to the use of different interaction pictures.
At future infinity the operators are related by the geodesic map:
\begin{align}\label{eq:geo_map}
	\hat{p}_1^{\mu\,(\rm SF)}(+\infty)
	=
	\exp\left(-\frac{i}{\hbar}\hat N^{(\rm geo)}\right)
	\hat{p}_1^{\mu\,(\rm PL)}(+\infty)
	\exp\left(\frac{i}{\hbar}\hat N^{(\rm geo)}\right)\,.
\end{align}
Thus, comparing \eqref{eq:geo_map} with \eqref{eq:bothExpansions}, the two Magnus
operators are related by composition of scattering maps,
\begin{align}
	\exp\left(\frac{i}{\hbar}\hat N^{(\rm PL)}\right)
	=
	\exp\left(\frac{i}{\hbar}\hat N^{(\rm geo)}\right)
	\exp\left(\frac{i}{\hbar}\hat N^{(\rm SF)>}\right).
	\label{eq:PLSFoperatorRelation}
\end{align}
Equivalently, the Baker--Campbell--Hausdorff expansion gives
\begin{align} \label{eq:PLSF_BCH}
&\frac{i}{\hbar}\hat N^{(\rm PL)}
=
\frac{i}{\hbar}\hat N^{(\rm geo)}
+
\frac{i}{\hbar}\hat N^{(\rm SF)>}
+
\frac12
\left[
	\frac{i}{\hbar}\hat N^{(\rm geo)},
	\frac{i}{\hbar}\hat N^{(\rm SF)>}
\right]
\\
&
+\frac1{12}
\left[
\frac{i}{\hbar}\hat N^{(\rm geo)},
\left[
\frac{i}{\hbar}\hat N^{(\rm geo)},
\frac{i}{\hbar}\hat N^{(\rm SF)>}
\right]
\right]
+\frac1{12}
\left[
\frac{i}{\hbar}\hat N^{(\rm SF)>},
\left[
\frac{i}{\hbar}\hat N^{(\rm SF)>},
\frac{i}{\hbar}\hat N^{(\rm geo)}
\right]
\right]
+\cdots . \nonumber
\end{align}
Equation~\eqref{eq:PLSF_BCH} should be read before projecting on the
worldline state. As a simple example, the PL diagrammatic expansion of $\langle0|\hat N^{(\rm geo)}|0\rangle$ at 0SF order is in \eqref{eq:0SFradialaction}. 

At 1SF order the matching is more subtle. In particular, the SF Magnus operator must be expanded both in bulk coherent-state components, as in \eqref{coherent}, and in worldline coefficients, as in \eqref{eq:worldline_zero_mode_expansion}. For the SF light-line contribution, the vertex $e_L\varphi(\bar x_L)$ should be viewed as the $z_L=0$ projection of the corresponding insertion  $e_L\varphi(\bar x_L+z_L)$ as in \eqref{eq:SF_light_lift}. In the matching one therefore keeps the $\hat z_L$ dependence, expands the light trajectory $\bar x_L$ in the PL expansion, and evaluates the BCH commutators using the PL worldline algebra in ~\eqref{zDecomp} before taking the worldline projection. The corresponding retarded PL expansion of the light-line vertex is illustrated in \eqref{eq:Lsubleading}, while the worldline commutators entering the matching are represented diagrammatically by cut worldline insertions, as in Fig.~\ref{diagBracket}. The heavy-line worldline coefficients are instead obtained from the explicit commutator-projected matrix elements computed in \sec{sec:Magnus_caseB}. Together, these ingredients provide the complete matching at 1SF order, which we have verified explicitly up to 3PL order.\footnote{We thank Dyako Arif for assisting us with these checks.}

\subsection{Classical factorisation and squeezing effects}

The coherent-state expansion \eqref{coherent} implies that, once non-linear interactions are included, the radiation sector of the outgoing state need not be purely coherent. The two-emission sector $N(t,t_0;k_1,k_2)$ provides the first place where departures from Poissonian statistics may enter,\footnote{Importantly, $N(t,t_0;k_1,k_2)\neq0$ is not, by itself, an invariant diagnostic of non-coherence. The invariant content is instead carried by moments of the radiation distribution. In particular, in classical electromagnetism the radiation state is exactly coherent: multi-emission sectors $N(t,t_0;k_1,\dots,k_n)$ can be non-vanishing but have a fully factorised structure determined by the single-emission amplitude, and all Poisson-subtracted invariants such as $W$ defined below vanish.} corresponding to a two-mode squeezed state.

To quantify these effects invariantly, we consider the number operator for quanta with energies above an infrared cutoff $\Lambda_{\rm IR}$,
\begin{align}
\hat N_{\Lambda_{\rm IR}}:= \frac{1}{\hbar}\int_{\mathbf{k},E>\Lambda_{\rm IR}} \hat{a}^\dagger(k)\,\hat{a}(k),
\end{align}
and define its mean and variance using \eqn{eq:observables}:
\begin{align}
\mu_{\Lambda_{\rm IR}}(t)&:=\langle \hat N_{\Lambda_{\rm IR}}(t)\rangle,
\\
\Sigma_{\Lambda_{\rm IR}}(t)&:=\langle \hat N_{\Lambda_{\rm IR}}(t)^2\rangle
-\langle \hat N_{\Lambda_{\rm IR}}(t)\rangle^2.
\end{align}
A convenient infrared-safe quantity is the variance minus the mean
\cite{Britto:2021pud}
\begin{align}
W(t):=\Sigma_{\Lambda_{\rm IR}}(t)-\mu_{\Lambda_{\rm IR}}(t),
\label{eq:W_def}
\end{align}
which vanishes identically for a strictly coherent state (Poisson statistics), but is sensitive to non-linear deviations. Although $\mu_{\Lambda_{\rm IR}}$ and
$\Sigma_{\Lambda_{\rm IR}}$ are individually IR-sensitive, the combination $W$ is IR-safe. The variation of $W$ between $t_0$ and $t$ is controlled by the same Dirac-bracket expansion as any other observable in \eqref{eq:observables},
\begin{align}
\Delta W(t,t_0)
:=\langle \hat W(t)\rangle-\langle \hat W(t_0)\rangle
=\sum_{n=1}^\infty\frac{(-i)^n}{n! \hbar^n} 
\Big\langle\big[\hat N(t,t_0)^{\odot n},\,\hat W\big]\Big\rangle_{t_0},
\label{eq:W_evolution_Magnus}
\end{align}
where $\langle\cdots\rangle_{t_0}$ denotes the reference expectation value $\langle0|\cdots|0\rangle$ at $t_0$. The leading contribution to $\Delta W$ takes the form
\begin{align}
\Delta W(t,t_0)
= \frac{i}{\hbar}\int_{k_1,k_2}
\Big[
N(t,t_0;k_1)\,N(t,t_0;k_2)\,N^*(t,t_0;k_1,k_2)
-\text{c.c.}
\Big] + \dots \,.
\label{eq:DeltaW_schematic}
\end{align}
It is worth emphasising that the presence of a connected two-emission contribution -- one not reducible to products of single-emission terms -- leads to $\Delta W(t,t_0)\neq 0$, signalling squeezing (or more generally departures from Poissonian statistics).

It is useful to contrast \eqref{eq:DeltaW_schematic} with the classical factorisation properties of physical radiative observables, such as the
waveform or energy flux. For two radiative observables $\mathcal{O}_1$ and $\mathcal{O}_2$ we define the connected correlator by
\begin{align}
\langle \mathcal{O}_1 \, \mathcal{O}_2\rangle_{\rm conn}
:=\langle \mathcal{O}_1\,\mathcal{O}_2\rangle-\langle \mathcal{O}_1\rangle\,\langle \mathcal{O}_2\rangle.
\label{eq:conn_def}
\end{align}
The uncertainty-principle argument of \rcite{Cristofoli:2021jas} implies that connected correlators of physical radiation observables are suppressed in the classical limit, which can also be verified with a simple $\hbar$ power counting in WQFT \cite{Jakobsen:2023oow}. We can now confirm that this happens for (on-shell) scattering and bound observables, and for instance both the waveform and the corresponding fluxes classically factorise 
\begin{align}
\langle \hat{\phi}\,\hat{\phi}\rangle_{\rm conn}=\mathcal O(\hbar),
\qquad
\langle \Delta E \Delta E\rangle_{\rm conn}=\mathcal O(\hbar)\,,
\label{eq:classical_factorisation}
\end{align}
reflecting the fact that quantum effects (such as graviton loops) are parametrically suppressed in powers of $\hbar$. There is no contradiction with \eqn{eq:DeltaW_schematic}: $W$ probes number fluctuations, whereas physical observables such as $\Delta E$ carry explicit energy weightings (and corresponding $\hbar$ factors) that ensure the  suppression of connected correlators in \eqn{eq:classical_factorisation}. Thus, non-Poissonian number statistics (including ``squeezing'') can coexist with factorisation for connected correlators of classical observables.

In the scalar model considered here, we expect non-factorisable multi-emission contributions to arise only when the coupling $g$ is non-vanishing.  For $g=0$ the theory is linear and the radiation state remains exactly coherent at the classical level: non-linearities responsible for squeezing effects therefore enter parametrically through powers of $g$. 
It would be interesting to study this in the present framework, where we expect classical bulk non-linearities 
to produce super-Poissonian deviations at higher orders, $\Delta W>0$, once non-factorisable multi-emission sectors appear. By contrast, genuinely quantum squeezing would require sub-Poissonian statistics ($\Delta W<0$), which cannot be generated by classical solutions of the field equations.

\section{Conclusions}\label{conclusions}

In this paper we have demonstrated how the WQFT formalism can be used to
describe classical two-body scattering events and bound orbits,
in both perturbative coupling and self-force (SF) expansions.
To do so we rebuilt WQFT from scratch,
avoiding path integrals in favour of canonical quantisation
and providing direct access to the time-evolution operator $\hat{U}(t,t_0)$.
Matrix elements of $\hat{N}(t,t_0)=-i \hbar\log\hat{U}(t,t_0)$
with external plane wave states $|k_1\cdots k_n\rangle$ ---
including the ones relevant for bound systems ---
are the WQFT amplitudes
$N(k_1,\ldots,k_n):=\langle k_1\cdots k_n|\hat{N}|0\rangle$,
encoding all the physical data needed to describe classical observables.
When expanded in the SF series,
which admits a non-zero background configuration in the bulk,
the amplitudes showcase an integrand-level link between scattering and bound orbits.
Besides the non-zero background,
the main subtlety for describing bound orbits is
that matrix elements of $\hat{N}(t,t_0)$ are computed over a finite time interval.
The analysis of this paper focused on a simple toy model involving a pair of
charged, massive particles interacting via a scalar field $\phi(x)$,
yet our approach generalises naturally to gravity and electromagnetism.

A number of innovations were crucial to our work.
Firstly, it is now clear that the $\hat{N}$ operator plays a vital role
in the pursuit of classical physics from QFT methods
\cite{Kim:2024svw,Kim:2025ebl,Kim:2025sey,Kim:2025olv,Kim:2025hwi,Kim:2025hpn,
Brandhuber:2025igz,Damgaard:2021ipf,Damgaard:2023ttc,Alessio:2025flu}.
The main advantage of $\hat{N}$ over $\hat{S}$
is that it is perturbatively given by the Magnus series~\eqref{eq:MagnusTerms},
rather than Dyson.
This means that it involves causal retarded and advanced propagators,
which are natural in a classical setting where
boundary conditions are fixed in the past or in the future.
When applying Feynman rules at tree level,
which in WQFT is the situation relevant for classical physics,
the Magnus series is straightforwardly applied by weighting causality
flows with Murua coefficients~\eqref{murua}.
This crucial insight is lost when adopting the path integral,
which implies a time ordering of fields:
\begin{align}
	\int\!{\cal D}\phi\,e^{iS[\phi]}\phi(x_1)\cdots\phi(x_n)=
	\langle0|T\{\hat\phi(x_1)\cdots\hat\phi(x_n)\}|0\rangle\,.
\end{align}
While the Schwinger-Keldysh formalism
\cite{Schwinger:1960qe,Keldysh:1964ud,Galley:2009px,Jakobsen:2022psy}
alleviates this problem in the context of one-point functions,
where the presence of two time-evolution operators
$\langle\cO(t,\mathbf{x})\rangle_{\rm in-in}:=
\langle0|\hat{U}(-\infty,t)\hat\cO(t,\mathbf{x})\hat{U}(t,-\infty)|0\rangle$
implies that the path integral is doubled,
this approach is not flexible enough to 
compute matrix elements of $\hat{N}$.

By adopting canonical quantisation,
we have now gained control over causality flows and can compute $\hat{N}$-matrix elements
--- WQFT amplitudes --- directly.
Yet in order to do so, we needed to overcome an obstacle:
that each worldline prefers its own proper time coordinate $\tau_i$
over the global time coordinate $t$ present in the bulk scalar field $\phi(t,\mathbf{x})$.
Adopting a common global time coordinate is crucial for defining the time-evolution operator $\hat U(t,t_0)$
via a suitable interaction picture.
Our solution was to identify $\tau_i$ as the proper times associated to the background
trajectories $\bar{x}_i^\mu(\tau_i)$, wherein $\dot{\bar{x}}_i^2=1$,
implying a simple nonperturbative relationship $t=\bar{x}_i^0(\tau_i)$ between time coordinates.
This constraint is maintained by writing the worldline Hamiltonian density $\cH_i(x)$
in terms of the usual worldline Hamiltonian $H_i(\tau_i)$ as
\begin{align}
	\cH_i(x)=\int\!\d\tau_i\,\delta^D(x-\bar{x}_i(\tau_i))H_i(\tau_i)\,,
\end{align}
which maintains Lorentz invariance.
In the context of bound orbits,
aligning our time coordinates with the background motion makes it
straightforward to define a common time interval
that can be applied both in the bulk and on the two worldlines.

Our quantisation procedure is designed to admit both the coupling (PL) and SF expansions,
the latter being necessary to describe bound orbits.
This necessitates the use of separate interaction pictures in each case.
In the coupling expansion, the background consists only of undeflected straight lines
$\bar{x}_i^\mu(\tau_i)=b_i^\mu+\tau_iv_i^\mu$ with a vanishing scalar field $\bar\phi(x)=0$;
in the SF expansion we expand around the integrable 0SF motion.
The result is two separate interaction Hamiltonians,
with corresponding Feynman rules used to build up different $\hat{N}$-matrix elements.
In the case of scattering, wherein both PL and SF expansions are applicable,
a correspondence between $\hat{N}$-matrices in each case is provided by considering
how each encodes the universal scattering observables.
This illustrates how the 0SF motion can be ``integrated out''
in a manner that resembles the path integral while preserving causality flows.
For bound orbits, we compute matrix elements of $\hat{N}(t,t_0)$ over a restricted time interval ---
yet each element has a precise analog for the case of scattering,
making it straightforward to see unbound-to-bound mappings
\cite{Kalin:2019rwq,Kalin:2019inp,Cho:2021arx,Saketh:2021sri,Gonzo:2023goe,Adamo:2024oxy,Khalaf:2025jpt}.
We have provided all matrix elements needed to describe the full 1SF motion
up to 3PL order for scattering, and corresponding matrix elements for bound orbits.

Finally, we have explained in detail how matrix elements of 
$\hat{N}$ encode the underlying physics,
and how observable quantities --- scattering and bound --- are extracted.
While $\hat{N}$'s dependence on bulk fields is captured by
the coherent state expansion on a plane-wave basis~\eqref{coherent},
the absence of such states on the worldline necessitates a different approach.
In the PL expansion, the background split symmetry allows commutators with
worldline fluctuation operators to be traded for derivatives with respect to the
background data. In the SF expansion this is no longer automatic, and new commutator matrix elements must be included as independent data.
The vacuum element $\langle0|\hat{N}|0\rangle$~\cite{Kim:2024svw,Kim:2025olv,Kim:2025gis}
also provides a useful working definition of conservative dynamics
that does not rely on a time-ordering of fields (i.e.~Feynman propagators).
This prescription differs from existing prescriptions already at 3PM order,
which instead relies on a separation via method of regions on loop integrals~\cite{Jakobsen:2022psy}.

The methodology developed in this paper opens up
several new avenues for follow-up work.
In the context of gravitational two-body scattering,
we now have an entire plethora of $\hat{N}$-matrix elements
that can be calculated at higher-PM orders.
In order to make use of the separation between conservative and radiative dynamics
provided by taking the vacuum element $\langle0|\hat{N}|0\rangle$
we will need to understand the infrared properties in more detail,
and see whether $\hat{N}$ admits factorisation properties analogous to the $\hat{S}$-matrix in QCD.
The $\hat{N}$-matrix will also be highly useful when considering spin,
as it captures information contained by both the momentum impulse $\Delta p_i^\mu$
and spin kick (change in the spin vector) $\Delta S_i^\mu$.
By focusing on a single worldline,
we can also now compute Compton-like matrix elements $N(k_1,k_2)$
that can be matched to solutions of the Teukolsky equation~\cite{Bautista:2026qse}.
More generic radiative matrix elements involving the full two-body scattering
will find application in matching to EOB-resummed
models of the full two-body dynamics~\cite{Buonanno:1998gg,Buonanno:2024vkx,Buonanno:2024byg}.

Most tantalisingly,
we would now like to perform direct calculations involving the bound
two-body motion, taking full advantage of the SF expansion and the $\hat{N}$ operator.
Previous efforts to learn about bound motion by first focusing on scattering
have been impeded at high perturbative orders by hereditary (tail) effects
\cite{Bini:2024tft,Dlapa:2024cje,Dlapa:2025biy},
in which back-reacted radiation re-enters the two-body system ---
implying a non-local-in-time underlying two-body Hamiltonian.
This prevents direct analytic continuation between bound and unbound
gauge-invariant quantities, such as the radial action, after integration.
However, by showcasing the link between $\hat{N}$-matrix elements at the \emph{integrand level},
we can now expose the correspondence prior to integration,
thus remaining agnostic to background trajectories.
Our final task is then to directly integrate over the bound trajectories,
thereby bypassing the obstruction that tails pose to an analytic continuation
---
an exciting prospect that we leave for the future.

\section*{Acknowledgments}

The authors would like to thank Dyako Arif, Gustav Jakobsen, Jung-Wook Kim,
Sangmin Lee, Donal O'Connell, Raj Patil, Jan Plefka and Canxin Shi for useful comments and discussions.
A special thanks also goes to Andreas Brandhuber, Graham Brown,
Paolo Pichini, Gabriele Travaglini and Pablo Vives Matasan
for collaboration on related topics and advance sight of their paper~\cite{Brandhuber:2025igz}.
This work is supported by The Royal Society under grant URF\textbackslash R1\textbackslash 231578,
``Gravitational Waves from Worldline Quantum Field Theory'',
and grant RF\textbackslash ERE\textbackslash 231084.

\appendix

\section{Worldline deflection algebra and background brackets}
\label{worldlinebrackets}

In this appendix we clarify the relation between two bracket structures used in the main text: the canonical algebra of the fluctuation operators $z_i^\mu,p_i^{\prime\mu}$, introduced in \sec{sec:quantisation}, and the background Poisson bracket \eqref{poissonBG} acting on $\bar x_i^\mu,\bar p_i^\mu$. The relation between them is not purely kinematical; it depends on how the interaction Hamiltonian $\hat H_I$ treats the split
\begin{align}
    x_i^\mu&=\bar x_i^\mu+z_i^\mu\,,&
    p_i^\mu&=\bar p_i^\mu+p_i^{\prime\mu}\,.
\end{align}
If $\hat H_I$ depends on the worldline variables only through the full combinations $x_i^\mu$ and $p_i^\mu$, then a fluctuation derivative can be traded for a derivative with respect to the corresponding background variable.
We refer to this as the \emph{background split symmetry}.
If instead the Hamiltonian depends separately on the background and fluctuation variables,
this identification no longer holds.
Since the Magnus operator is constructed from nested commutators of $\hat H_I$ as in \eqn{eq:MagnusTerms},
\begin{align}
    \hat N=\hat N[\hat H_I],
\end{align}
it inherits the same dependence on the split variables. The question addressed
below is therefore whether the $z_i,p_i'$ dependence of $\hat N$ can be
absorbed into a shift of the background data $\bar x_i,\bar p_i$. This is true
in the PL expansion, but not in the SF expansion.

\paragraph{PL expansion}

The worldline interaction Hamiltonian in eq.~\eqref{eq:HintA_def} is
\begin{align}
	H^{(\rm PL)}_{I,i}(\tau_i)=e_i\,\varphi(\bar{x}_i+z_i)\,.
\end{align}
Thus the PL interaction depends on the split variables only through the full worldline coordinate $x_i^\mu=\bar x_i^\mu+z_i^\mu$. This implies that
\begin{align}
    \frac{\partial H_{I,i}^{(\rm PL)}}{\partial z_i^\alpha}
    =
    \frac{\partial H_{I,i}^{(\rm PL)}}{\partial \bar x_i^\alpha}.
    \label{eq:split_symmetry}
\end{align}
In theories such as gravity, where the coupling also depends on the momentum,
the analogous statement also holds for $p_i^\mu=\bar p_i^\mu+p_i^{\prime\mu}$.

Given the straight-line background and free deflection operator,
\begin{align}
    \bar x_i^\mu(\tau_i)
    &=
    b_i^\mu+\frac{\tau_i}{m_i}\bar p_i^\mu\,, &
    \hat z_i^\mu(\tau_i)
    &=
    \hat b_i^{\prime\mu}
    +
    \frac{\tau_i}{m_i}\hat p_i^{\prime\mu}\,,
\end{align}
the full position operator is then
\begin{align}
    \hat x_i^\mu(\tau_i)
    =
    b_i^\mu+\hat b_i^{\prime\mu}
    +
    \frac{\tau_i}{m_i}
    \left(
        \bar p_i^\mu+\hat p_i^{\prime\mu}
    \right)\,.
\end{align}
Therefore, after projecting only on bulk states,
the Magnus operator depends on the worldline operators only through the shifted data $b_i+\hat b'_i$ and $\bar p_i+\hat p'_i$.
It also follows that commutators with worldline
fluctuations are reproduced by background derivatives:
\begin{align}
\begin{aligned}
    \frac{1}{i\hbar}
    \left[
            \hat N^{(\rm PL)},\hat z_i^\alpha(\tau)
        \right]
    &=
    \frac{1}{i\hbar}
    \left[
            \hat N^{(\rm PL)},\hat b_i^{\prime\alpha}
        \right] 
    +
    \frac{\tau}{m_i}
    \frac{1}{i\hbar}
    \left[
            \hat N^{(\rm PL)},\hat p_i^{\prime\alpha}
        \right] 
    \\
    &=
    -\frac{\partial\,\hat N^{(\rm PL)} }
    {\partial \bar p_{i,\alpha}}
    +
    \frac{\tau}{m_i}
    \frac{\partial\, \hat N^{(\rm PL)} }
    {\partial b_{i,\alpha}}
    \\
    &=
    \{
    \hat N^{(\rm PL)},
    \bar x_i^\alpha(\tau)
    \}.
    \label{eq:PL_comm_to_bg_bracket}
\end{aligned}
\end{align}
Thus, any matrix element of $\hat{N}^{(\rm PL)}$ will obey this property.
Hence, in the PL expansion bulk-projected commutators such as $[\hat N_\varphi,\hat b'_{i,\alpha}]$ and $[\hat N_\varphi,\hat p'_{i,\alpha}]$ are not independent basis elements: their coefficients are fixed by background derivatives of the corresponding projected Magnus elements.

\paragraph{SF expansion}

In the SF expansion the same identification is no longer possible in general.
Although one still decomposes $x_i^\mu=\bar x_i^\mu+z_i^\mu$,
the fluctuation Hamiltonian is obtained by expanding around a non-trivial background solution where the light-body trajectory $\bar x_L^\mu$ is the orbit in the field sourced by the heavy body. 
As a result, the SF Hamiltonian contains explicit dependence on the background field and on the expansion point, not only on the full combination $\bar x_i^\mu+z_i^\mu$.
This is apparent in the worldline interaction Hamiltonians~\eqref{eq:HintB_def}:
\begin{subequations}
\begin{align}
H_{I,L}^{(\rm SF)}(\tau_L) &= e_L
\Big[ e^{z_L\cdot\partial}\varphi(\bar{x}_L) + \bigl( e^{z_L\cdot\partial}\bar{\phi}(\bar{x}_L) -\bar{\phi}(\bar{x}_L) - z_L \cdot \partial \bar{\phi}(\bar{x}_L) \bigr) \Big]\,,\\
H_{I,H}^{(\rm SF)}(\tau_H) &= e_H\Big[ e^{z_H\cdot\partial}\varphi(\bar{x}_H)-\varphi(\bar{x}_H) \Big]\,.
\end{align}
\end{subequations}
Thus, differentiation with respect to $z_i$ is not
equivalent to differentiation with respect to $\bar x_i$.
More explicitly,
\begin{subequations}
\begin{align}
\frac{\partial H_{I,L}^{(\rm SF)}}{\partial z_L^\alpha} -
\frac{\partial H_{I,L}^{(\rm SF)}}{\partial \bar x_L^\alpha}
&=
e_Lz_L^\mu\partial_\alpha\partial_\mu\bar\phi(\bar x_L)\neq0\,,
\\
\frac{\partial H_{I,H}^{(\rm SF)}}{\partial z_H^\alpha} -
\frac{\partial H_{I,H}^{(\rm SF)}}{\partial \bar x_H^\alpha}
&=
e_H\partial_\alpha\varphi(\bar x_H)\neq0\,.
\end{align}
\end{subequations}
This is in contrast with the PL split-symmetry identity \eqref{eq:split_symmetry}. 

This has a direct consequence for the choice of matrix-element basis.
In the PL expansion, matrix elements of $[\hat N,\hat b'_{i,\alpha}]$
and $[\hat N,\hat p'_{i,\alpha}]$
are fixed by background-field derivatives of $\hat N$.
In the SF expansion this is no longer true in general,
so these commutator-projected quantities must be included as
independent worldline-deflection matrix elements.
At 1SF this applies in particular to the heavy line.
The reduced SF Hamiltonian contains
\begin{align}
    e_H\int\!{\rm d}\tau_H\,\delta^D(x-\bar x_H)\,
    z_H^\alpha\partial_\alpha\varphi(x),
\end{align}
while the reduced light-worldline sector contains no corresponding explicit $z_L$ vertex.
In order to take matrix elements of the term above,
we therefore take brackets with $\hat b'_{H,\alpha}$ and $\hat p'_{H,\alpha}$
before projecting on external states.
The necessary matrix elements were accordingly introduced in \sec{sec:Magnus_caseB}.

\section{Example WQFT-Magnus calculation}
\label{exampleMagnus}

To illustrate how the Magnus series is used
to explicitly produce $\hat{N}$-matrix elements at sub-sub-leading order,
we compute the $e_1^2e_2$ component of $N^{(3)}(k)$ in the PL expansion.
This example highlights how the Magnus expansion
grows in complexity at higher perturbative orders,
and the convenience of instead using Murua coefficients in combination with
Feynman rules to directly assemble tree-level $\hat{N}$-matrix elements.
The steps taken here are closely inspired by \rcite{Brandhuber:2025igz}.

Starting with \eqn{eq:Magnus3} we have
\begin{align}
	\left.i\hat N^{(3)}\right|_{e_1^2e_2}&=\frac{(-i)^3}{6}\int\d^Dx \d^Dy \d^Dz\,(\theta_{xy}\theta_{yz}+\theta_{zy}\theta_{yx})
	\big([\hat\cH_{I,1}(x),[\hat\cH_{I,1}(y),\hat\cH_{I,2}(z)]]\nn\\
	&\qquad+
	[\hat\cH_{I,1}(x),[\hat\cH_{I,2}(y),\hat\cH_{I,1}(z)]]+
	[\hat\cH_{I,2}(x),[\hat\cH_{I,1}(y),\hat\cH_{I,1}(z)]]\big)\, .
\end{align}
First, we integrate out $x$, $y$ and $z$ by inserting
$\hat{\cH}_{I,i}(x)=\int\d\tau\,\delta^D(x-\bar{x}_i(\tau))\hat H_{I,i}(\tau)$.
We are left with
\begin{align}\label{N3exp}
	&\left.i\hat N^{(3)}\right|_{e_1^2e_2}=\frac{(-i)^3}{6}\int\d\tau_A\d\tau_B\d\tau_C\,
	(\theta_{AB}\theta_{BC}+\theta_{CB}\theta_{BA})
	\big([\hat{H}_{I,1}(\tau_A),[\hat{H}_{I,1}(\tau_B),\hat{H}_{I,2}(\tau_C)]]\nn\\
	&\qquad\qquad+
	[\hat{H}_{I,1}(\tau_A),[\hat{H}_{I,2}(\tau_B),\hat{H}_{I,1}(\tau_C)]]+
	[\hat{H}_{I,2}(\tau_A),[\hat{H}_{I,1}(\tau_B),\hat{H}_{I,1}(\tau_C)]]\big)\,,
\end{align}
where $\theta_{AB}=\theta(t_A-t_B)$.
When a step function $\theta_{AB}$
connects two points on the same worldline $i=1,2$,
then $\theta(t_A-t_B)=\theta(\bar{x}^0_i(\tau_A)-\bar{x}^0_i(\tau_B))=
\theta(\tau_A-\tau_B)$.
This allows us to recover retarded worldline propagators~\eqref{wlProps}.

Plugging in the relevant contributions to the interaction Hamiltonian~\eqref{eq:HintA_def},
the three contributions in eq.~\eqref{N3exp}
carrying $\theta_{AB}\theta_{BC}$ are determined to be
\begin{subequations}
\begin{align}
	(1)&=\,\theta_{AB}\theta_{BC}[\hat{H}_{I,1}(\tau_A),[\hat{H}_{I,1}(\tau_B),\hat{H}_{I,2}(\tau_C)]]\\
	&=e_1^2e_2\Delta_{1,R}^{\mu\nu}(\tau_A-\tau_B)
	\partial_\mu\Delta_R(\bar{x}_1(\tau_B)-\bar{x}_2(\tau_C))
	\partial_\nu\hat\varphi(\bar{x}_1(\tau_A))+\cdots\,,\nn\\
	(2)&=\,\theta_{AB}\theta_{BC}[\hat{H}_{I,1}(\tau_A),[\hat{H}_{I,2}(\tau_B),\hat{H}_{I,1}(\tau_C)]]\\
	&=-e_1^2e_2\theta_{AB}\Delta_{1,R}^{\mu\nu}(\tau_A-\tau_C)
	\partial_\mu\Delta_R(\bar{x}_2(\tau_B)-\bar{x}_1(\tau_C))
	\partial_\nu\hat\varphi(\bar{x}_1(\tau_A))+\cdots\,,\nn\\
	(3)&=\,\theta_{AB}\theta_{BC}[\hat{H}_{I,2}(\tau_A),[\hat{H}_{I,1}(\tau_B),\hat{H}_{I,1}(\tau_C)]]\\
	&=e_1^2e_2\Delta_{1,R}^{\mu\nu}(\tau_B-\tau_C)
	\big[
	-\partial_\mu\Delta_R(\bar{x}_2(\tau_A)-\bar{x}_1(\tau_B))
	\partial_\nu\hat\varphi(\bar{x}_1(\tau_C))\nn\\
	&\qquad\qquad\qquad\qquad
	-\theta_{AB}\partial_\mu\Delta_R(\bar{x}_2(\tau_A)-\bar{x}_1(\tau_C))
	\partial_\nu\hat\varphi(\bar{x}_1(\tau_B))\big]+\cdots\,.\nn
\end{align}
\end{subequations}
While in each case we obtained retarded propagators from brackets between fields,
in instances (2) and (3) we needed to insert
$\theta_{AB}\theta_{BC}=\theta_{AB}\theta_{BC}\theta_{AC}$.
This is valid because $t_A>t_B$ and $t_B>t_C$ implies $t_A>t_C$.
Thus, we have a leftover $\theta_{AB}$
that spoils Lorentz invariance of our final answer.
To cancel it,
we relabel the second line of (3) with $\tau_A\leftrightarrow\tau_B$,
allowing us to insert $\theta_{AB}+\theta_{BA}=1$ when combined with (2):
\begin{align}
	(1)+(2)+(3)=
	e_1^2e_2\big[\Delta_{1,R}^{\mu\nu}(\tau_A-\tau_B)
	\partial_\mu\Delta_R(\bar{x}_1(\tau_B)-\bar{x}_2(\tau_C))
	\partial_\nu\hat\varphi(\bar{x}_1(\tau_A))&\\
	-
	\Delta_{1,R}^{\mu\nu}(\tau_B-\tau_C)
	\partial_\mu\Delta_R(\bar{x}_2(\tau_A)-\bar{x}_1(\tau_B)
	\partial_\nu\hat\varphi(\bar{x}_1(\tau_C))&\nn\\
	-
	\Delta_{1,R}^{\mu\nu}(\tau_A-\tau_C)
	\partial_\mu\Delta_R(\bar{x}_2(\tau_B)-\bar{x}_1(\tau_C)
	\partial_\nu\hat\varphi(\bar{x}_1(\tau_A))&\big]+\cdots\,.\nn
\end{align}
The other three terms in eq.~\eqref{N3exp} carrying $\theta_{CB}\theta_{BA}$ give the same result but with causality flow reversed,
i.e.~advanced propagators in place of retarded.

Putting everything together, we have
\begin{align}
\begin{aligned}
	\left.i\hat N^{(3)}\right|_{e_1^2e_2}&=
	\frac{(-i)^3}{6} e_1^2e_2 \int\d\tau_A\d\tau_B\d\tau_C\,
	\partial_\mu\hat\varphi(\bar{x}_1(\tau_A))\\
	&\qquad\qquad\times\big[
		2\Delta_{1,R}^{\mu\nu}(\tau_A-\tau_B)
		\partial_\nu\Delta_R(\bar{x}_1(\tau_B)-\bar{x}_2(\tau_C))
		\\&\qquad\qquad\,\,+
		2\Delta_{1,A}^{\mu\nu}(\tau_A-\tau_B)
		\partial_\nu\Delta_A(\bar{x}_1(\tau_B)-\bar{x}_2(\tau_C))
		\\&\qquad\qquad\,\,\,+
		\Delta_{1,R}^{\mu\nu}(\tau_A-\tau_B)
		\partial_\nu\Delta_A(\bar{x}_1(\tau_B)-\bar{x}_2(\tau_C))
		\\&\qquad\qquad\,\,\,+
		\Delta_{1,A}^{\mu\nu}(\tau_A-\tau_B)
		\partial_\nu\Delta_R(\bar{x}_1(\tau_B)-\bar{x}_2(\tau_C))
	\big]+\cdots \, .
\end{aligned}
\end{align}
The weightings of retarded and advanced propagators correspond
exactly with those obtained from using the Murua coefficients~\eqref{murua3pt}.
Taking the matrix element $iN^{(3)}(k)=\langle k|i\hat{N}^{(3)}|0\rangle$,
and using $\langle k|\partial_\mu\hat\varphi(x)|0\rangle=ik_\mu e^{ik\cdot x}$,
we recover our result from the main text~\eqref{N3e1e1e2Explicit} that
was derived using Feynman rules and Murua coefficients.

\bibliographystyle{JHEP}
\bibliography{refs.bib}

\end{document}